\pdfoutput=1
\documentclass[final, journal, doublecolumn, 10pt, letterpaper]{IEEEtran}
\usepackage{amsmath,amssymb,amsfonts}
\usepackage{amsbsy}
\usepackage{graphicx}
\usepackage{graphics}
\usepackage{algorithm}
\usepackage[noend]{algorithmic}
\usepackage{subfigure}
\usepackage{cite}
\usepackage{slashbox}
\usepackage{multirow}
\usepackage{pdfsync}
\usepackage{epstopdf}
\usepackage{color}
\usepackage{enumerate}
\usepackage{paralist}
\usepackage{cite}
\usepackage{textcomp}
\usepackage{wrapfig}
\usepackage{lipsum}
\usepackage{hyperref}
\usepackage{boldline}
\usepackage{balance}

\DeclareMathAlphabet\mathbfcal{OMS}{cmsy}{b}{n}
\DeclareMathOperator{\sgn}{sgn}
 \DeclareMathOperator*{\argmin}{argmin}
\newtheorem{theorem}{{Theorem}}

\newtheorem{example}{{Example}}
\newcommand{\Rbb}{\mathbb{R}}
\newcommand{\V}{{\cal V}}
\newcommand{\E}{{\cal E}}
\newcommand{\G}{{\cal G}}
\renewcommand{\l}{\ell}
\renewcommand{\L}{{\mathbfcal L}}

\newcommand{\Indicator}{1\!\!1}
\definecolor{lightkhaki}{RGB}{250,250,210}
\definecolor{darkkhaki}{RGB}{79,79,47}
\definecolor{darkkhaki}{RGB}{139,129,76}
\newcommand{\change}[1]{{\color{black}#1}}
\newcommand{\lpipe}{\rule[-16ex]{0.7pt}{70ex}}

\begin{document}

\title{{\huge Localized Spectral Graph Filter Frames}  \\ \emph{\Large A Unifying Framework, Survey of Design Considerations, and Numerical Comparison \\ \vspace{-.16in} (Extended Cut)} } 
\author{David I Shuman \\
Macalester College, Department of Mathematics, Statistics, and Computer Science \\ 
dshuman1@macalester.edu
\thanks{MATLAB code for all figures and numerical experiments in this paper is available at \url{http://www.macalester.edu/\textasciitilde dshuman1/publications.html}. }}

\maketitle

\thispagestyle{empty}

\begin{abstract}
Representing data residing on a graph as a linear combination of building block signals can enable efficient and insightful visual or statistical analysis of the data, and such representations prove useful as regularizers in signal processing and machine learning tasks. Designing collections of building block signals -- or more formally, dictionaries of atoms -- that specifically account for the underlying graph structure as well as any available representative training signals has been an active area of research over the last decade. In this article, we survey a particular class of dictionaries called localized spectral graph filter frames, whose atoms are created by localizing spectral patterns to different regions of the graph. After showing how this class encompasses a variety of approaches from spectral graph wavelets to graph filter banks, we focus on the two main questions of how to design the spectral filters and how to select the center vertices to which the patterns are localized. Throughout, we emphasize computationally efficient methods that ensure the resulting transforms and their inverses can be applied to data residing on large, sparse graphs. We demonstrate how this class of transform methods can be used in signal processing tasks such as denoising and non-linear approximation, and provide code for readers to experiment with these methods in new application domains. 
\end{abstract}

\section{Introduction: Dictionaries of Graph Signals}

A major line of work in graph signal processing \cite{shuman2013emerging,ortega2018graph} over the past ten years has been to design new transform methods that account for the underlying graph structure in order to identify and exploit structure in data residing on a connected, weighted, undirected graph. The most common approach is to construct a dictionary of atoms (building block signals), and represent the graph signal of interest as a linear combination of these atoms. Such representations enable visual analysis of data, statistical analysis of data, and data compression, and can also be leveraged as regularizers in machine learning and ill-posed inverse problems such as inpainting, denoising, and classification.

In general, desirable properties when designing dictionaries for graph signals include: (i) the atoms have an interpretable form that accounts for the underlying graph structure, so that the inner products between a graph signal and each atom are informative; (ii) the dictionary comprises an orthonormal basis or tight frame for the signal space, so that the contribution of each atom can be computed via an inner product with the graph signal, and the energy of the graph signal is equal to a constant multiple of the energy of the transform coefficients; (iii) it is numerically efficient to apply the dictionary analysis and synthesis operators (forward and inverse transforms); and (iv) signals of certain mathematical classes can be represented exactly or approximately as \emph{sparse} linear combinations of a subset of the dictionary atoms. 

By our count, approximately 100 conference and journal articles written in the last decade have introduced new dictionaries for graph signals. These include designs for analytic dictionaries that are adapted to the graph structure but not any specific training data, as well as techniques for learning dictionaries from training data. Some broad classes of dictionaries include graph Fourier transforms; windowed graph Fourier transforms (e.g., \cite{shuman2015vertex}); vertex domain designs including spatial wavelets (e.g., \cite{Crovella2003,wang}), hierarchical trees (e.g., \cite{gavish}), lifting transforms (e.g., \cite{narang_lifting_graphs}), and top-down approaches (e.g., \cite{szlam,irion}); diffusion-based designs (e.g., \cite{diffusion_wavelets}); spectral domain designs (e.g., \cite{hammond2011wavelets}); pyramid transforms (e.g., \cite{shuman_TSP_multiscale}); and generalized filter banks (e.g., \cite{narang_icip,narang_bipartite_prod,ekambaram2013critically,li_mcsfb_2018}). Despite, or perhaps because of, the number of new dictionary designs for graph signals, 
it remains difficult to identify which dictionary might be best suited for a specific task, or to understand subtle qualitative tradeoffs when specifying the parameters of a given dictionary construction.

%
%
%
%
%
%

\begin{figure*}[tb]
\centering
\begin{minipage}{.18\linewidth}
\centering
\includegraphics[width=.9\linewidth,page=1]{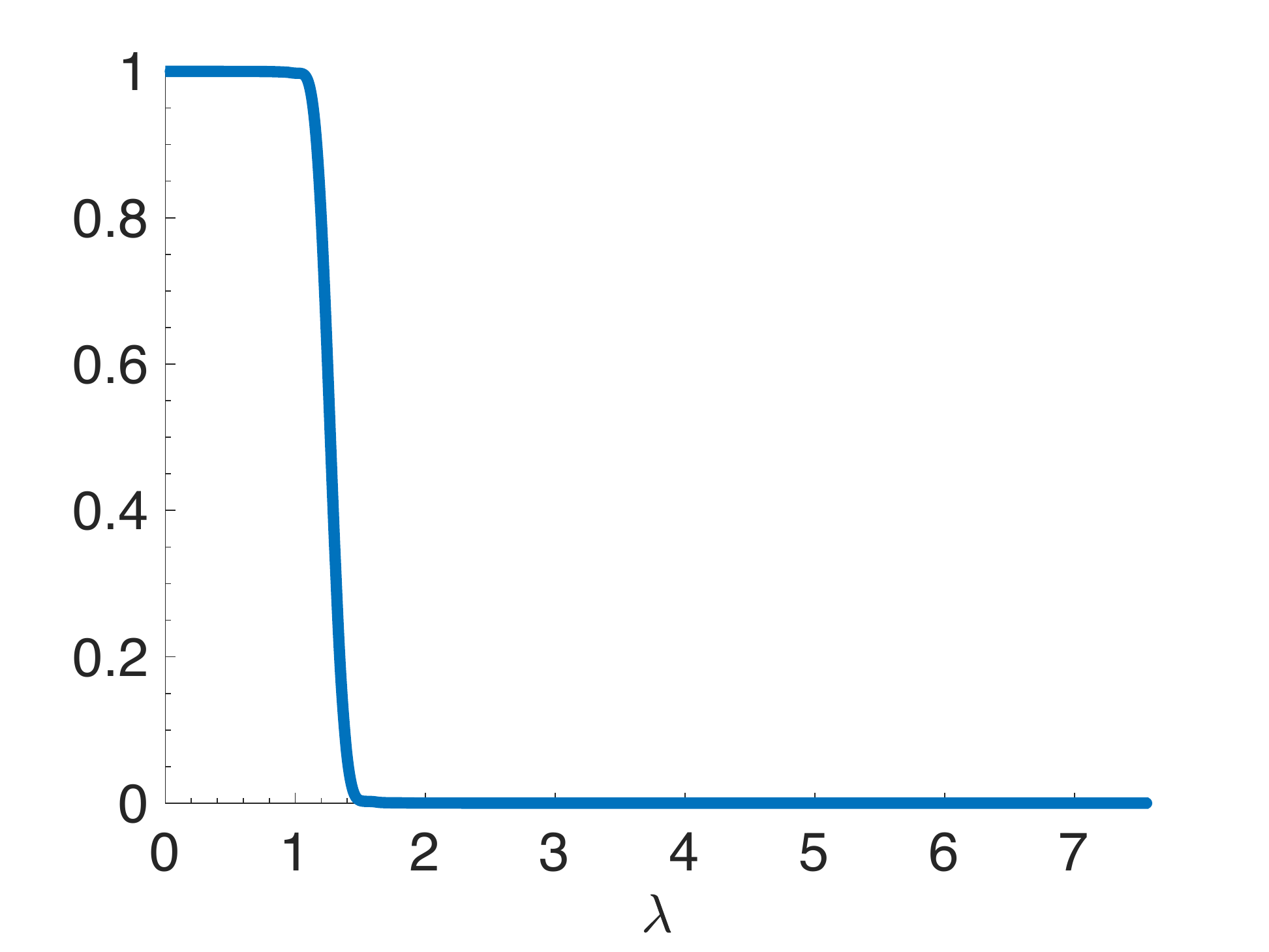} \\
\vspace{.1in}
 
\includegraphics[width=.9\linewidth,page=1]{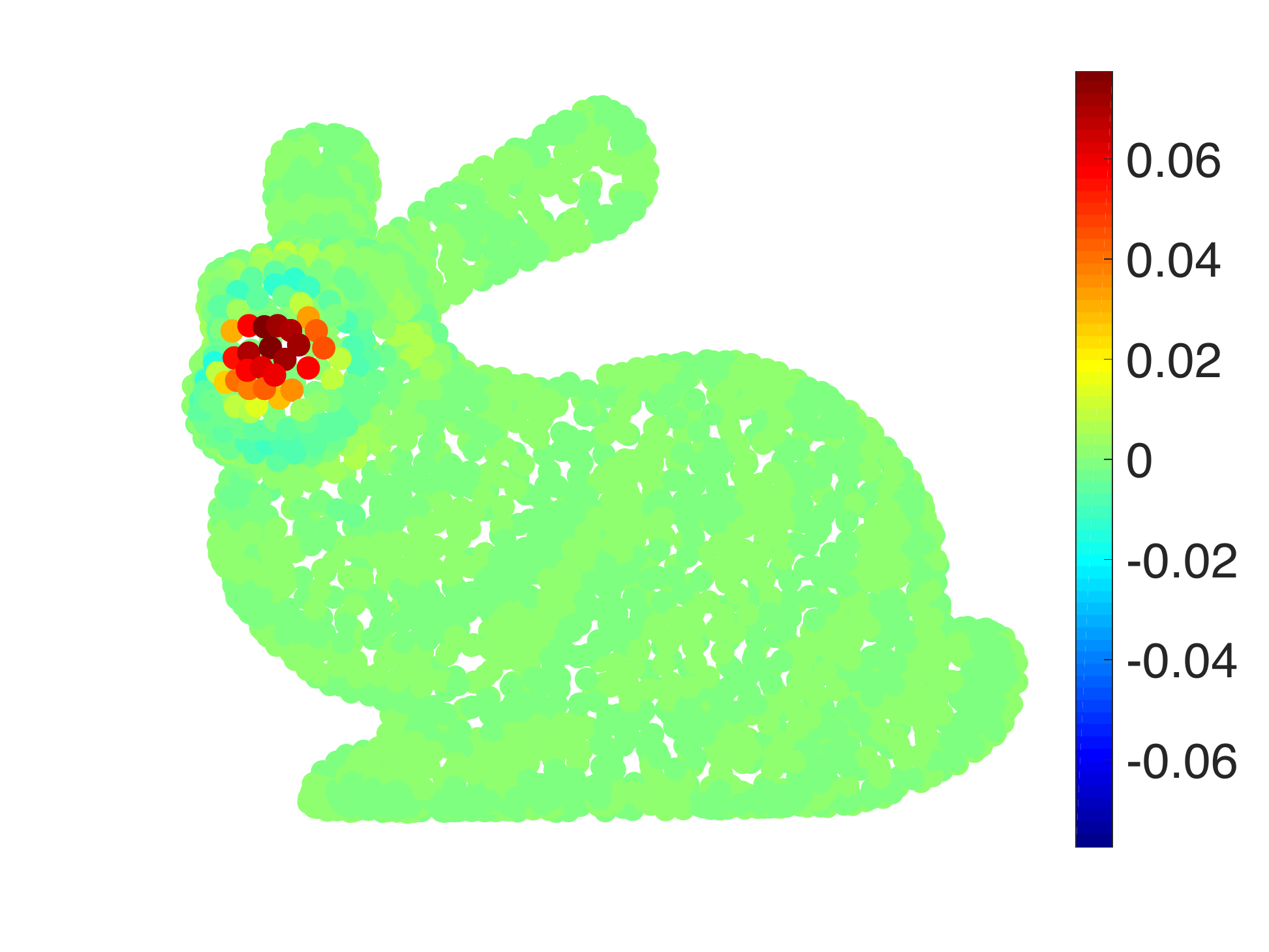} 
{\small{(a)~}}
\end{minipage}
\hspace{.001in}
\begin{minipage}{.18\linewidth}
\centering
\includegraphics[width=.9\linewidth,page=2]{figures/filters} \\
\vspace{.1in}
 
\includegraphics[width=.9\linewidth,page=2]{figures/atoms} 
{\small{(b)~}}
\end{minipage}
\hspace{.001in}
\begin{minipage}{.18\linewidth}
\centering
\includegraphics[width=.9\linewidth,page=3]{figures/filters} \\
\vspace{.1in}
 
\includegraphics[width=.9\linewidth,page=3]{figures/atoms} 
{\small{(c)~}}
\end{minipage}
\hspace{.001in}
\begin{minipage}{.18\linewidth}
\centering
\includegraphics[width=.9\linewidth,page=3]{figures/filters} \\
\vspace{.1in}
 
\includegraphics[width=.9\linewidth,page=4]{figures/atoms} 
{\small{(d)~}}
\end{minipage}
\hspace{.001in}
\begin{minipage}{.18\linewidth}
\centering
\includegraphics[width=.9\linewidth,page=3]{figures/filters} \\
\vspace{.1in}
 
\includegraphics[width=.9\linewidth,page=5]{figures/atoms} 
{\small{(e)~}}
\end{minipage}
\caption{Localized spectral graph filter frame atoms. (a)-(c) Three different filters/patterns localized to the same center vertex. (c)-(e) The same filter/pattern localized to three different center vertices.}\label{Ex:atoms}
\end{figure*}

In this survey, we restrict our attention to \emph{localized spectral graph filter frames}, whose atoms are created by localizing patterns (spectral filters) to different regions of the graph. The seminal example of a dictionary of such atoms is spectral graph wavelets\cite{hammond2011wavelets}. However,  
localized spectral graph filter frames are \emph{broad in their scope}, including more recently proposed methods such as single-level filter banks for graph signals \cite{narang_bipartite_prod,narang_bior_filters,ekambaram2013critically,tay2015design,chen2015discrete,sakiyama2016spectral,anis2017critical,tay2017critically,tay2017almost,teke2016extending,li_mcsfb_2018,kotzagiannidis2017splines}, variational or interpolating splines \cite{pesenson_splines,erb2019graph}, frames adapted to training data \cite{thanou_learning_TSP_2014,behjat2016signal}, frame constructions for general graph signals \cite{leonardi_multislice,shuman2013spectrum,gobel2018construction,dong2017sparse}, frame constructions tailored to specific applications such as fMRI data analysis \cite{behjat2013statistical,behjat2014canonical,behjat2015anatomically} or community mining \cite{tremblay2014graph}, ``natural'' wavelets \cite{saitoNatural}, and even some vertex domain constructions \cite{Crovella2003,wang}.\footnote{While diffusion wavelets \cite{diffusion_wavelets} likely inspired many of these dictionaries, diffusion wavelets do not technically fit into the localized spectral graph filter dictionary framework outlined in Sec. \ref{Se:dict} due to the additional step of numerically orthogonalizing the atoms.}  \change{Our motivations for examining these dictionaries include (i) the design framework is flexible - it can yield highly redundant dictionaries to sparsely represent graph signals or new bases to efficiently extract structure from data on graphs, and it can also incorporate representative training signals when they are available; 
(ii) the atoms have a physically interpretable structure and their closed form 
definition opens the door to formal mathematical analysis;
and (iii) fast numerical approximations exist to efficiently apply these dictionary transforms and their inverses to data residing on large, sparse graphs, which are increasingly common in signal processing and machine learning applications. Due to the multiscale and localized structure of their atoms, these dictionaries are particularly relevant for applications where interesting phenomena are expressed in discontinuities or quick changes in signal values in smaller regions of the graph, analogous to edges in images.} 

The organization of the article is as follows. In the next section, we detail a unifying framework for localized spectral graph filter dictionaries. We survey the key design considerations for this class of dictionaries in Section \ref{Se:filters} (design of the spectral filters) and Section \ref{Se:centers} (selection of the center vertices). In Section \ref{Se:theory}, we review recent work 
on theoretical considerations and metrics that inform the design of these dictionaries. In Sections \ref{Se:experiments} and \ref{Se:conclusion}, we pose specific high-level questions that get to the heart of the ``where do we start when specifying a dictionary?'' issue, attempt to answer these questions with new numerical comparisons and qualitative insights, and discuss how these comparisons inform future work in the area. 

\section{Dictionaries of Localized Spectral Patterns}\label{Se:dict}
Keeping with the notation of \cite{shuman2013emerging}, we consider data residing on a connected, weighted, undirected graph $\G = \{\V, \E, {\bf W}\}$ characterized by a finite set of vertices $\V$ with $|\V|=N$, a set of edges $\E$, and a weighted adjacency matrix ${\bf W}$.
A signal or function $f: \V \rightarrow \Rbb$ defined on the vertices of the graph may be represented as a vector ${\bf f} \in \Rbb^N$, where the $i$th element of the vector ${\bf f}$ represents the graph signal value at vertex $i$ in $\V$.

The dictionaries we consider feature atoms of the form 
\begin{eqnarray} \label{Eq:atom_form}
{\boldsymbol \varphi}_{i,j} :=T_i g_j := \hat{g}_j({\L}){\boldsymbol \delta}_i = {\bf U}\hat{g}_j({\boldsymbol \Lambda}){\bf U}^{*} {\boldsymbol \delta}_i.
\end{eqnarray}
In \eqref{Eq:atom_form},  ${\boldsymbol \delta}_i$ is a graph signal with a value of 1 at vertex $i$ and 0 elsewhere, ${\L}={\bf D}-{\bf W}={\bf U}{\boldsymbol \Lambda}{\bf U}^{*}$ is the (combinatorial) graph Laplacian,  
the columns of ${\bf U}$ are the orthogonal eigenvectors of ${\L}$, the $^*$ symbol denotes conjugate transpose, and ${\boldsymbol \Lambda}$ is a diagonal matrix whose $\l$th diagonal element $\lambda_{\l}$ is the eigenvalue of ${\L}$ associated with the eigenvector ${\bf u}_{\l}$, the $\l$th column of ${\bf U}$.\footnote{While we use the combinatorial (non-normalized) graph Laplacian ${\L}$ throughout, the ideas we discuss apply to dictionaries comprised of atoms of the form \eqref{Eq:atom_form} with the graph Fourier basis ${\bf U}$ chosen as the eigenvectors of other symmetric generalized graph Laplacian operators such as the normalized graph Laplacian ${\L}_{\hbox{norm}}={\bf D}^{-\frac{1}{2}}\L {\bf D}^{-\frac{1}{2}}$.} Each \emph{spectral graph filter} or \emph{kernel} $\hat{g}_j(\cdot)$ is a function from $\sigma(\L)=\{\lambda_0,\lambda_1,\ldots,\lambda_{N-1}\}$, the set of Laplacian eigenvalues, to the real numbers. Thus, $\hat{g}_j({\boldsymbol \Lambda})$ is a diagonal matrix with the $\l$th diagonal entry equal to $\hat{g}_j(\lambda_\l)$. In practice, these filter functions are often defined on the continuous range $[0,\bar{\lambda}]$, where $0=\lambda_0 \leq \lambda_1 \leq \ldots \leq \lambda_{N-1}=\lambda_{\max} \leq \bar{\lambda}$.

At a more intuitive level, we can think of each spectral graph filter  $\hat{g}_j(\cdot)$ as defining a spectral pattern that is localized to different regions of the graph, with vertex $i$ being the center of the localized pattern ${\boldsymbol \varphi}_{i,j}=T_i g_j$. As an extreme example, if the spectral pattern is $\hat{g}(\lambda_{\l})=1$ for all $\lambda_{\l}$, then the localized pattern centered at vertex $i$ is $T_i g={\bf U}{\bf U}^{*} {\boldsymbol \delta}_i={\boldsymbol \delta}_i$. 
Fig. \ref{Ex:atoms} displays more localized spectral patterns on the Stanford bunny graph \cite{bunny}. 
 
We refer to a collection of 
atoms of the form \eqref{Eq:atom_form} as a \emph{localized spectral graph filter dictionary} (LSGFD), denoted by 
\begin{align}\label{Eq:LSGFD}
{\cal D}=\{T_i g_j\}_{j=1,2,\ldots,J; i\in \V_j}.
\end{align}
In \eqref{Eq:LSGFD}, $\V_j \subseteq \V$ is the set of center vertices to which the $j$th spectral pattern $\hat{g}_j(\cdot)$ is localized, \change{and each atom ${\boldsymbol \varphi}_{i,j} =T_i g_j$ is a graph signal in $\Rbb^N$.} Therefore, to fully specify a  LSGFD ${\cal D}$, we need to answer the following questions, which are the focus of the next two sections, respectively: 
\begin{enumerate} 
\item How many spectral patterns should we use, and what should those patterns be? That is, we must specify the number of filters, $J$, and the form of the filters $\{\hat{g}_1(\cdot),\hat{g}_2(\cdot),\ldots,\hat{g}_J(\cdot)\}$.
\item For each spectral pattern $\hat{g}_j(\cdot)$, how many center vertices should the pattern be localized to, and which vertices should those be? That is, we must specify the sets $\V_j$ for each $j$. 
\end{enumerate}

In specifying the spectral patterns and sets of center vertices for LSGFDs, it is also important to keep in mind (i) what information is available, and (ii) the graph size. In all cases in this survey, we assume the underlying graph structure $\G$ is known, although learning graph structures is a vibrant area of ongoing research (see, e.g., \cite{dong2019learning,mateos2019connecting} and references therein). In terms of data available in the design of the dictionary, there are three possibilities: (i) no data is available (the default unless otherwise specified), (ii) the design of the dictionary atoms may also take into account a set of one or more specific graph signals that are to be analyzed by the dictionary (we refer to the resulting dictionaries as \emph{signal-adapted}), and (iii) a set of training data is available to learn parameters of the LSGFD, but the dictionary is then used to analyze a different set of (presumably similar) graph signals. 

For small to medium sized graphs (say on the order of 10,000 or fewer vertices), the full Laplacian eigendecomposition $\L={\bf U}{\boldsymbol \Lambda}{\bf U}^{*}$ can be computed, and therefore the exact Laplacian eigenvectors and eigenvalues can be used in the dictionary design. For larger graphs, however, it may not be tractable to perform this decomposition, and we therefore put an emphasis in the next two sections on methods that do not require these quantities. Without the Laplacian eigenvectors, we almost always need an estimate of the maximum eigenvalue $\lambda_{\max}$ via, e.g., a few steps of the Lanczos algorithm \cite{lanczos_bound} or a closed form upper bound on it. For example, $\bar{\lambda}$ can be taken to be the maximum sum of the degrees of any two vertices connected by an edge, $\lambda_{\max} \leq \max_{\{(m,n) \in \E\}}\{d(m)+d(n)\}$, where $d(n)$ is the degree of vertex $n$ \cite{anderson_morley}, \cite[Cor. 3.2]{das}. 

\begin{figure}[bt] 
\begin{minipage}[m]{0.32\linewidth}
\centerline{\small{~~~~gnp}}
\centerline{~~\includegraphics[width=1\linewidth,page=1]{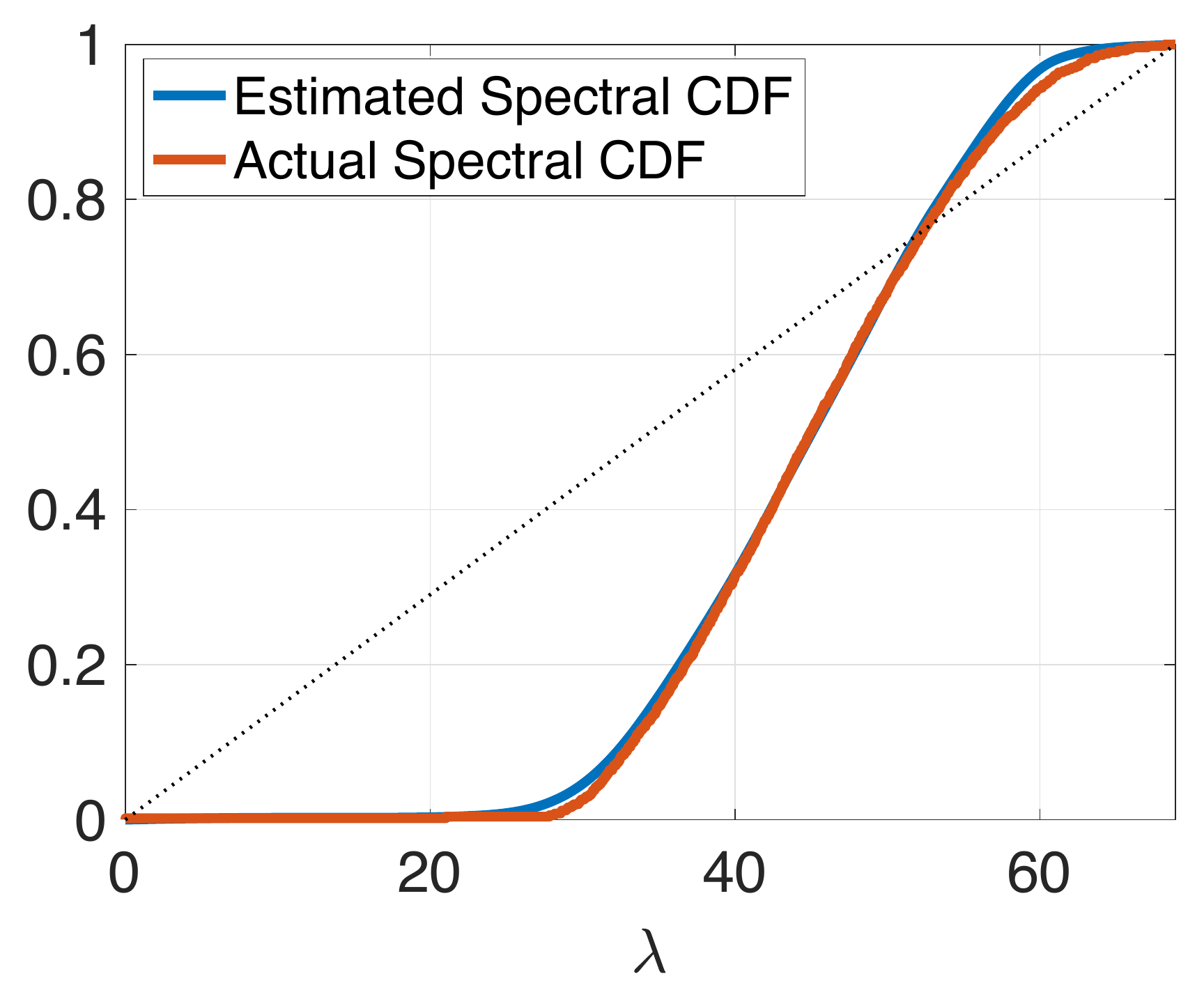}}
\end{minipage}
\begin{minipage}[m]{0.32\linewidth}
\centerline{\small{~~~~minnesota}}
\centerline{~~\includegraphics[width=1\linewidth,page=2]{figures/cdfs}}
\end{minipage}
\begin{minipage}[m]{0.32\linewidth}
\centerline{\small{~~~~net25}}
\centerline{~~\includegraphics[width=1\linewidth,page=3]{figures/cdfs}}
\end{minipage}
\\
\vspace{0.03\linewidth}

\begin{minipage}[m]{0.32\linewidth}
\centerline{\small{~~~~bunny}}
\centerline{~~\includegraphics[width=1\linewidth,page=4]{figures/cdfs}}
\end{minipage} 
\begin{minipage}[m]{0.32\linewidth}
\centerline{\small{~~~~eastern mass}} 
\centerline{~~\includegraphics[width=1\linewidth,page=5]{figures/cdfs}}
\end{minipage}
\begin{minipage}[m]{0.32\linewidth}
\centerline{\small{~~~~cerebellum}} 
\centerline{~~\includegraphics[width=1\linewidth,page=6]{figures/cdfs}}
\end{minipage}
\caption{Estimated and actual cumulative spectral density functions \eqref{Eq:spectral_cdf} for six graph Laplacians: a random Erd\"{o}s-Renyi graph with $N=500$ vertices and edge probability 0.2; the Minnesota traffic network ($N=2642$) \cite{gleich}; the Laplacian of the Andrianov net25 matrix ($N=9520$) from the  SuiteSparse Matrix Collection \cite{suitesparse}; the Stanford bunny graph ($N=2503$) \cite{bunny}; an 8-neighbor local graph for 
Eastern Massachusetts 
($N= 877$) \cite{li_mcsfb_2018}; and a graph of the cerebellum region of the brain ($N=4465$) \cite{behjat2015anatomically}.
}\label{Fig:cdf}
\vspace{-.3in}
\end{figure}

In addition to obtaining a fast estimate for the spectral range $[0,\lambda_{\max}]$, it is often also beneficial to estimate the distribution of the Laplacian eigenvalues over the spectral range.
Specifically, the \emph{cumulative spectral density function} or \emph{empirical spectral cumulative distribution} of $\L$, defined as 
\begin{align}\label{Eq:spectral_cdf}
P_\lambda(z):=\frac{1}{N} \sum_{\l=0}^{N-1} \Indicator_{\{\lambda_\l \leq z\}}, 
\end{align}
can be efficiently estimated by different methods \cite{lin_spectral_density}. We use a variant of the kernel polynomial ethod \cite{silver1996kernel} detailed in \cite[Alg. 2]{li_mcsfb_2018} that leverages Hutchinson's stochastic trace estimator to estimate the number of eigenvalues below linearly spaced points between 0 and $\lambda_{\max}$, and then interpolates these values via monotonic piecewise cubic interpolation to generate an estimate of the cumulative spectral density function \eqref{Eq:spectral_cdf}. The computational cost is proportional to the number of edges in the graph. As an order of magnitude example, for a sparse graph with more than 469,000 vertices, estimates for the maximum eigenvalue and the density function can be computed on a laptop in approximately 1 second and 16 seconds, respectively. 
In Fig. \ref{Fig:cdf}, we show examples of exact and approximate cumulative spectral density functions on six different graphs. 
In summary, while the full Laplacian eigendecomposition is necessary to exactly compute the atoms in \eqref{Eq:atom_form} and their inner products with a graph signal, the spectral range and density function can be computed inexpensively and leveraged in the design of the filters, the selection of the center vertices, and the \emph{approximate} computation of the inner products between the graph signal and each dictionary atom. We discuss these details further in the next two sections.

We represent the \emph{synthesis operator} with the matrix ${\boldsymbol \Phi} \in \Rbb^{N \times M}$, where the columns of ${\boldsymbol \Phi}$ are the \change{$M=\sum_{j=1}^J |\V_j|$} dictionary atoms in ${\cal D}$. We refer to its adjoint ${\boldsymbol \Phi}^*$ as the \emph{analysis operator}; this conjugate transpose matrix maps a graph signal to the analysis coefficients $\{\langle {\bf f}, {\boldsymbol \varphi}_{i,j} \rangle\}$.     
If the dictionary ${\cal D}$ satisfies the \emph{frame condition} for all vectors ${\bf f}$ in some subspace ${\cal S}$ of $\Rbb^N$ (or all of $\Rbb^N$),
\begin{align}\label{Eq:frame}
A ||{\bf f}||_2^2 \leq ||{\boldsymbol \Phi}^* {\bf f}||_2^2 = \sum_{j=1}^{J} \sum_{i\in \V_j}  \left|\langle {\bf f}, {\boldsymbol \varphi}_{i,j} \rangle \right|^2 \leq B ||{\bf f}||_2^2,
\end{align}
then any graph signal in the subspace can be exactly recovered from its analysis coefficients ${\boldsymbol \Phi}^* {\bf f}$. Moreover, if $A=B$ in \eqref{Eq:frame}, the dictionary is said to be a \emph{tight frame}, and
\begin{align*}
{\bf f}=\frac{1}{A} \sum_{j=1}^{J} \sum_{i\in \V_j} \langle {\bf f}, {\boldsymbol \varphi}_{i,j} \rangle  {\boldsymbol \varphi}_{i,j}  = \frac{1}{A}{\boldsymbol \Phi}{\boldsymbol \Phi}^* {\bf f}.
\end{align*}
A tight frame with frame bounds $A=B=1$ is called a \emph{Parseval frame}, and has the added benefit that $||{\boldsymbol \Phi}^* {\bf f}||_2=||{\bf f}||_2$; i.e., the energy of the analysis coefficients is the same as the energy of the graph signal. For more properties of frames, see \cite{kovacevic_frames1,frames}.
 
\begin{figure*}
\centering
\includegraphics[width=6in]{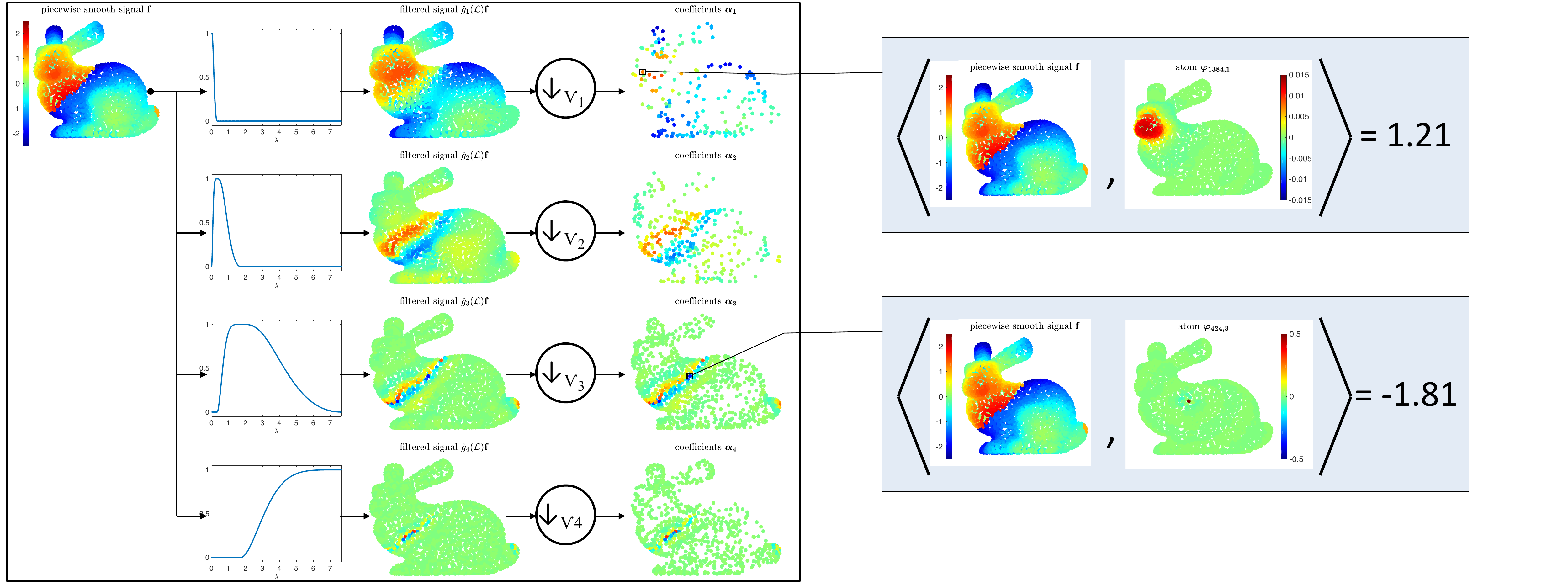}
\caption{\change{Equivalence between single-level graph spectral filter banks and localized spectral graph filter dictionaries. In the 4-channel  filter bank on the left, the graph signal is filtered by each of the $J=4$ spectral filters and then downsampled on the corresponding vertex sets $\{\V_j\}$ to yield the filter bank coefficients $\{{\boldsymbol \alpha}_j\}$. Each coefficient
$\alpha_{i,j}=[\hat{g}_j(\L){\bf f}](i) =  {\boldsymbol \delta}_i^* \hat{g}_j(\L){\bf f}  =  {\bf f}^* \hat{g}_j(\L){\boldsymbol \delta}_i$ in the final column of the left box corresponds exactly to the inner product between the graph signal and the dictionary atom ${\boldsymbol \varphi}_{i,j}$, as
$ \langle {\bf f},{\boldsymbol \varphi}_{i,j} \rangle = \langle {\bf f}, T_i g_j \rangle= \langle {\bf f}, \hat{g}_j(\L){\boldsymbol \delta}_i \rangle = {\bf f}^* \hat{g}_j(\L){\boldsymbol \delta}_i$.}}\label{Fig:equivalence}
\end{figure*}

Finally, we mention the connection between the aforementioned analysis coefficients and graph spectral filter banks. \change{As shown in Fig. \ref{Fig:equivalence},} in a $J$-channel graph filter bank (e.g., \cite{narang_icip,narang_bipartite_prod,ekambaram2013critically,li_mcsfb_2018}), $J$ different filters are applied to the signal, and the values of $\hat{g}_j(\L){\bf f}$, the filtered signal in the $j$th channel,
at a specified set of downsampled vertices $\V_j$ are stored. The set of analysis coefficients $\{\langle{\bf f}, {\boldsymbol \varphi}_{i,j} \rangle \}_{i \in \V_j}$ 
derived from the atoms 
generated by localizing the filter $\hat{g}_j(\cdot)$ to each of the center vertices in $\V_j$ corresponds exactly to the downsampled values in the $j$th channel of the filter bank. 

\section{Design of the Spectral Filters}\label{Se:filters}
\change{Three are three broad classes of spectral filter designs: (i) those adapted only to the spectral range $[0,\lambda_{\max}]$ (e.g., \cite{hammond2011wavelets,narang_bipartite_prod,leonardi_multislice,tay2015design,sakiyama2016spectral,tay2017almost,dong2017sparse,gobel2018construction})\footnote{The tight wavelet frames of \cite{leonardi_multislice} are also adapted to the maximum degree of the graph.}; (ii) those adapted to an estimate of the cumulative spectral density function of the graph (e.g., \cite{shuman2013spectrum,li_mcsfb_2018}); and (iii) those adapted to both the graph and training signals residing on the graph (e.g., \cite{thanou_learning_TSP_2014,behjat2016signal}). 
Examples of all three of these classes are shown in Examples \ref{Ex:filters}-\ref{Ex:poly}.} In this section, we survey a number of considerations when designing the filters for localized spectral graph filter dictionaries. All but the final of the following  considerations apply to all three of the aforementioned classes of filter designs.

\subsection{Localization of the atoms in the vertex domain}
Whether the underlying graph represents a traffic network, a social network, a biological network, or some other type of network, interesting phenomena in the data often occur at a \emph{local scale}, particularly for extremely large graphs. To find or make inferences about such localized patterns, it can be helpful to have dictionary atoms whose energies are concentrated in 
smaller regions of the graph. One method to guarantee that each atom's energy is strictly localized in a small neighborhood of radius $K$ around its center vertex $i$ is to choose the spectral filters to be degree $K$ polynomials. 
\begin{theorem}[Lemma 5.2 of \cite{hammond2011wavelets}, Lemma 2 of \cite{shuman2015vertex}]\label{Th:strict_loc}
Let $d_{\G}(i,n)$ be the \emph{geodesic} or \emph{shortest path} distance between vertices $i$ and $n$; i.e., the minimum number of edges in any path connecting the two vertices. Let $\widehat{p_K}$ be a polynomial kernel with degree $K$; i.e., \begin{align}\label{Eq:poly_kern}
\widehat{p_K}(\lambda)=\sum_{k=0}^K a_k \lambda^k 
\end{align}
for some coefficients $\{a_k\}_{k=0,1,\ldots,K}$. If $d_{\G}(i,n)>K$, then $(T_i p_K)(n) = 0$.
\end{theorem}
More generally, the localization of the dictionary atoms in the vertex (spatial) domain is closely related to the smoothness of the filters. There are multiple ways to measure both ``localization'' and ``smoothness,'' but one is to examine how the magnitude of the localized pattern at vertex $n$ decays as the distance between $n$ and the center vertex $i$ increases, depending on how close the filter $\hat{g}_j(\cdot)$ is to a degree $K$ polynomial.
\begin{theorem}[Theorem 1 of \cite{shuman2015vertex}, Theorem 5.16 of \cite{handscomb}, Theorem 8.2 of \cite{atap}] \label{Th:loc}
Let $\hat{g}:[0,\lambda_{\max}]\rightarrow \Rbb$ be a spectral filter 
 and define 
 $K_{in}:=d_{\G}(i,n)-1.$  Then 
\begin{align} \label{Eq:loc_bound0}
|(T_i g_j)(n)|  
&\leq \inf_{\widehat{p_{K_{in}}}}\left\{\sup_{\lambda \in \sigma(\L)} \left|\hat{g}_j(\lambda)-\widehat{p_{K_{in}}}(\lambda)
\right|\right\} \nonumber \\
& \leq \inf_{\widehat{p_{K_{in}}}}\left\{\sup_{\lambda \in [0,\lambda_{\max}]} \left|\hat{g}_j(\lambda)-\widehat{p_{K_{in}}}(\lambda)
\right|\right\}, 
\end{align}
where the infimum in \eqref{Eq:loc_bound0} is taken over all polynomial kernels of degree $K_{in}$, as defined in \eqref{Eq:poly_kern}. If $\hat{g}_j(\cdot)$ is real analytic on $[0,\lambda_{\max}]$, the upper bound in \eqref{Eq:loc_bound0} converges geometrically to 0 as $d_{\G}(i,n)$ increases.
\end{theorem}
In short, and less precisely mathematically, the smoother the filter $\hat{g}_j(\cdot)$ is in the spectral domain, the more concentrated is the energy of the atom ${\boldsymbol \varphi}_{i,j}=T_i g_j$ around the center vertex $i$; compare, e.g., the first two atoms shown in 
Fig. \ref{Ex:atoms}. 

\subsection{Eigenvector groupings}
Recall from the introduction that in order for the inner products between a graph signal and each atom to be informative, the atoms should have interpretable structural features that account for the underlying graph. The localization in the vertex domain described above is one such structural feature. The shape of the filter $\hat{g}(\cdot)$ in the graph spectral domain leads to another: smoothness in terms of how much the atom's values vary between neighboring vertices, particularly those connected by a high edge weight. The unit-norm Laplacian eigenvectors satisfy 
\begin{align}\label{Eq:smoothness}
\lambda_\l
={\bf u}_{\l}^*\L {\bf u}_{\l} = \sum_{(m,n)\in \E} W_{m,n} [u_\l(m)-u_\l(n)]^2,
\end{align}
and therefore the eigenvectors associated with the lower eigenvalues vary less from vertex to neighboring vertex. Moreover, the eigenvectors are roughly ordered in terms of the number of zero crossings, defined as edges where the values of the eigenvector at the two connected vertices 
have opposite signs \cite[Fig. 3]{shuman2013emerging}.   

\begin{figure}[tb] 
\hspace{-.1in}
\fboxsep=3mm
\fboxrule=2pt
\fcolorbox{darkkhaki}{lightkhaki}{
\begin{minipage}{.92\linewidth}
\begin{example}[Variational/interpolating splines] \label{Ex:variational}
\emph{Variational or interpolating splines on graphs}, pioneered in \cite{pesenson_splines} and further studied in \cite{shuman_TSP_multiscale,ward2018interpolating,erb2019graph}, are atoms of the form \eqref{Eq:atom_form} with a single low pass filter $\hat{g}(\cdot)$ localized to a subset $\V_1$ of the vertices. They are used as an interpolation basis to interpolate an entire graph signal from its sample values at the  vertices in $\V_1$. In the images below, we show three examples of interpolating kernels and an atom generated from each on the Stanford bunny graph. The three filters are (a) a Green's kernel \cite{pesenson_splines} $\hat{g}(\lambda_\l)=\frac{\epsilon}{(\lambda_\l+\epsilon)^s}$ with $\epsilon=.05$ and $s=1$; (b) a diffusion kernel \cite{kondor2002diffusion} $\hat{g}(\lambda_\l)=e^{-\tau \lambda_\l}$ with $\tau=10$; and (c) a polynomial decay kernel \cite{erb2019graph} $\hat{g}(\lambda_\l)=\frac{1}{(\l+1)^s}$ with $s=1$.
\end{example}

\vspace{.08in}

\begin{minipage}{\linewidth}
\includegraphics[width=.24\linewidth,page=1]{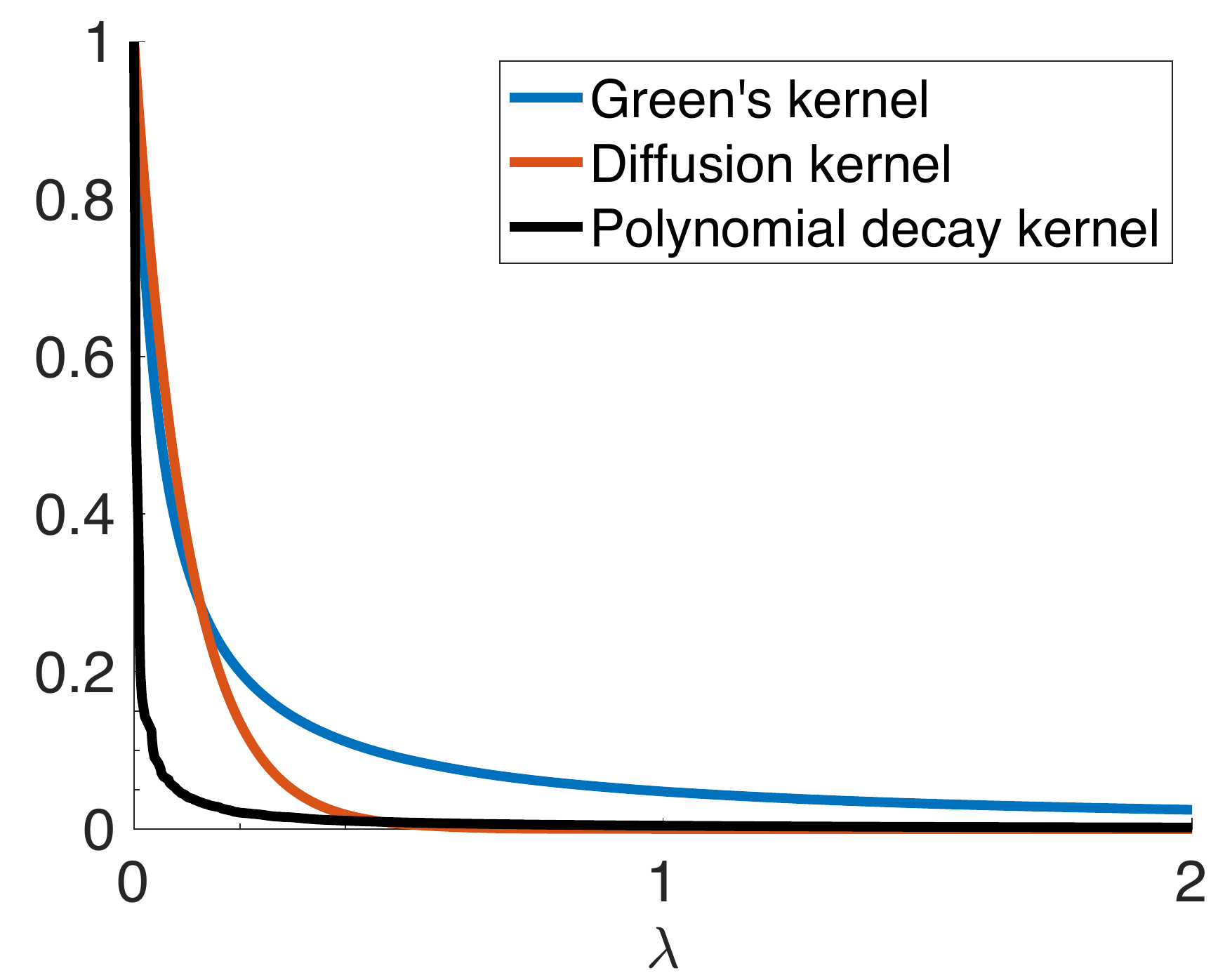}
\includegraphics[width=.24\linewidth,page=2]{figures/variational}
\includegraphics[width=.24\linewidth,page=3]{figures/variational}
\includegraphics[width=.24\linewidth,page=4]{figures/variational} \\
\vspace{-.2in}

\hspace{.33\linewidth}
{\small{(a)}}
\hspace{.18\linewidth}
{\small{(b)}}
\hspace{.18\linewidth}
{\small{(c)}}
\end{minipage}

\end{minipage}
}
\vspace{-.3in}
\end{figure}

Based off the analogy between this smoothness of Laplacian eigenvectors and the frequency of complex exponentials in one-dimensional signal processing, the most common spectral design approach in the graph signal processing literature is to choose filters concentrated on one part of the graph spectrum, grouping together eigenvectors with similar levels of total variation with respect to the graph, as defined in \eqref{Eq:smoothness}. In particular, when localized to different center vertices via \eqref{Eq:atom_form}, 
filters whose support is concentrated on the eigenvectors associated with small eigenvalues lead to \emph{scaling functions} or \emph{windows} around the center vertex (c.f., 
Fig. \ref{Ex:atoms}(a) and Example \ref{Ex:variational}). The inner products between such atoms and a graph signal provide information about the trend or local average of the signal in the neighborhood of the center vertex. On the other hand, all Laplacian eigenvectors associated with eigenvalues greater than 0 sum to zero, because they are orthogonal to ${\bf u}_0$, which is constant across all vertices. Thus, any 
filter with $\hat{g}_j(0)=0$ yields atoms ${\boldsymbol \varphi}_{i,j}$ that have a mean of zero and feature some oscillation (c.f., 
Fig. \ref{Ex:atoms}(b)-(c)).       

\begin{figure}[b] 
\begin{minipage}{.31\linewidth}
\centering
\includegraphics[width=1\linewidth,page=1]{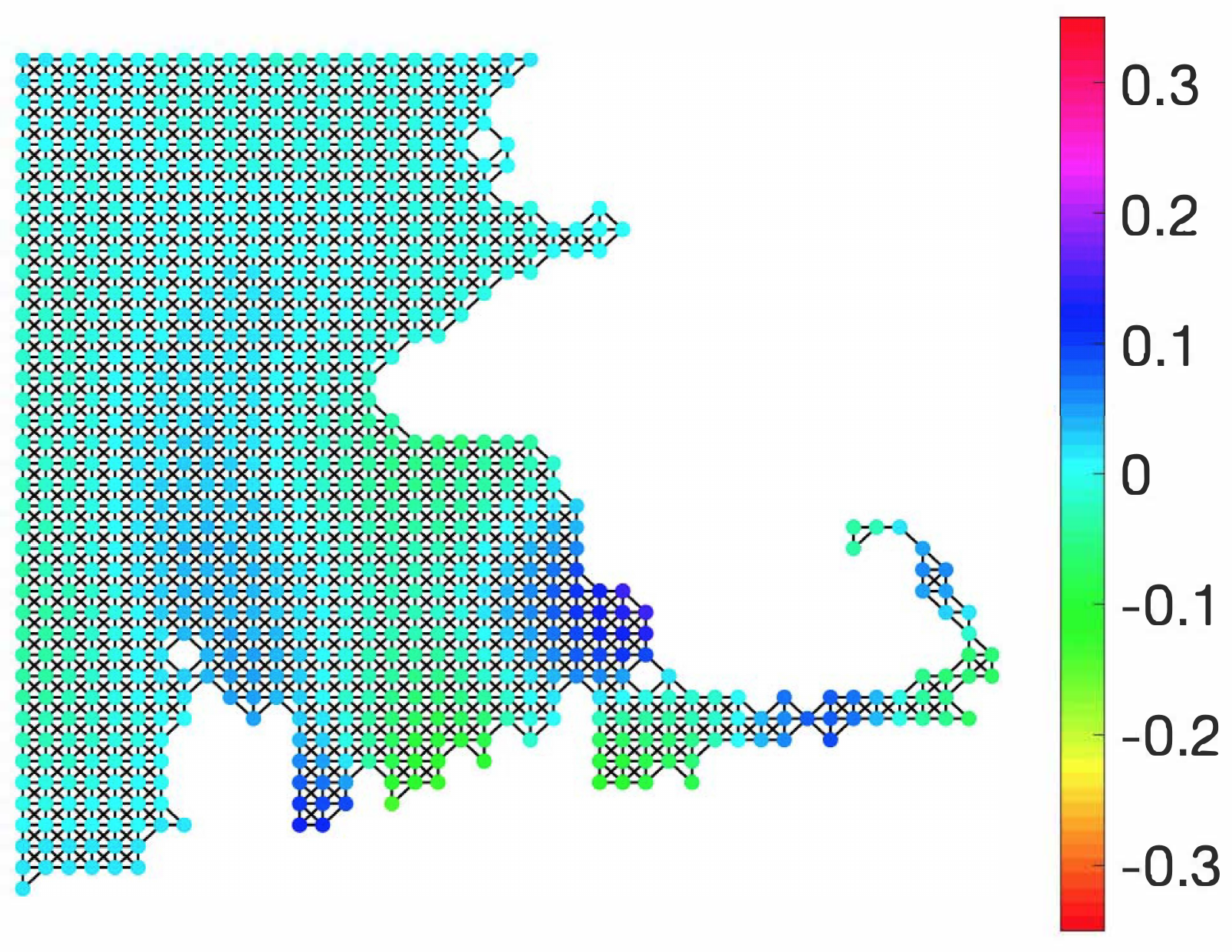}
\vspace{-.25in}

{\small{$\lambda=0.39$~~}}\\ 
\vspace{.1in}
 
\includegraphics[width=1\linewidth,page=4]{figures/evecsd} 
\vspace{-.25in}

{\small{$\lambda=1.07$~~}}
\end{minipage}
\hspace{.001in}
\begin{minipage}{.31\linewidth}
\centering
\includegraphics[width=1\linewidth,page=2]{figures/evecsd}
\vspace{-.25in}

{\small{$\lambda=0.45$~~}}\\
\vspace{.1in}
 
\includegraphics[width=1\linewidth,page=5]{figures/evecsd} 
\vspace{-.25in}

{\small{$\lambda=1.10$~~}}
\end{minipage}
\hspace{.001in}
\begin{minipage}{.31\linewidth}
\centering
\includegraphics[width=1\linewidth,page=3]{figures/evecsd}
\vspace{-.25in}

{\small{$\lambda=0.46$~~}}\\
\vspace{.1in}
 
\includegraphics[width=1\linewidth,page=6]{figures/evecsd} 
\vspace{-.25in}

{\small{$\lambda=1.14$~~}}
\end{minipage}
\caption{Six Laplacian eigenvectors of the Eastern Massachusetts 8-neighbor graph, labeled by the corresponding eigenvalues. While the three in the top row and the three in the bottom row are more similar to each other in terms of total variation (captured by the eigenvalue), the directions of the oscillations and regions where the eigenvectors' energies are concentrated are not necessarily in linear order. For example, both ${\bf u}_{0.39}$ and ${\bf u}_{1.07}$ have more of their energies concentrated on the vertices corresponding to Cape Cod.} \label{Fig:evecs}
\end{figure}

Noting that the Laplacian eigenvalues correspond to the total variation of the eigenvectors but not necessarily the directions of their oscillations on the graph (see Fig. \ref{Fig:evecs}), some more recent works \cite{saitoNatural},\cite{saito2018can,cloninger2018dual,li2019metrics}
investigate other ways to group the eigenvectors. For example, \cite{saitoNatural} suggests to view the eigenvectors as probability distributions on the graph, quantify the distances between eigenvectors using optimal transport theory, cluster the eigenvectors 
based on their distances, and construct a spectral filter for each cluster, with the support of the filter set to match the eigenvectors of that cluster. 

\subsection{Orthogonality or near orthogonality}
To reduce the correlation between atoms (and in turn improve the discriminatory power of taking inner products between each atom and a target signal, as discussed below in Sec.
\ref{Se:theory}), it may be desirable for all atoms that are generated from a single filter to be orthogonal or near orthogonal to all atoms that are generated from all other filters; i.e., $\langle T_i g_j , T_{i^{\prime}}g_{j^{\prime}} \rangle \approx 0$ for all $j^{\prime} \neq j$). This can be ensured via the filter design, with a sufficient condition for the orthogonality of atoms generated from different patterns being that $\hat{g}_j(\lambda_\l)\hat{g}_{j^{\prime}}(\lambda_\l)=0$ for all $j^{\prime} \neq j$ and all $\lambda_\l$. 

For the specific case when $J=2$ and $\G$ is a bipartite graph with the normalized Laplacian eigenvectors as the graph Fourier basis\footnote{These conditions can be adapted for a regular bipartite graph with the non-normalized Laplacian eigenvectors as the graph Fourier basis.}, it is possible to go a step further and generate $N$ atoms that are not only orthogonal to atoms generated from the other filter, but also from the same filter.
\begin{theorem}[\cite{narang_bipartite_prod}]
Let $\G$ be a bipartite graph with a bipartition $\{\V_1,\V_2\}$, and consider an LGSFD ${\cal D}$ of the form \eqref{Eq:LSGFD} with $J=2$ (i.e., $\hat{g}_i(\cdot)$ is localized to the center vertices in $\V_i$ for $i=1,2$), using the normalized Laplacian graph Fourier basis. Then necessary and sufficient conditions on the filters to ensure that the $N$ atoms of ${\cal D}$ form an orthogonal basis for $\Rbb^N$ are that $\hat{g}_1(\lambda_\l)\hat{g}_1(2-\lambda_\l)=\hat{g}_2(\lambda_\l)\hat{g}_2(2-\lambda_\l)$ and $|\hat{g}_1(\lambda_\l)|^2 + |\hat{g}_2(\lambda_\l)|^2 = c^2$
for all $\lambda_\l$ and any constant $c$.
\end{theorem}

\subsection{Coverage of the spectrum}
A necessary 
condition for ${\cal D}$ to be a frame for all graph signals in $\Rbb^N$ is that for all $\lambda \in \sigma(\L)$, $\hat{g}_j(\lambda) \neq 0$ for some $j \in \{1,2,\ldots,J\}$. If this is not true for some $\lambda_\l$, then $\langle {\bf u}_{\l},{\boldsymbol \varphi}_{i,j}\rangle=0$ for all $i$ and $j$ and thus $||{\boldsymbol \Phi}^* {\bf u}_{\l}||_2^2=0$, contradicting the frame condition \eqref{Eq:frame}. Thus, choosing a set of filters that covers the full spectral range $[0,\lambda_{\max}]$ (or at least the portion of it whose Laplacian eigenvectors span the subspace of signals of interest) is a good place to start. 

\begin{figure*}
\vspace{-.3in}
\fboxsep=3mm
\fboxrule=2pt
\fcolorbox{darkkhaki}{lightkhaki}{
\begin{minipage}{6.75in}
\begin{example} [Spectral filter designs that only use the spectral range] \label{Ex:filters}
We show 11 different sets of 
six filter patterns for the cerebellum graph \cite{behjat2015anatomically}, whose spectral range is $[0,\lambda_{\max}]=[0,32.4]$. In all  
images, the vertical axis represents the value of the filter,
and the shaded gray/black circles represent the values of 
$G(\lambda)=\sum_{j=1}^J |\hat{g}_j(\lambda)|^2$ at each of the Laplacian eigenvalues, with darker areas denoting regions of higher spectral density. 
 For each of the filters $\hat{g}_j(\lambda)$ in the set of uniform translates, 
 the corresponding filter in the log-warped set (right) is given by 
$\hat{h}_j(\lambda)= \hat{g}_j\left(\frac{\lambda_{\max}}{\omega(\lambda_{\max})}\omega(\lambda)\right),$
 where the warping function is $\omega(\lambda)=\log(1+\nu \lambda)$
 for a parameter $\nu>0$ ($\nu=10$ here).  
\vspace{.05in}

\vspace{.07in}

\begin{minipage}{.45\linewidth}
  \centerline{\underline{\small{\bf Uniform Translates}}}
  \vspace{.05in}
     
   \centerline{\small{Ideal filters \cite{li_mcsfb_2018}}}
   \vspace{.01in}
   
   \centerline{\includegraphics[width=\linewidth,page=1]{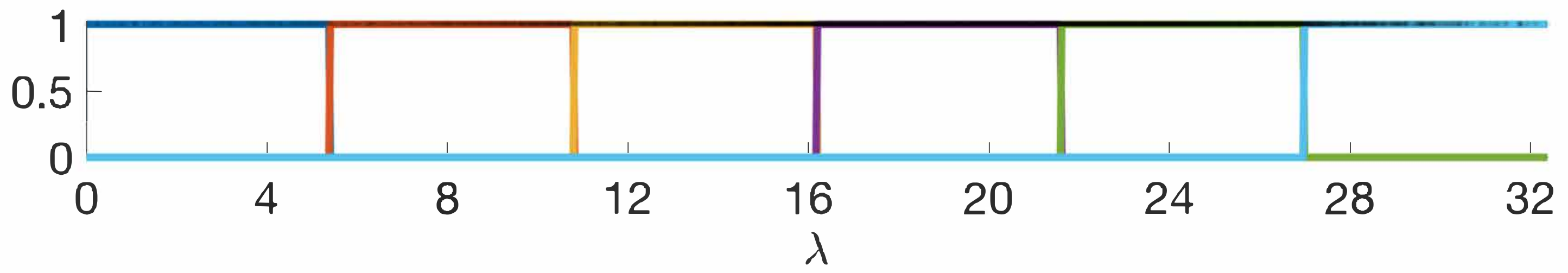}} 
      \vspace{.03in} 
   
   \centerline{\small{Uniform translates: Meyer-type}} 
   \vspace{.01in}
   
         \centerline{\includegraphics[width=\linewidth,page=2]{figures/uniform}} 
               \vspace{.03in} 
               
                  \centerline{\small{Uniform translates: Itersine kernel \cite{shuman2013spectrum,gspbox}}}
   \vspace{.01in}
   
            \centerline{\includegraphics[width=\linewidth,page=3]{figures/uniform}} 
                  \vspace{.03in} 
                          
                                    \centerline{\small{DCT with frequency conversion \cite{sakiyama2016spectral}}}
   \vspace{.01in}
   
            \centerline{\includegraphics[width=\linewidth,page=5]{figures/uniform}} 
                 \vspace{.1in} 
 
\change{Let us highlight some of the design considerations mentioned in this section: 
\begin{enumerate}
\item All of these designs cover the entire spectrum; i.e., $G(\lambda)>0$ for  all $\lambda \in [0,\lambda_{\max}]$. Thus, each design yields a frame when every filter is localized to be centered at every vertex. 
\item The last two sets of wavelet filters are the only two amongst those shown that do not satisfy the Parseval frame condition, $G(\lambda)=1$ for all $\lambda \in \sigma(\L)$, from Theorem \ref{Th:frame}. 
\item Because the ideal filters in the top row do not overlap, the atoms generated from a filter are orthogonal to the atoms generated from any other
\end{enumerate}
}
\end{minipage}
\hfill
\begin{minipage}{.45\linewidth}
   \centering
  \underline{\small{\bf Wavelets (Octave-Band)}}
  \vspace{.05in}
     
   {\small{Ideal octave-band filters \cite{li_mcsfb_2018}}}
   \vspace{.01in}
   
   \centerline{\includegraphics[width=\linewidth,page=1]{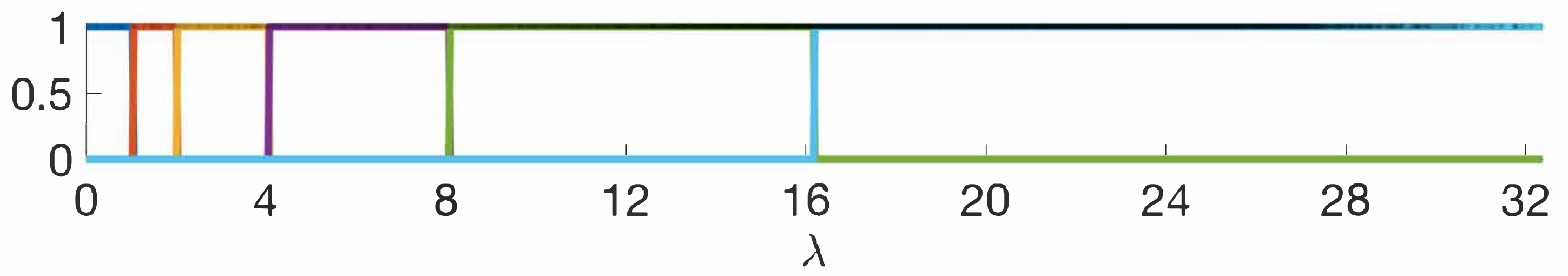}} 
      \vspace{.03in} 
   
   {\small{Meyer-type wavelets \cite{leonardi_multislice}}}
   \vspace{.01in}
   
         \centerline{\includegraphics[width=\linewidth,page=2]{figures/wavelet}} 
               \vspace{.03in} 
               
                  {\small{Log-warped translates: Itersine kernel \cite{shuman2013spectrum,gspbox}}}
   \vspace{.01in}
   
            \centerline{\includegraphics[width=\linewidth,page=3]{figures/wavelet}} 
                  \vspace{.03in} 
                          
                                    {\small{Log-warped DCT with frequency conversion \cite{sakiyama2016spectral}}}
   \vspace{.01in}
   
            \centerline{\includegraphics[width=\linewidth,page=5]{figures/wavelet}} 
                  \vspace{.03in} 
                  
                                                      {\small{Fast tight wavelet frame \cite{dong2017sparse}}}
   \vspace{.01in}
   
               \centerline{\includegraphics[width=\linewidth,page=6]{figures/wavelet}} 
                  \vspace{.03in} 
                  
                                                      {\small{Spectral graph wavelets \cite{hammond2011wavelets}}}
   \vspace{.01in}

            \centerline{\includegraphics[width=\linewidth,page=7]{figures/wavelet}} 
                  \vspace{.03in} 
                  
                                                      {\small{Cohen-Daubechies-Feauveau 9/7 filters \cite{sakiyama2016spectral}}}
   \vspace{.01in}
   
            \centerline{\includegraphics[width=\linewidth,page=8]{figures/wavelet}} 
                  \vspace{.02in} 
                  
\end{minipage}
\vspace{.02in}

\hspace{.22in} \change{filter. On the other hand, the atoms generated 
by localizing
filters with significant overlap (e.g., the yellow and red \\
\vspace{-.17in}

\hspace{.22in} filters of the fast tight wavelet frame)  
to the same center vertex are likely to be highly correlated. \\
\vspace{-.19in}

\begin{enumerate}
\setcounter{enumi}{3}
\item As 
detailed in Theorem \ref{Th:loc}, filters that are smooth (well approximated by low order polynomials) yield dictionary atoms that are more localized around the center vertex; i.e., most of their energy is close to the center. In particular, for the octave-band designs such as the 
Meyer-type wavelets, log-warped translates, and spectral graph wavelets, the filters that cover the upper end of the spectrum yield more localized atoms than the scaling and wavelet filters at the low end of the spectrum. 
\end{enumerate}
}
\end{example}
\end{minipage}
}
\vspace{-.2in}
\end{figure*}

In the case that $\V_j=\V$ for all $j$ (i.e., every spectral pattern is localized to every vertex), which is often referred to as \emph{complete sampling} or an \emph{undecimated} filter bank, it is possible to design the spectral filters such that ${\cal D}$ is a (tight) Parseval frame.
\begin{theorem}[Theorem 5.6 of \cite{hammond2011wavelets}, \cite{leonardi_multislice}, and Lemma 1 of \cite{shuman2013spectrum}] \label{Th:frame}
Let $\V_j=\V$ for all $j$ and ${\cal D}:=\left\{{\boldsymbol \varphi}_{i,j}\right\}_{i=1,2,\ldots,N;~j=1,2,\ldots,J}$ be a dictionary of atoms with ${\boldsymbol \varphi}_{i,j}:=T_i g_j$. Define $G(\lambda):=\sum_{j=1}^{J} \bigl|\hat{g}_j(\lambda)\bigr|^2$.
If $G(\lambda)>0$ for all $\lambda \in \sigma(\L)$, 
then the frame condition \eqref{Eq:frame} is satisfied  
for all ${\bf f} \in \Rbb^N$, with frame bounds $A=\min_{\lambda \in \sigma(\L)} G(\lambda)$ and $B=\max_{\lambda \in \sigma(\L)} G(\lambda)$.
In particular, if $G(\lambda)$ is constant on $\sigma(\L)$, 
${\cal D}$ is a \emph{tight frame} with $A=B$. Moreover, if $\sum_{j=1}^{J} |\hat{g}_j(\lambda)|^2 =1,~\forall \lambda \in \sigma(\L)$, then ${\cal D}$ is a \emph{Parseval frame}; i.e., $||{\boldsymbol \Phi}^*{\bf f}||_2^2=\sum_{j=1}^J \sum_{i=1}^N | \langle {\bf f},{\boldsymbol \varphi}_{i,j} \rangle |^2 =  ||{\bf f}||_2^2,~\forall {\bf f} \in \Rbb^N.$
\end{theorem}
Example \ref{Ex:filters} contains 
filters that satisfy the sufficient condition of Theorem \ref{Th:frame} for a Parseval frame. 

\begin{figure*}
\fboxsep=3mm
\fboxrule=2pt
\fcolorbox{darkkhaki}{lightkhaki}{
\begin{minipage}{6.75in}
\begin{minipage}{.67\linewidth}
\begin{example} [Spectral filter designs that are adapted to the spectral density of the graph or ensemble energy density of training signals]\label{Ex:adapted} 
\change{The ideal filters and uniform translates from Example \ref{Ex:filters} are designed to cover equal portions of the spectral range $[0,\lambda_{\max}]$, but are the same for any two graphs with the same value of $\lambda_{\max}$. One option to further adapt the filters to the specific graph structure is to warp them so that each filter contains roughly the same number of Laplacian eigenvalues \cite{shuman2013spectrum}. 
This is accomplished by defining spectrum-adapted filters $\tilde{\hat{g}}_j(\lambda) = \hat{g}_j(\lambda_{\max}P_\lambda(\lambda))$, where $\{\hat{g}_j(\cdot)\}$ are the uniform translates from Example \ref{Ex:filters} and $P_\lambda(\cdot)$ is an estimate of the cumulative spectral density defined in \eqref{Eq:spectral_cdf}. For the cerebellum graph, the Laplacian eigenvalues are concentrated in the middle of the spectral range (right), and therefore the spectrum-adapted uniform 
Meyer-type filters shown in the middle row below are narrower in this region of}
\end{example}
\end{minipage}
\hfill
\begin{minipage}{.33\linewidth}
\center{\includegraphics[width=.95\linewidth,page=1]{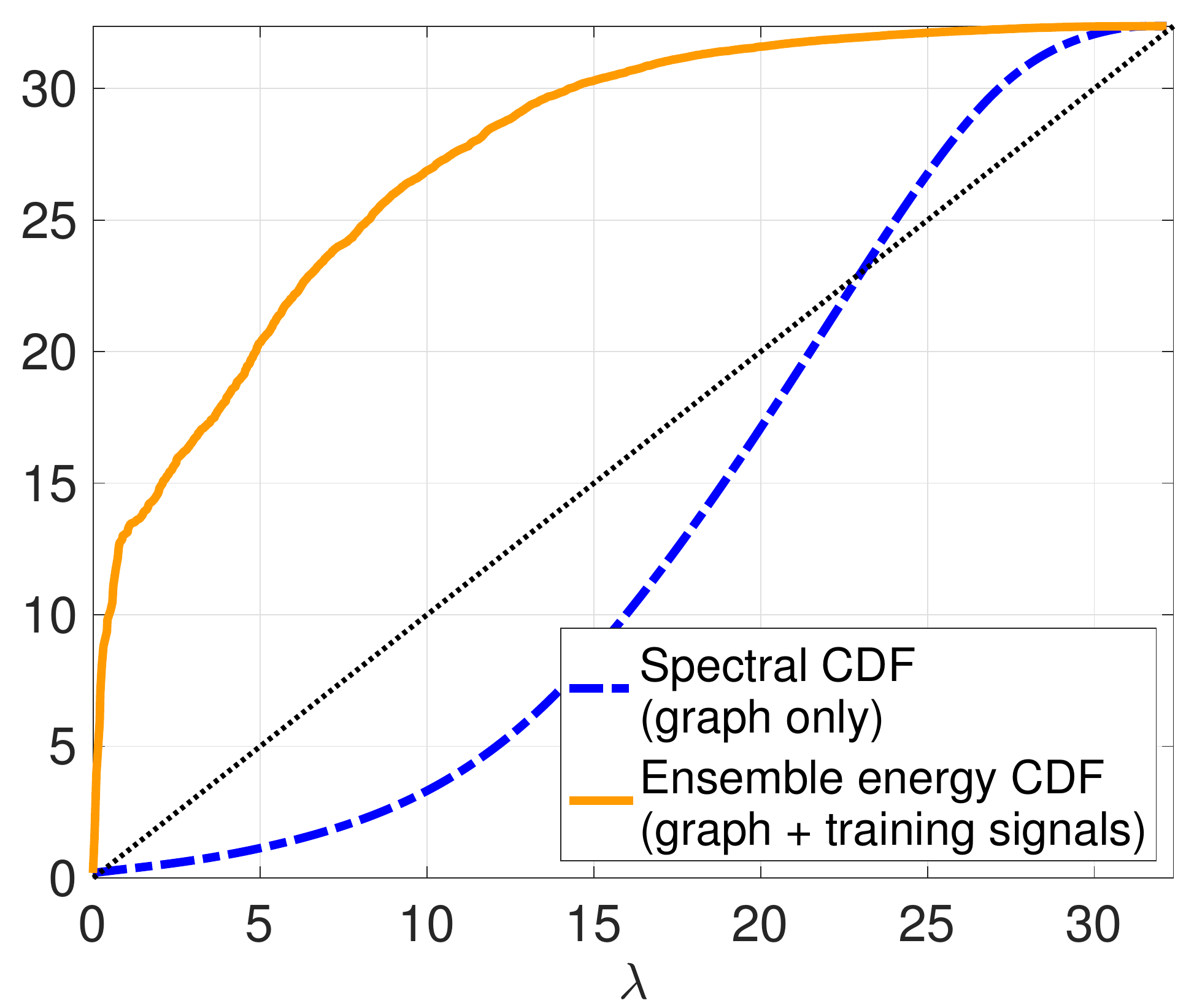}}
\end{minipage}
\vspace{.015in}

\change{high spectral density. The same idea can be used to generate spectrum-adapted wavelets $\tilde{\hat{h}}_j(\lambda)= \hat{g}_j\left(\frac{\lambda_{\max}}{\omega(\lambda_{\max})}\omega(\lambda)\right),$ by using a warping function $\omega(\lambda)=\log(1+\nu \lambda_{\max}P_{\lambda}(\lambda))$.}
\medskip

\begin{minipage}{.52\linewidth}
\change{When training data is available on the graph, a second option is to adapt the filters to be narrower in the regions of the spectrum where the energy of the training signals is concentrated \cite{behjat2016signal}. This can again be achieved via warping, using an estimate of the ensemble energy cumulative spectral density function $P_{\cal Y}(z)$ defined in \eqref{Eq:energy_cdf} in place of the spectral density estimate $P_\lambda(z)$ in the warping function. The plot of the density functions (above right) shows that despite the Laplacian eigenvalues being more heavily concentrated in the middle of the spectrum, the energies of 292 fMRI training signals on the cerebellum graph are heavily concentrated in the lower end of the spectrum.} Therefore, the signal-adapted design that aims to have roughly the same signal energy in each filter band (bottom row, right) features narrower filters at lower eigenvalues.
\end{minipage}
\hfill
\begin{minipage}{.45\linewidth}
 \vspace{-.1in}
 
   \centering
  \vspace{.12in}
   
   {\small{Uniform translates: Meyer-type \cite{leonardi_multislice}}}
   \vspace{.01in}
   
   \centerline{\includegraphics[width=\linewidth,page=1]{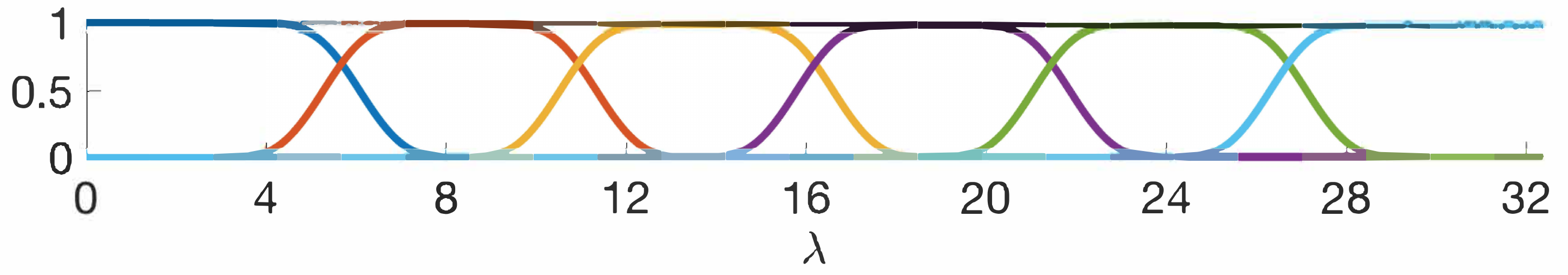}} 
      \vspace{.03in} 
   
   {\small{Spectrum-adapted Meyer-type \cite{shuman2013spectrum}}}
   \vspace{.01in}
   
         \centerline{\includegraphics[width=\linewidth,page=2]{figures/sig_adapteda}} 
               \vspace{.03in} 
                            
                  {\small{Signal-adapted Meyer-type \cite{behjat2016signal}}}
   \vspace{.01in}
   
            \centerline{\includegraphics[width=\linewidth,page=3]{figures/sig_adapteda}} 
                  \vspace{.03in} 
 \vspace{.1in}
                  
\end{minipage}
\end{minipage}
}
\vspace{-.2in}
\end{figure*}

It is important to distinguish between coverage of the spectral range and coverage of the spectrum. One subtlety about Theorem \ref{Th:frame} is that while the filters are often designed over the interval $[0,\lambda_{\max}]$ or $[0,\bar{\lambda}]$, the condition for a tight frame is that $G(\lambda)$ only needs to be constant on the set of actual Laplacian eigenvalues $\sigma(\L)$, as these are the values of the filter that contribute to the definition of the atom in \eqref{Eq:atom_form}. Related to this point, if a filter is defined on the interval $[0,\lambda_{\max}]$, but $\hat{g}_j(\lambda)=0$ for all $\lambda$ in $\sigma(\L)$, then any atoms $T_i g_j$ derived from this filter are equal to the zero vector, and therefore do not provide any useful information about the graph signal. One way to avoid such non-informative atoms is to adapt the filter design not only to the spectral width, but also to the estimated spectral density function \eqref{Eq:spectral_cdf}. By leveraging the spectral density approximation, the \emph{spectrum-adapted} designs in \cite{shuman2013spectrum} warp a set of non-adapted filters in order that the support of each filter approximately contains a desired number of Laplacian eigenvalues (e.g., each filter has the same number of eigenvalues or they satisfy a dyadic structure with twice as many in each subsequent filter, moving from the low end of the spectrum to the high end). See \change{Example \ref{Ex:adapted}} 
for examples of spectrum-adapted filter designs.

\subsection{Computational efficiency and approximation}
As mentioned in Sec. \ref{Se:dict}, exactly computing the graph Laplacian eigenvectors is only feasible for small to medium graphs, implying that for large graphs, the computation of the analysis coefficients $\langle {\bf f}, {\boldsymbol \varphi}_{i,j} \rangle = {\boldsymbol \delta}_i^* \hat{g}_j(\L){\bf f}$ must be efficiently approximated.  Methods for approximating a \emph{matrix function} times a vector (i.e., $\hat{g}_j(\L){\bf f}$) include Krylov subspace methods such as the Lanczos method, contour integral methods, conjugate gradient, algebraic multigrid methods, rational approximations (also referred to as \emph{infinite impulse response filters} in the graph signal processing community \cite{shi2015infinite,liu2018filter}), spline approximations, and polynomial approximations (see \cite{higham,frommer,shuman_distributed_sipn} for surveys of these methods in centralized and distributed settings).

\begin{figure*}
\vspace{-.2in}
\fboxsep=3mm
\fboxrule=2pt
\fcolorbox{darkkhaki}{lightkhaki}{
\begin{minipage}{6.75in}
\begin{example} [Fast transforms and inverse transforms via polynomial approximation]\label{Ex:poly}
\change{Approximating each spectral filter $\hat{g}_j(\cdot)$ by a degree $K$ polynomial $\hat{p}_{j,K}(\cdot)$ reduces the complexity of applying the dictionary analysis 
operator from ${\cal O}(N^3)$ to ${\cal O}(K|\E|)$, which for large, sparse graphs 
is ${\cal O}(N)$. The drawback of this scalable approximation is that the tight Parseval frame condition $G(\lambda)=1$ for all $\lambda \in \sigma(\L)$ of Theorem \ref{Th:frame} is not typically satisfied by the polynomial approximants. Let $\tilde{{\boldsymbol \Phi}}^*$ be the approximate analysis operator with the polynomial filters, and ${\boldsymbol \alpha}= \tilde{{\boldsymbol \Phi}}^*{\bf f}$ be the resulting analysis coefficients. There are three common options for fast, approximate inverse transforms. 
The first is to solve $\tilde{{\boldsymbol \Phi}}\tilde{{\boldsymbol \Phi}}^*{\bf f}_{\hbox{rec}}=\tilde{{\boldsymbol \Phi}}{\boldsymbol \alpha}$ via the conjugate gradient method \cite{hammond2011wavelets}; and the second is the frame inversion algorithm \cite[Ch. 3]{daubechies1992ten} that sets ${\bf f}_{\hbox{rec}}^{(0)}=\frac{2}{A+B}\tilde{{\boldsymbol \Phi}}{\boldsymbol \alpha}$ and iterates ${\bf f}_{\hbox{rec}}^{(t)}={\bf f}_{\hbox{rec}}^{(0)}+{\bf f}_{\hbox{rec}}^{(t-1)}-\frac{2}{A+B}\tilde{\boldsymbol \Phi}\tilde{\boldsymbol \Phi}^*{\bf f}_{\hbox{rec}}^{(t-1)}$. Both of these iterative methods have complexity ${\cal O}(2TK|\E|)$, where the number of iterations $T$ is typically small (5-10), and the speed of convergence depends on how close the ratio of frame bounds $\frac{B}{A}$ is to 1 (recall that when $\V_j=\V$ for all $j$, 
the lower frame bound is $A=\min_{\lambda \in \sigma(\L)}\sum_{j=1}^J |\hat{p}_{j,K}(\lambda)|^2$ and the upper frame bound is $B=\max_{\lambda \in \sigma(\L)}\sum_{j=1}^J |\hat{p}_{j,K}(\lambda)|^2$). Thus, for the non-tight frame generated from the polynomial filters, near perfect reconstruction is still possible at the same ${\cal O}(N)$ complexity, but the inverse transform may require 10-20 times the number of computations as the fast analysis operator. A third, faster (${\cal O}(K|\E|)$) but less accurate option is to just take ${\bf f}_{\hbox{rec}}=\frac{2}{A+B}\tilde{{\boldsymbol \Phi}}{\boldsymbol \alpha}$ (i.e., stop the frame inversion algorithm after the initial guess). The high-level intuition is that $\frac{2}{A+B}\tilde{\boldsymbol \Phi}\tilde{\boldsymbol \Phi}^*$ is close to the identity matrix ${\bf I}_N$ if $\frac{B}{A}$ is close to 1 \cite[Ch. 3]{daubechies1992ten}. For this faster synthesis operator, the squared reconstruction error can be upper bounded by
\begin{align*}
||{\bf f}-{\bf f}_{\hbox{rec}}||_2^2&=\Bigl|\Bigl|{\bf f}-\frac{2}{A+B}\tilde{\boldsymbol \Phi}{\boldsymbol \alpha}\Bigr|\Bigr|_2^2\\
&=\Bigl|\Bigl|\Bigl({\bf I}_N-\frac{2}{A+B}\sum_{j=1}^J \hat{p}_{j,K}^2(\L)\Bigr){\bf f} \Bigr|\Bigr|_2^2 = \sum_{\l=0}^{N-1} |\hat{f}(\lambda_\l)|^2\Bigl[1-\frac{2}{A+B}\sum_{j=1}^J|\hat{p}_{j,K}(\lambda_\l)|^2\Bigr]^2 \leq \left(\frac{r}{2+r}\right)^2 ||{\bf f}||_2^2, 
\end{align*}    
where $r=\frac{B}{A}-1$ \cite{hammond2011wavelets}, \cite[Ch. 3]{daubechies1992ten}.
\emph{So, regardless of the choice of fast inverse transform, it desirable for $\tilde{G}(\lambda):=\sum_{j=1}^J |\hat{p}_{j,K}(\lambda_\l)|^2$ to be close to 1 for each $\lambda_\l$, yielding a ratio of frame bounds $\frac{B}{A}$ close to 1 and a small value of $r$.}
\medskip

In the figures below, we show degree 40 Chebyshev polynomial approximations to three different sets of five filters on 
the net25 graph, which features many repeated eigenvalues and therefore has areas of the spectrum with high density. 
The polynomial approximants to the ideal filter bank in the top row yield a $\tilde{G}(\lambda)$ (black points in the right column of images) that fluctuates across $\lambda \in \sigma(\L)$, and the ratio of frame bounds $\frac{B}{A}$
is equal to
2.78.} By shifting the filter end points slightly to be in regions of lower spectral density (middle row), the frame bound ratio drops to 1.85. The smoother spectrum-adapted translates (bottom row) are more amenable to polynomial approximation; 
the polynomial filters for this design have a frame bound ratio of 1.16. 
\vspace{.08in}

\begin{minipage}{.45\linewidth}
  \centerline{\underline{\small{\bf Initial Design}}}
   \centering
  \vspace{.05in}
   
   {\small{Spectrum-adapted uniform ideal \cite{li_mcsfb_2018}}}
   \vspace{.01in}
   
   \centerline{\includegraphics[width=\linewidth,page=1]{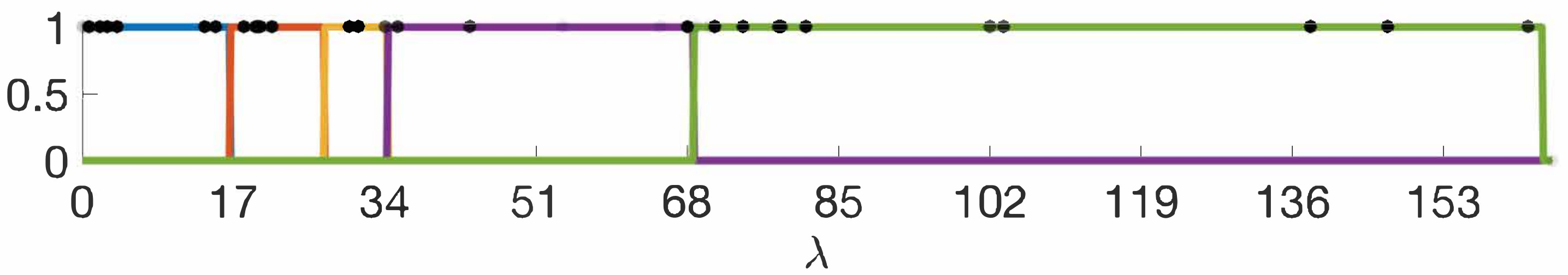}} 
      \vspace{.03in} 
   
   {\small{Spectrum-adapted and shifted uniform ideal \cite{li_mcsfb_2018}}}
   \vspace{.01in}
   
   \centerline{\includegraphics[width=\linewidth,page=3]{figures/frame_poly}} 
               \vspace{.03in} 
               
                  {\small{Spectrum-adapted uniform translates: Itersine kernel}}
   \vspace{.01in}
   
               \centerline{\includegraphics[width=\linewidth,page=5]{figures/frame_poly}} 
                  \vspace{.03in} 
                  
\end{minipage}
\hfill
\begin{minipage}{.45\linewidth}
  \centerline{\underline{\small{\bf Degree 40 Chebyshev Polynomial Approximation}}}
   \centering
  \vspace{.05in}
   
   {\small{~~}}
   \vspace{.01in}
   
      \centerline{\includegraphics[width=\linewidth,page=2]{figures/frame_poly}} 
      \vspace{.06in} 
   
 {\small{~~}}
   \vspace{.01in}
   
            \centerline{\includegraphics[width=\linewidth,page=4]{figures/frame_poly}} 
               \vspace{.06in} 
               
 {\small{~~}}
   \vspace{.01in}
   
               \centerline{\includegraphics[width=\linewidth,page=6]{figures/frame_poly}} 
                  
\end{minipage} \\
\vspace{.15in}

\begin{minipage}{1\linewidth}
While the Chebyshev polynomial approximations \cite{druskin,hammond2011wavelets}
are good general choices as they are near optimal in terms of minimizing the maximum approximation error across the spectrum, they may not be 
\begin{wrapfigure}{r}{7.5cm}
\vspace{-.07in}
\includegraphics[width=7.5cm]{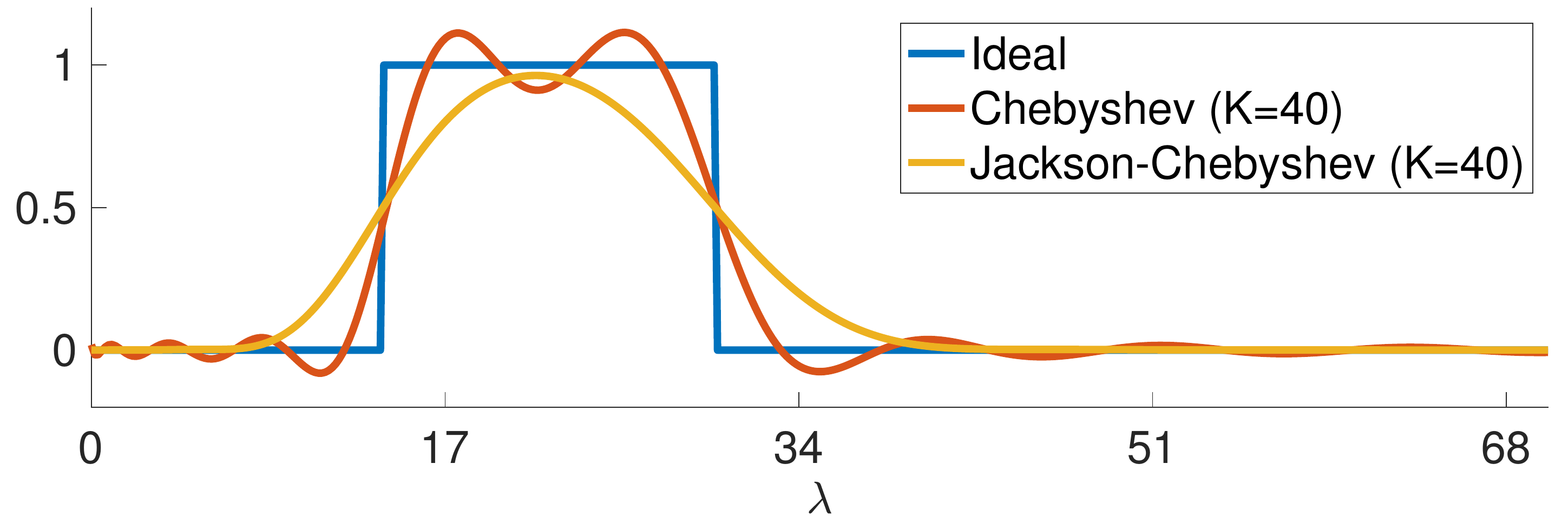}
\end{wrapfigure}
the most desirable in certain situations, such as approximating a series of ideal bandpass filters. 
This is because the oscillations of the Chebyshev polynomials may lead to more energy at the spectral values farther from the  
bandpass region. The Jackson-Chebyshev polynomial approximations  \cite{di2016efficient}  damp these Gibbs oscillations, resulting in less energy farther from the bandpass region (right).

\end{minipage}

\end{example} 
\end{minipage}
}
\end{figure*}

We focus our attention and numerical experiments in this survey on degree $K$ polynomial approximations $\hat{p}_{j,K}(\lambda)$ of the form \eqref{Eq:poly_kern} 
to each filter $\hat{g}_j(\cdot)$ (also referred to as \emph{finite impulse response (FIR) filters} in the graph signal processing community). The approximation ${\bf x}_j^{(K)}=\hat{p}_{j,K}(\L){\bf f}$ to $\hat{g}_j(\L){\bf f}$ can be computed recursively, either through a three-term recurrence for specific types of polynomials (e.g., Chebyshev) or through the nested multiplication iteration \cite[Section 9.2.4]{golub}
\begin{align}\label{Eq:poly_iter}
{\bf x}_j^{(l)}=a_{j,K-l}{\bf f} + \L {\bf x}_j^{(l-1)},~l=1,2,\ldots,K,
\end{align}
with ${\bf x}_j^{(0)}=a_{j,K}{\bf f}$. The computational complexity of computing ${\bf x}_j^{(K)}=\hat{g}_j(\L){\bf f}$ through \eqref{Eq:poly_iter} or a three-term recurrence is ${\cal O}(K |\E|)$, which for a large, sparse graph is approximately linear in the number of vertices, as opposed to the  ${\cal O}(N^3)$ required to naively compute the full eigendecomposition of $\L$. Additional advantages of the polynomial approximations include (i) the atoms are strictly localized as described in Theorem \ref{Th:strict_loc}, (ii) in addition to the analysis operator, the synthesis operator ${\boldsymbol \Phi}$ can be applied efficiently, and (iii) both the analysis and synthesis computations can be performed in a distributed setting where each vertex only knows its own signal value and can only communicate with its neighboring vertices \cite{shuman_distributed_sipn,loukas2015distributed}. Polynomial approximation methods commonly used in the graph signal processing literature include Chebyshev \cite{druskin,hammond2011wavelets} and  Jackson-Chebyshev 
\cite{di2016efficient}. 
\change{Example \ref{Ex:poly}} shows filters resulting from these polynomial approximation methods. For the specific case of approximating an ideal low pass filter with a small degree polynomial, \cite{teke2018energy} introduces energy compaction filters that 
maximize the energy of the polynomial filter that is concentrated on the specified 
band.

However, there are also tradeoffs to using polynomial filters. First, polynomial approximations to filter designs that meet the orthogonality or tight frame criteria may no longer satisfy these conditions. In fact, \cite{tay2017almost} shows that it is not possible to find $J$ polynomial filters with the property that $G(\lambda)$ is constant for all $\lambda$ in the interval $[0,\lambda_{\max}]$. Although it may be possible to satisfy this condition for all $\lambda$ in $\sigma(\L)$, it is not usually tractable to do so.  When using ideal filters, one option to mitigate the approximation error at the Laplacian eigenvalues (recall that the filter values at these eigenvalues are the only filter values that actually affect the form of the dictionary atoms) is to attempt to place the endpoints of the subband filters in areas of the spectrum with low density (or even better, in spectral gaps), as the error is typically highest near the endpoints 
\change{(c.f., middle row of Example \ref{Ex:poly})} \cite{li_mcsfb_2018}. A second option for mitigating the approximation error is to choose polynomials that control the error in specific parts of the spectrum, such as transformed linear phase multirate filters \cite{tay2017almost}, which reduce the error near the eigenvalue 0 (no DC leakage) or spectrum-adapted polynomial approximation \cite{fan_saop_ICASSP_2019}, which can reduce the error in high density areas of the spectrum. A third option is to directly choose the initial set of filters to be polynomials, or at least choose them to be smoother functions that are more accurately approximated by polynomials (e.g., \cite{sakiyama2016spectral}).   

\subsection{The number of filters}
For sets of filters that cover the whole spectrum, typical choices of $J$ in the literature are in the 4-8 range; however, we are not aware of theoretical analysis concerning the choice of $J$. In general, increasing $J$ may lead to sparser representations by increasing the number of atoms and the ability to distinguish between signals by capturing the behavior of the signal across smaller spectral regions. On the other hand, at some point, the benefits may saturate as the atoms become more correlated (see, e.g., \cite[Fig. 13]{behjat2016signal}). Moreover, as the filters become narrower, they are more difficult to approximate by polynomials. If the dictionary is critically sampled ($M=\sum_{j=1}^J |\V_j|=N$), increasing $J$ also leads to fewer center vertices for each filter, making accurate reconstruction from the analysis coefficients more difficult.
Narrower filters, especially those whose support is at the high end of the spectrum, can also lead to dictionary atoms that are more sensitive to small perturbations in the graph weights, an important consideration in applications where the graph is estimated. We explore the choice of $J$ further in Section \ref{Se:experiments}.

\subsection{Available data}
The graph spectral filters can also be adapted to an ensemble of $T$ training signals, $\{{\bf y}_t\}_{t=1,2,\ldots,T}$, that are exemplary of the data to be analyzed by the dictionary ${\cal D}$, when such training signals are available. We briefly review two \emph{parametric dictionary learning} approaches, 
both of which set $\V_j=\V$ for all $j$ and adapt the filters $\{\hat{g}_j(\cdot)\}$ to the training data. The first approach, presented in \cite{behjat2016signal} and shown in Example \ref{Ex:adapted}, is to design the filters so that (i) each filter captures a roughly equal amount of the energy of the ensemble of training signals, and (ii) the filters satisfy the tight frame condition $G(\lambda)=1$ for all $\lambda \in [0,\lambda_{\max}]$. 
Similar to the spectrum-adapted filter design of \cite{shuman2013spectrum}, this \emph{signal-adapted filter design} constructs a set of prototype filters that uniformly cover the spectrum, and then warps/transforms the filters. Whereas the warping function in \cite{shuman2013spectrum} is an estimate of the cumulative distribution of the Laplacian eigenvalues \eqref{Eq:spectral_cdf}, the warping function in \cite{behjat2016signal} estimates the \emph{ensemble energy cumulative spectral density} 
\begin{align}\label{Eq:energy_cdf}
P_{\cal Y}(z):=\frac{\sum_{\{\l :~0<\lambda_\l \leq z\}}\frac{1}{T}\sum_{t=1}^T \left| \left\langle \frac{{\bf y}_t}{||{\bf y}_t||_2},{\bf u}_\l\right\rangle \right|^2}{
\sum_{\l=1}^{N-1}\frac{1}{T}\sum_{t=1}^T \left| \left\langle \frac{{\bf y}_t}{||{\bf y}_t||_2},{\bf u}_\l\right\rangle \right|^2},
\end{align} 
which can also be efficiently approximated \cite{behjat2019spectral}.
An example where a signal-adapted spectral design is particularly beneficial is in functional magnetic resonance imaging (fMRI), where the energy of the data 
tends to be highly concentrated at the low end of the spectrum of the cerebellum graph even though there are more eigenvalues at the upper end of the spectrum \cite{behjat2016signal}. 

A second approach to incorporate the training data, presented in \cite{thanou_learning_TSP_2014}, is to force the $J$ spectral filters to be polynomials, and through optimization, find polynomial coefficients that (i) lead to sparse representations of the training data, and (ii) yield filters that cover the spectrum so that the frame is close to being tight (i.e., the ratio of frame bounds $\frac{B}{A}$ is close to 1).  
Specifically, \cite{thanou_learning_TSP_2014} suggests to alternate between (a) a sparse approximation step that fixes the dictionary (i.e., fixes the polynomial filters) and uses orthogonal matching pursuit to find the coefficient matrix ${\bf X}$ that minimizes $||{\bf Y}-\tilde{{\boldsymbol \Phi}}{\bf X}||_F^2$ subject to $||{\bf x}_t||_0 \leq K_0$ for all $t$, where the columns of ${\bf Y}$  are the training signals, $\tilde{{\boldsymbol \Phi}}$ is the current dictionary with normalized atoms, and $K_0$ is a fixed sparsity level; and (b) a dictionary update step that fixes that coefficient matrix ${\bf X}$ and updates the polynomial filter coefficients by minimizing $||{\bf Y}-{{\boldsymbol \Phi}}{\bf X}||_F^2+\mu\sum_{j=1}^J ||{\bf a}_j||_2^2$, where ${\bf a}_j$ is a vector of the $K+1$ polynomial coefficients in \eqref{Eq:poly_kern} for the $j$th filter, subject to constraints ensuring that the learned polynomial filters are nonnegative and cover the whole spectrum ($c-\epsilon \leq \sum_{j=1}^J \hat{g}(\lambda) \leq c + \epsilon$ for some constants $c$ and $\epsilon$).  

\section{Selection of the Center Vertices}\label{Se:centers}
When selecting the center vertices $\V_j$ for the $j$th filter, four broad options are most commonly used: (i) take $\V_j =\V$ for every filter (i.e., localize every pattern to every vertex \change{as done in \cite{hammond2011wavelets}}); (ii) select the center vertex sets such that $\sum_{j=1}^J |\V_j| = N$ (i.e., the total number of atoms is equal to $N$, the number of vertices in the graph), 
which is typically referred to as \emph{critical sampling} \change{(e.g., \cite{li_mcsfb_2018})}; (iii) do not localize every filter to every vertex but do not restrict the total number of atoms, 
resulting in a frame (overcomplete dictionary) with more than $N$ atoms, but fewer than $JN$ atoms; and (iv) localize a single filter $\hat{g}_1(\cdot)$ to a strict subset $\V_1 \subset \V$ of vertices to generate an interpolation basis for a subspace of graph signals, as discussed in Example \ref{Ex:variational}. \change{For option (i), there is no choice to be made regarding the selection of center vertices; for options (ii)-(iv), the specific center vertices for each filter $\hat{g}_j(\cdot)$ must be chosen, and this selection process may also involve deciding how many center vertices to use for each filter. We now outline the main considerations when choosing between these four options and/or selecting the specific center vertices.}

\change{
\subsection{Frame Bounds and Reconstruction Error}
Recall that when $\V_j =\V$ for every $j$, Theorem \ref{Th:frame} outlines the computation of the frame bounds and provides a sufficient condition on the filters to yield a (tight) Parseval frame. When ${\cal D}$ is a Parseval frame, 
\begin{align} \label{Eq:pframe_rec}
{\bf f} = \sum_{j=1}^J \sum_{i=1}^N \langle {\bf f}, {\boldsymbol \varphi}_{i,j}\rangle {\boldsymbol \varphi}_{i,j}=\sum_{j=1}^J \hat{g}_j(\L){\boldsymbol \alpha}_j, 
\end{align}
and so the graph signal ${\bf f}$ can be perfectly reconstructed from the vectors of analysis coefficients, ${\boldsymbol \alpha}_j = \{\langle {\bf f}, {\boldsymbol \varphi}_{i,j}\rangle \}_{i \in \V_j}$.

When $\V_j =\V$ for every $j$ and $G(\lambda)>0$ for all $\lambda \in \sigma(\L)$, the atoms form a frame, but not necessarily a tight frame. This is also the case when each filter is not centered to each vertex, as long as the dictionary atoms span the space of graph signals under consideration (typically $\Rbb^N$). Example \ref{Ex:poly} details three options for inverse transforms in these situations where the dictionary atoms form a frame, but not a tight frame. The least squares solution via conjugate gradient and iterative frame inversion algorithm are accurate but converge slower when the frame is farther from being tight ($\frac{B}{A}>>1$).  
We are not aware of any investigations into how to select the center vertices from general weighted graphs in a manner that explicitly controls the ratio of frame bounds.}

\change{
\subsection{Band-By-Band Reconstruction and Connections to Sampling and Interpolation Theory}\label{Se:band_by_band}
When the filters are not localized to every vertex, but (i) do not overlap too much and (ii) evenly cover the whole spectrum ($\sum_{j=1}^J \hat{g}_j(\lambda) \approx 1$ for all $\lambda$), an alternative approach to synthesis is to try to reconstruct the signal from each band separately and add them up, similar to \eqref{Eq:pframe_rec}, except that we replace $\hat{g}_j(\L)$ with a different synthesis operator  
for each band. The main idea is that  ${\bf f}\approx \sum_{j=1}^J \hat{g}_j(\L){\bf f}$, where
each filtered signal $\hat{g}_j(\L){\bf f}$ belongs to the subspace ${\cal U}_j$ spanned by the eigenvectors $\{{\bf u}_\l\}_{\{\l:~\hat{g}_j(\lambda_\l) \neq 0\}}$. Thus, $\hbox{dim}({\cal U}_j)$ provides an estimate for the number of center vertices, $|\V_j|$, required to recover  $\hat{g}_j(\L){\bf f}$ from the analysis coefficients ${\boldsymbol \alpha}_j$.

The problem of first selecting the center vertices $\V_j$ and then specifying a method to recover $\hat{g}_j(\L){\bf f}$ from ${\boldsymbol \alpha}_j$ falls into}
the broader category of \emph{sampling and interpolation of graph signals}.
Generally speaking, algorithm development in this area proceeds as follows: (i) define a signal model and, if appropriate, a measurement noise model; (ii) specify a reconstruction method that maps a given set of (possibly noisy) sample values of the signal to the entire signal in a way that optimizes a specified error criteria, accounting for the signal model (and noise model) and graph structure; and (iii) given the signal model and reconstruction method, select vertices (often constrained to a fixed number of them) on which to sample the graph signal values. See \cite{anis2016efficient,sakiyama2019eigendecomposition,lorenzo2018sampling,tanaka2020sampling} for more detailed literature reviews and theoretical results on sampling and interpolation of graph signals. \change{The majority of sampling selection methods (i) focus on smooth or lowpass graph signals, and (ii) require the Laplacian eigenvectors associated with the eigenvalues $\lambda_\l$ for which $\hat{g}_j(\lambda_\l)\neq 0$. We focus our review in the remainder of this section on sampling strategies where at least one of these conditions is not met, as these strategies are particularly relevant for LSGFD design.} 

\begin{figure}[t]
\includegraphics[width=.48\linewidth]{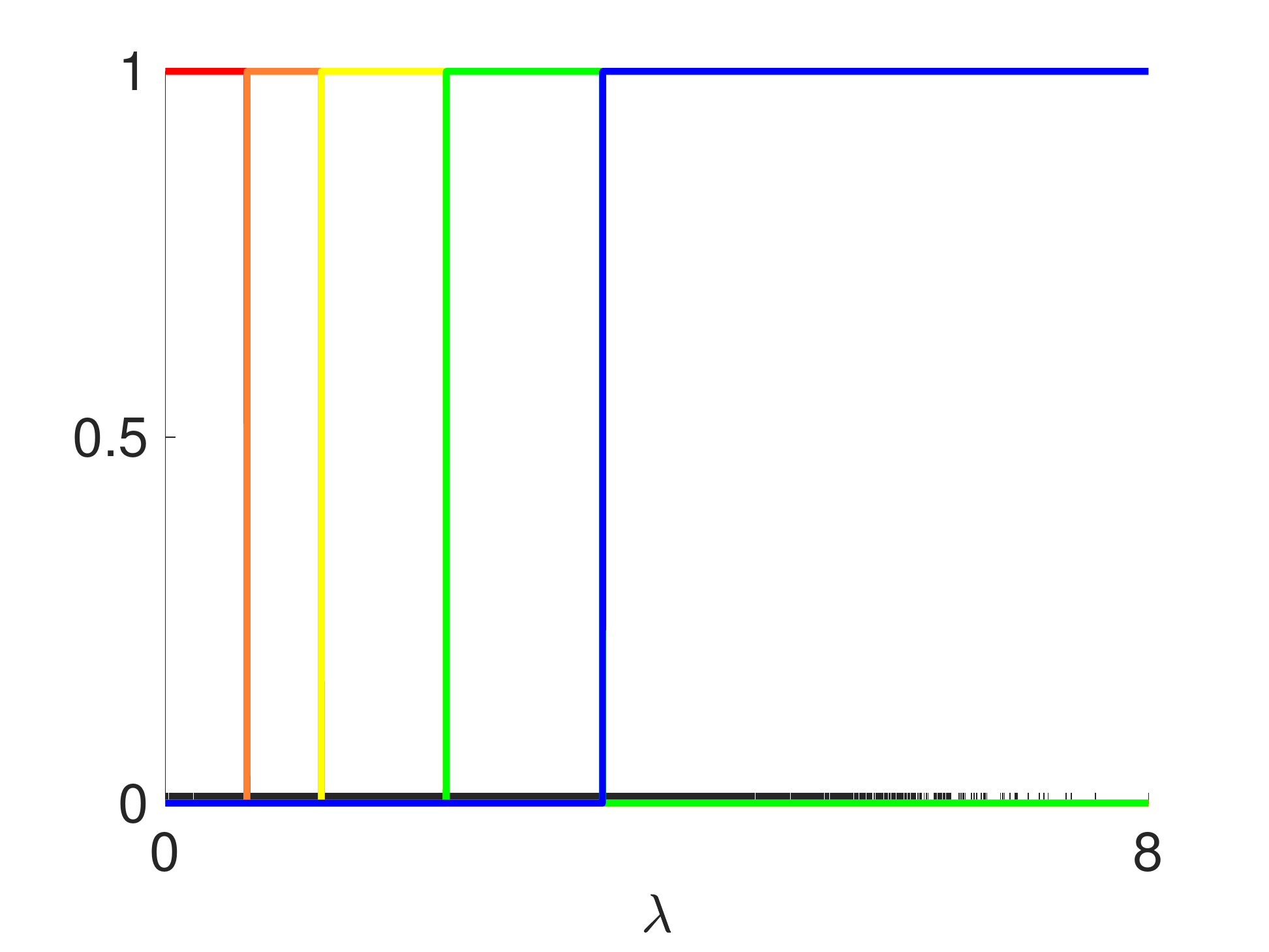} 
\includegraphics[width=.43\linewidth]{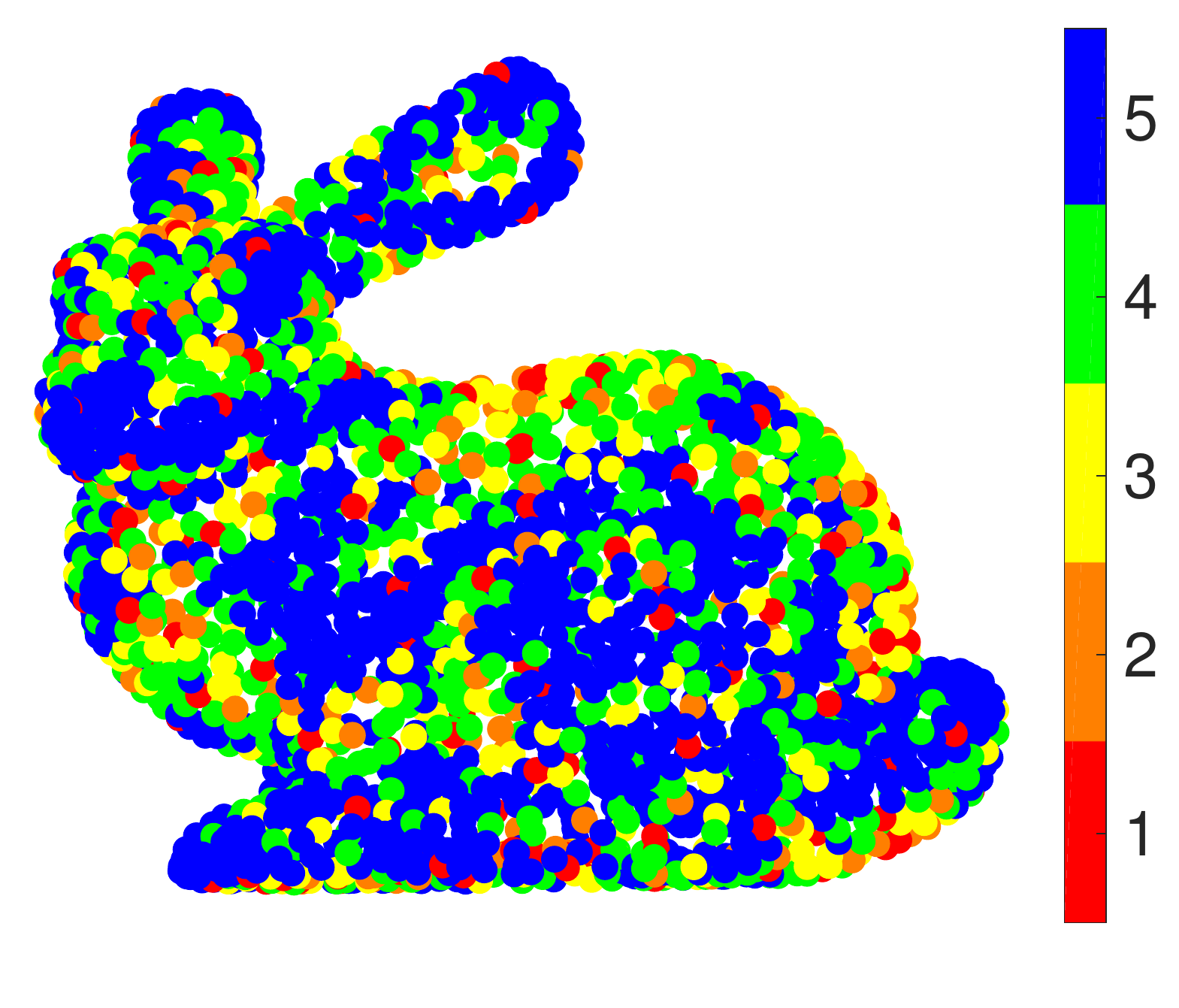} 
\vspace{-.1in}
\caption{Selection of center vertices for a critically sampled LSGFD with five spectrum-adapted ideal octave-band filters on the bunny graph. Exactly one of the five filters is localized to each vertex $i$, according to the mapping shown in the right image, yielding a total of $N$ atoms \cite{li_mcsfb_2018}.}\label{Fig:uniqPart}
\vspace{-.3in}
\end{figure}

For the special case of a set of ideal filters that partition the spectrum, as shown, 
\change{e.g., in Example \ref{Ex:filters}, Example \ref{Ex:poly},} and 
Fig. \ref{Fig:uniqPart}, the critically sampled sets of center vertices can be constructed so as 
to ensure the signal can be recovered perfectly from the $N$ analysis coefficients. 
\begin{theorem}[Prop. 2 and Cor. 1 of \cite{li_mcsfb_2018}]\label{Th:uniquenessPart}
Let $\hat{g}_1(\cdot),\hat{g}_2(\cdot),\ldots,\hat{g}_J(\cdot)$ be a set of spectral filters with the properties that for all $\lambda \in \sigma(\L)$, $G(\lambda)=1$ and $\hat{g}_j(\lambda)\hat{g}_{j^{\prime}}(\lambda)=0$  for all $j \neq j^{\prime}$ (i.e., the filters form a partition of the spectrum $[0,\lambda_{\max}]$ such that each eigenvalue is in the support of exactly one filter). Then there exists a partition $\{\V_1,\V_2,\ldots,\V_J\}$ of $\V$ with $|\V_j|=\sum_{\l=0}^{N-1} \Indicator_{\{\hat{g}_j(\lambda_\l)=1\}}$ such that the resulting dictionary ${\cal D}$ of the form \eqref{Eq:LSGFD} is a basis. Each atom in the basis is orthogonal to all atoms generated from a different filter.   
\end{theorem}
Ref. \cite{li_mcsfb_2018} provides a constructive algorithm for finding the center vertex sets $\{\V_1,\V_2,\ldots,\V_J\}$ in Theorem \ref{Th:uniquenessPart}, the choice of which is not unique. 
This algorithm requires a full eigendecomposition and therefore is only applicable to small or medium graphs. 
Each set $\V_j$ is a \emph{uniqueness set} \cite{pesenson_paley} for the subspace ${\cal U}_j$ spanned by the eigenvectors $\{{\bf u}_\l\}_{\{\l:~\hat{g}_j(\lambda_\l)=1\}}$. That is, any graph signal in ${\cal U}_j$ can be uniquely recovered from its values at the vertices in $\V_j$. In Fig. \ref{Fig:uniqPart}, we display a set of five spectrum-adapted ideal octave-band filters on the bunny graph, and the corresponding partition of the vertices into the uniqueness sets $\V_1$ to $\V_5$. 

Which sampling and interpolation techniques are applicable to \change{smoother approximations to ideal filter banks (e.g., middle row, right in Example \ref{Ex:poly}) for} large, sparse graphs, where computation of the Laplacian eigenvectors is not tractable? The fastest methods are random, not necessarily uniform, sampling methods. For example, leveraging the literature on compressed sensing,\cite{PuyTGV15,li_mcsfb_2018} propose to sample according to a non-uniform distribution with the weight at vertex $i$ proportional to an approximation of $||{\bf U}_{{\cal R}_j}^*{\boldsymbol \delta}_i||_2^2$, where ${\bf U}_{{\cal R}_j}$ is the submatrix of ${\bf U}$ containing the columns corresponding to the support of the ideal filter $\hat{g}_j$.
The $j$th filtered signal 
is then reconstructed 
via the optimization problem
\begin{align}\label{Eq:approx_rec_opt}
\min_{{\bf z}_j \in \Rbb^N} \left\{{\bf z}_j^{*}\phi_j(\L){\bf z}_j + \kappa ||{\boldsymbol \Omega}_{j}^{-\frac{1}{2}}\left({\bf M}_j {\bf z}_j-{\boldsymbol \alpha}_j \right) ||_2^2\right\},
\end{align} 
where ${\boldsymbol \Omega}_{j}$ is a $|\V_j| \times |\V_j|$  diagonal matrix with the $j^{th}$ channel sampling weights of $\V_j$ along the diagonal, ${\bf M}_j \in \Rbb^{|\V_j| \times N}$ is a downsampling matrix that maps a full graph signal to a vector of its values on $\V_j$, and $\kappa>0$ is a parameter to trade off the two optimization objectives: the regularization term ${\bf z}_j^{*}\phi_j(\L){\bf z}_j$ in \eqref{Eq:approx_rec_opt} penalizes reconstructions with support outside of the desired spectral band, and the data fidelity term $||{\boldsymbol \Omega}_{j}^{-\frac{1}{2}}\left({\bf M}_j {\bf z}_j-{\boldsymbol \alpha}_j \right) ||_2^2$ penalizes reconstructions that do not match the analysis coefficients (filtered signal values). From the first-order optimality conditions, the solution to \eqref{Eq:approx_rec_opt} is the solution to the linear system of equations
\begin{align}\label{Eq:approx_rec_sol}
\Bigl(\kappa {\bf M}_j^{*}{\boldsymbol \Omega}_{j}^{-1}{\bf M}_j+\phi_j(\L)\Bigr){\bf  z}_j=\kappa {\bf M}_j^{*}{\boldsymbol \Omega}_{j}^{-1}{\boldsymbol \alpha}_j,
\end{align}
which can be solved, for example, with the preconditioned conjugate gradient method (see \cite{li_mcsfb_2018} for more on the choice of the preconditioner). 

\begin{figure*}
\vspace{-.32in}
\fboxsep=3mm
\fboxrule=2pt
\fcolorbox{darkkhaki}{lightkhaki}{
\begin{minipage}{6.75in}
\begin{example}[Scalable sampling strategies for LSGFD design] \label{Ex:sampling}
We compare the sampling and reconstruction of a lowpass  
and a bandpass  
filtered signal using (i) 
greedy eigendecomposition-free (ED-free) sampling 
\cite{sakiyama2019eigendecomposition}; (ii) uniform random sampling; (iii)  
non-uniform random sampling 
\cite{PuyTGV15,li_mcsfb_2018}; and (iv) 
signal-adapted non-uniform random sampling  
\cite{li_mcsfb_2018}. 
For each polynomial filter type, we show the random sampling distributions, examples of sets of center vertices selected by these methods, and the normalized mean square reconstruction error (NMSE) $\frac{||{\bf z}_j^*-\hat{g}_j(\L){\bf f}||_2^2}{||\hat{g}_j(\L){\bf f}||_2^2}$ between the filtered signal and the 
reconstruction ${\bf z}_j^*$, computed via \eqref{Eq:approx_rec_sol}, averaged over 50 trials of choosing the random center locations (samples). 
The greedy ED-free method explicitly prioritizes choosing samples 
that are not too close to the previously chosen samples. 
The non-uniform random sampling weights  
are derived from a computationally-efficient approximation to the ideal 
distribution, for which the probability of sampling vertex $i$
is proportional to $||{\bf U}_{{\cal R}_j}^*{\boldsymbol \delta}_i||_2^2$.  
This approach has a close connection to \emph{leverage score} sampling in the statistics and numerical linear algebra literature 
\cite{drineas2012fast,mahoney2009cur}. The signal-adapted non-uniform random sampling distribution (shown in the third column of the right group of images) is computed 
 by multiplying the initial non-uniform weight
 associated with vertex $i$ (shown in the second column of the right group of images) by $\log(1+|(\hat{g}_j(\L){\bf f})(i)|)$ and then renormalizing. The intuition is that it is beneficial to take additional samples in regions of the graph where the filtered signal has the most energy. For this particular bandpass signal, the regions of highest energy are around the midsection and tail of the bunny. Indeed, the signal-adapted method leads to more samples in these areas, and, in turn, to better reconstruction performance \cite{li_mcsfb_2018}. The average NMSE of the signal-adapted sampling method at the dashed vertical line  
 represents a 79\%, 78\%, and 83\% reduction of the errors of the ED-free, non-uniform random sampling, and uniform random sampling methods, respectively.
\vspace{.15in}

\begin{minipage}{\linewidth}
\begin{minipage}{.395\linewidth}
\centering
   {\small{Lowpass polynomial filter}}
      \vspace{.01in}
            \centerline{\includegraphics[width=.9\linewidth,page=1]{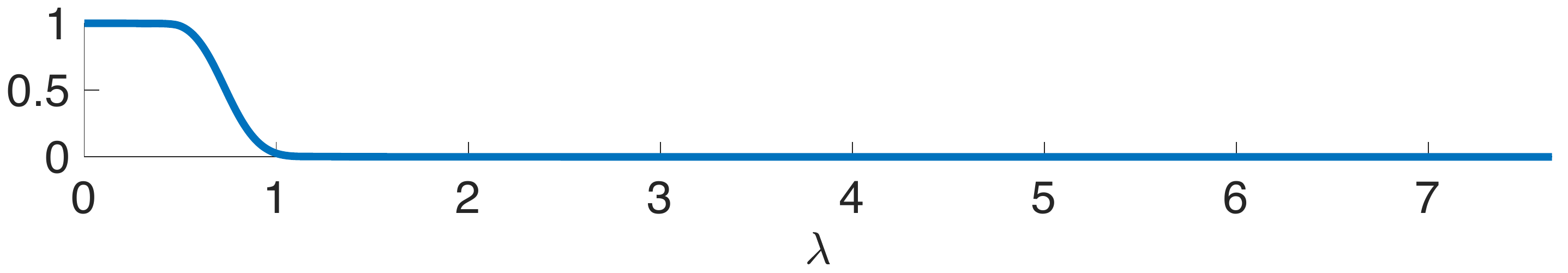}} 
               \vspace{-.05in} 
               
\begin{minipage}{.48\linewidth}
\centering
   {\small{Lowpass}}
   
   {\small{filtered signal}}
   \vspace{.01in}
\includegraphics[width=\linewidth,page=8]{figures/sampling}

   {\small{ED-free greedy centers \cite{sakiyama2019eigendecomposition}}}
   \vspace{.01in}
\includegraphics[width=\linewidth,page=1]{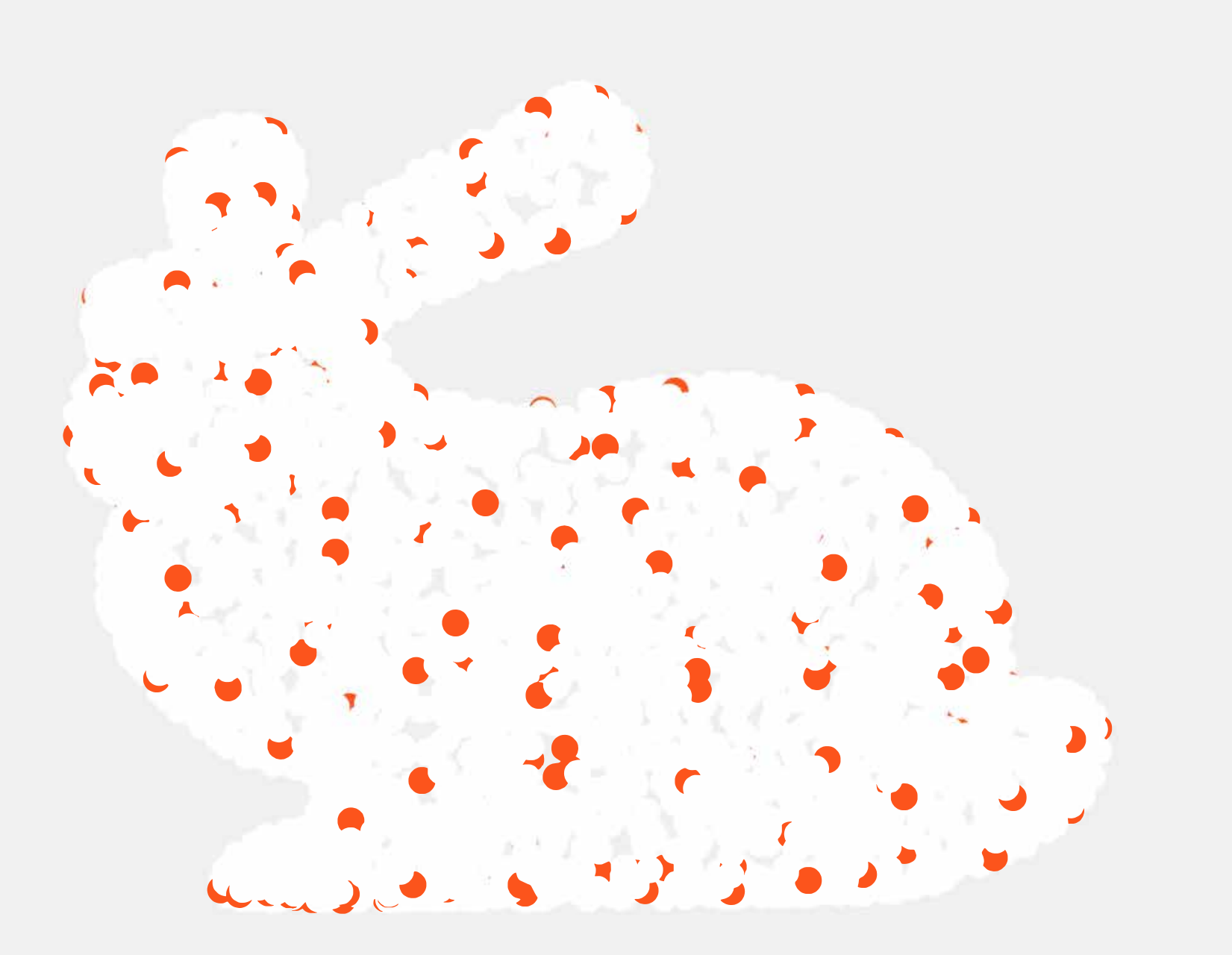}

\end{minipage}
\hfill
\begin{minipage}{.48\linewidth}
\centering
   {\small{Non-uniform random sampling weights \cite{PuyTGV15}}}
   \vspace{.01in}
\includegraphics[width=\linewidth,page=9]{figures/sampling}

   {\small{Non-uniform random centers}}
         \vspace{.01in}
\includegraphics[width=\linewidth,page=10]{figures/sampling}

\end{minipage}

      \vspace{.02in}
      
  {\small{Lowpass reconstruction error}}
      \vspace{.02in}
      
            \centerline{\includegraphics[width=1\linewidth,page=1]{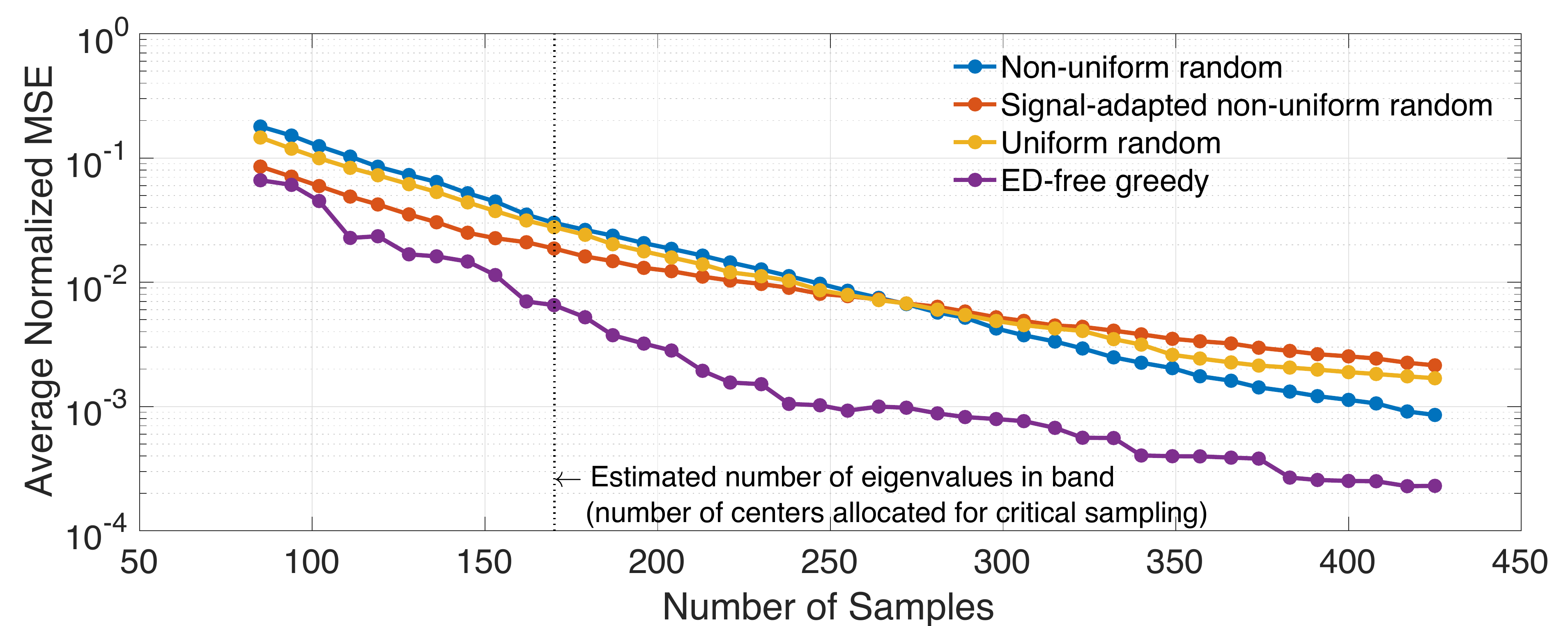}} 

\end{minipage}
\hfill
\begin{minipage}{.005\linewidth}
\centering
\vspace{.1in}

\lpipe
\end{minipage}
\hfill
\begin{minipage}{.585\linewidth}
\centering
   {\small{Bandpass polynomial filter}}
      \vspace{.01in}
            \centerline{\includegraphics[width=.608\linewidth,page=2]{figures/filters50}} 
               \vspace{-.05in} 
               
\begin{minipage}{.324\linewidth}
\centering
   {\small{Bandpass}}
   
   {\small{filtered signal}}
   \vspace{.01in}
\includegraphics[width=\linewidth,page=2]{figures/sampling}

   {\small{ED-free greedy centers \cite{sakiyama2019eigendecomposition}}}
   \vspace{.01in}
\includegraphics[width=\linewidth,page=2]{figures/edfree_selected}

\end{minipage}
\hfill
\begin{minipage}{.324\linewidth}
\centering
   {\small{Non-uniform random sampling weights \cite{PuyTGV15}}}
   \vspace{.01in}
\includegraphics[width=\linewidth,page=3]{figures/sampling}

   {\small{Non-uniform random centers}}
      \vspace{.01in}
\includegraphics[width=\linewidth,page=4]{figures/sampling}

\end{minipage}
\hfill
\begin{minipage}{.324\linewidth}
\centering
   {\small{Signal-adapted sampling weights \cite{li_mcsfb_2018}}}
   \vspace{.01in}
\includegraphics[width=\linewidth,page=5]{figures/sampling}

   {\small{Signal-adapted random centers}}
   \vspace{.01in}
\includegraphics[width=\linewidth,page=6]{figures/sampling}

\end{minipage}

      \vspace{.02in}
      
  {\small{Bandpass reconstruction error}}
      \vspace{.02in}
      
            \centerline{\includegraphics[width=.675\linewidth,page=2]{figures/sampling_error}} 

\end{minipage}
\vspace{.02in}

\change{
Key insights highlighted by this example include:
\begin{enumerate}
\item For the lowpass filter, the results are consistent with the common insight from the graph sampling literature that the scalable greedy methods are slower than random sampling, but can yield better reconstruction performance by forcing the samples to be more spread out \cite{tanaka2020sampling}.
\item For bandpass or highpass filters, however, forcing or incentivizing the samples to be more spread out does not necessarily (or even usually) improve reconstruction performance.
\item When performing random sampling to choose the center vertices, using a non-uniform sampling distribution is more important for bandpass and highpass filtered signals. For lowpass filtered signals, the non-uniform random sampling distribution is much closer to uniform as the energy distributions of the eigenvectors at the low end of the spectrum tend to be more evenly spread across the graph than those associated with higher Laplacian eigenvalues.
\item In the classical sampling and interpolation problem, the complete signal is not usually available when selecting the sample locations. However, in the context of subsampled LSGFDs, it is often feasible and highly beneficial to adapt the non-uniform random selection of the center vertices to the energy distribution of each filtered signal.
\end{enumerate}
\vspace{-.1in}
}
\end{minipage}
\end{example}
\end{minipage}
}
\end{figure*}

The other more scalable options are efficient greedy methods that do not rely on the Laplacian eigendecomposition, such as  
\cite{sakiyama2019eigendecomposition,bai2019fast}. For example, for each $j$, the eigendecomposition-free (ED-free) method of \cite{sakiyama2019eigendecomposition} attempts to select center vertices $i$ such that (i) $||T_i g_j ||_1=||{\boldsymbol \varphi}_{i,j}||_1$ is large, and (ii) the atoms ${\boldsymbol \varphi}_{i,j}$ do not overlap too much in the vertex domain. The binary search Gershgorin disc alignment (BS-GDA) method of \cite{bai2019fast}  aims to select vertices in a way that reduces the condition number of the matrix $\kappa {\bf M}_j^{*}{\boldsymbol \Omega}_{j}^{-1}{\bf M}_j+\phi_j(\L)$ in \eqref{Eq:approx_rec_sol} (with ${\boldsymbol \Omega}_{j}={\bf I}_N$ and a regularization term of $\phi(\L)=\L$), improving the reconstruction stability. \change{These methods may lead to better reconstruction for lowpass signals, but are slower than random sampling and are either not applicable to or may lead to worse reconstruction for bandpass signals (see Sec. \ref{Se:samp_comp} and Example \ref{Ex:sampling}).}

\change{
\subsection{Allocation of Center Vertices Across the Filters}
When the sum of the cardinalities $\{|V_j|\}$ is constrained (e.g., to $N$), how should we allocate the center vertices (samples) across the filters?}
One option is to estimate the spectral cumulative density and match the number of samples to the estimated number of eigenvalues in each band. Another option 
is to adjust the distribution of samples to the signal ${\bf f}$ 
by multiplying the initial allocation of samples to $\V_j$ by a factor that
increases with the amount of energy in the filtered signal $\hat{g}_j(\L){\bf f}$ \cite{li_mcsfb_2018}.

\change{
\subsection{Computational Complexity} \label{Se:samp_comp}
We briefly analyze the computational complexity implications of the choice of method for selecting the center vertices on large, sparse graphs, 
the most typical and important case for complexity considerations. We assume the filters are degree $K$ polynomials throughout this analysis. First, for selecting the center vertices, the costs of the random sampling methods (either signal-adapted or not) are negligible if the spectral density of the Laplacian has been estimated (or ${\cal O}(K|\E|)$ if it has not), at least an order of magnitude faster than the ${\cal O}(KN|\E|)={\cal O}(N^2)$ complexity of the greedy ED-free method. Second, regardless of the selection of center vertices, applying the analysis operator (forward transform) has the same complexity ${\cal O}(K|\E|)$ as the case where we localize each filter to every vertex; i.e., 
there is not a significant computational savings in the analysis step from subsampling the center vertices. Third, as detailed in Example \ref{Ex:poly}, the complexity of the inverse transform is  ${\cal O}(2TK|\E|)$ if either the least squares
least squares solution via conjugate gradient or iterative frame inversion algorithm is used, and ${\cal O}(K|\E|)$ if the inverse is approximated by a constant multiple of the analysis operator. The band-by-band reconstruction method of \eqref{Eq:approx_rec_sol} has a similar complexity of  ${\cal O}(TJ\tilde{K}|\E|)$, where $\tilde{K}$ is the degree of the polynomial penalty function $\phi$. These inverse transform complexities do not depend on either the method used to select the center vertices or the number of center vertices for each filter. In summary, for large, sparse graphs, the complexities of the design, forward transform, and inverse transform are all no more than linear in the number of vertices, as long as each filter is localized to every vertex or the center vertices are selected through non-uniform random sampling.  

\subsection{Memory}
The total number of analysis coefficients to store for a graph signal of length $N$ is $\sum_{j=1}^J |\V_j|$. Thus, the complete sampling of option (ii) yields $NJ$ coefficients to store, while the critical sampling of option (iii), e.g., reduces that number to $N$. Except for special filters such as the heat kernel, the ED-free method requires storing the entire matrices $\{\hat{g}_j(\L)\}$; thus, when memory is an important concern and a critically sampled dictionary is desired, random sampling should be used to select the center vertex sets.}

\section{Theoretical Considerations and Metrics} \label{Se:theory}
Whether the dictionary of atoms is critically sampled (a basis for the vector space of signals) or redundant (a frame for the vector space of signals), different mathematical characteristics can be beneficial for different applications. Desirable characteristics of dictionaries may include, for example:
\begin{itemize}
\item The atoms are not too correlated, in order to enhance the discriminatory power of taking inner products between each atom and a target signal; 
\item The atoms are jointly localized in the vertex domain and the spectral domain. 
\item Classes of signals on the graph (e.g., globally smooth or locally smooth signals), can be represented as sparse combinations of the dictionary atoms.
\end{itemize} 
We briefly describe each of these considerations, in order. 

One common metric for capturing the correlation between dictionary atoms is the \emph{cumulative coherence} \cite{tropp} of the dictionary ${\cal D}$, which, for a given sparsity level $k$, is defined as 
\begin{align*}
\mu_1(k):= \max_{\{\Theta \subset {\cal D}:~|\Theta|=k\}}~\max_{{\boldsymbol \psi} \in {\cal D} \setminus \Theta}~\sum_{{\boldsymbol \theta} \in \Theta} \frac{|\langle {\boldsymbol \psi},{\boldsymbol \theta} \rangle|}{ ||{\boldsymbol \psi}||_2 ||{\boldsymbol \theta}||_2}.
\end{align*}

Uncertainty principles for signals on graphs (e.g., \cite{tsitsvero2016signals,perraudin_uncertainty_APSIPA_2018,van2017slepian,teke2017uncertainty,erb2019shapes}) characterize the degree to which graph signals can be jointly localized in both the vertex (spatial) domain and the spectral (frequency) domain. Particularly interesting for guiding the selection of the center vertices of LGSFDs are the uncertainty principles developed in \cite{tsitsvero2016signals,erb2019shapes}. Let ${\boldsymbol \gamma}\in \Rbb^N$ be a  \emph{spatial filter or window function}; i.e., a set of weights assigned to the vertices, with $0\leq \gamma_i \leq 1$ for all $i \in \V$ and $\max_{i \in \V} \gamma_i = 1$. Typical examples of such spatial filters include (i) setting the weights equal to 1 for all vertices in a certain region of the graph and 0 elsewhere, or (ii) choosing each $\gamma_i$ to be a measure of the distance from vertex $i$ to a fixed center vertex $n$. Then for any graph signal ${\bf f}$, the quantity ${\bar{\bf m}}_{\boldsymbol \gamma}({\bf f}):=\frac{{\bf f}^*{\boldsymbol \Gamma}{\bf f}}{||{\bf f}||_2^2}$, where ${\boldsymbol \Gamma}$ is a diagonal matrix with diagonal elements equal to the weights ${\boldsymbol \gamma}$, captures the portion of the energy of ${\bf f}$ that is located in the specified region of the graph (i.e., those vertices with high weights $\gamma_i$). Similarly, for each filter satisfying $0 \leq \hat{g}_j(\lambda_\l) \leq 1$ for all $\l$ and $\max_\l \hat{g}_j(\lambda_\l)=1$, the quantity ${\bar{\bf c}}_{\hat{g}_j}({\bf f}):=\frac{{\bf f}^*\hat{g}_j(\L){\bf f}}{||{\bf f}||_2^2}$ captures the portion of the energy of ${\bf f}$ that is located in the region of the spectrum specified by the filter. The uncertainty principles in \cite{tsitsvero2016signals,erb2019shapes} characterize and provide algorithms to approximate the sets ${\cal W}_{{\boldsymbol \gamma},\hat{g}_j} := \left\{\bigl({\bar{\bf m}}_{\boldsymbol \gamma}({\bf f}), {\bar{\bf c}}_{\hat{g}_j}({\bf f})\bigr) : ||{\bf f}||_2=1 \right\} \subseteq [0,1]^2.$ For example, for a given filter pattern, such a principle can inform how localized a dictionary atom of the form \eqref{Eq:atom_form} can be in the vertex domain (typically around the center vertex $i$). 
Due to the irregularity of general graphs and the possibility of highly localized Laplacian eigenvectors, uncertainty does not always exist, in which case  ${\cal W}_{{\boldsymbol \gamma},\hat{g}_j}$ may be equal to $[0,1]^2$. 

With regard to the third desirable characteristic above, nearly a decade after being listed as an open issue in \cite{shuman2013emerging}, relatively little progress has been made in developing a mathematical theory of approximation linking structural properties of graph signals and their underlying graphs to the sparsity of the analysis coefficients $\{\langle {\bf f}, {\boldsymbol \varphi}_{i,j} \rangle\}$, analogous to the theory for wavelet transform coefficients in Euclidean domains (see, e.g., \cite{donoho_theory}). In \cite{chen2018multiresolution}, vertex domain dictionary designs are proposed that sparsely represent defined classes of piecewise constant and piecewise smooth graph signals. For the special case of signals on circulant graphs, \cite{kotzagiannidis2017splines} defines a family of complex exponential polynomial graph signals and designs a class of filters that annihilate graph signals from this class; i.e., $\langle {\bf f }, T_i g_j \rangle = 0$ for all $i$. Reference \cite{ricaud_sparsity_SPIE_2013} defines notions of global and local regularity for graph signals, and begins to connect the regularities of the signals and the degree of the polynomial filters to the decay of the magnitudes of  
spectral graph wavelet analysis coefficients.    

\section{Application Examples and Comparison Via Numerical Experiments}\label{Se:experiments}
In this section, we \change{first describe LSGFD transform methods for two signal processing tasks -- denoising and non-linear approximation (compression) -- and then} perform 
a set of targeted numerical experiments that attempt to answer high-level design questions and help focus the community's research going forward. \change{Our objective is not to determine} whether a specific dictionary outperforms other dictionaries in a specific task; \change{for that type of analysis, we encourage readers to experiment on their own data with the publicly available code used to generate all figures and tables in this article.}

\change{
\subsection{Denoising}
We consider the denoising problem of recovering a graph signal ${\bf f}$ from a noisy observation, ${\bf y} = {\bf f} + {\boldsymbol \xi}$, where ${\boldsymbol \xi} \in \Rbb^N$ is a white Gaussian noise vector whose entries are independent and identically distributed (i.i.d.) normal random variables with mean 0 and known variance $\sigma^2$. We use the common wavelet denoising method of performing soft thresholding on the LSGFD transform coefficients, and then resynthesizing the signal with the inverse transform. 
Specifically, we take 
${\bf f}_{\hbox{denoised}}={{\Phi^*}^{-1}}(\bar{\boldsymbol \alpha})$, 
where each soft thresholded coefficient in the vector $\bar{\boldsymbol \alpha}$ is set to 
\begin{align}\label{Eq:soft}
\bar{\alpha}_{i,j}=\sgn(\langle {\bf y}, {\boldsymbol \varphi}_{i,j}\rangle)\cdot\max\{0,|\langle {\bf y}, {\boldsymbol \varphi}_{i,j}\rangle|-\varUpsilon_{i,j}\}.
\end{align} 
If the dictionary used to transform the noisy signal is a Parseval frame, then ${{\Phi^*}^{-1}}(\bar{\boldsymbol \alpha})={\boldsymbol \Phi}\bar{\boldsymbol \alpha}$; otherwise (e.g., not every filter is localized to every vertex or the filters do not satisfy the tight frame condition of Theorem \ref{Th:frame} due to polynomial approximation), ${{\Phi^*}^{-1}}$ can be taken to be any of the three approximate inverse transforms discussed in Example \ref{Ex:poly}. 

\begin{table*}[t]
\centering
{\footnotesize
\tabcolsep=0.11cm
\begin{tabular}{l|cccc|cccc|cccc|cccc}
\cline{2-17}
 & \multicolumn{4}{ c| }{Sensor Network} & \multicolumn{4}{ c| }{Bunny} & \multicolumn{4}{ c| }{Minnesota} & \multicolumn{4}{ c }{Cerebellum}\\ 
\cline{2-17}
\multicolumn{1}{r|}{$\sigma$/$\sigma_{\bf f}$} & 1/8 & 1/4 & 1/2 & 1 & 1/8 & 1/4 & 1/2 & 1& 1/8 & 1/4 & 1/2 & 1 & 1/8 & 1/4 & 1/2 & 1 \\ 
\clineB{1-17 \vspace{.02in} }{2.5}
\multicolumn{1}{ l| }{Spectral graph wavelets \cite{hammond2011wavelets}} &  1.56 & 2.29 & {\bf 4.04}  & 6.37  & {\bf 4.99}  & {\bf 5.92} & {\bf 7.69}  & {\bf 10.46} &  5.81 &  6.69 & 8.50 & 10.93  &  0.53 & 1.74 & 3.76  & 6.85 \\
\multicolumn{1}{ l| }{Cohen-Daubechies-Feauveau 9/7 filters \cite{sakiyama2016spectral}} & 1.50 & 2.04  & 3.83  & 5.85  & 4.76 & 5.68 & 7.44  & 10.42  &  5.51 & 6.52 & 8.61  & 11.86  & 0.50 & 1.60  & 4.03  & 6.86 \\ 
\cline{1-17}
\multicolumn{1}{ l| }{Uniform ideal filters} & 1.00  & 1.89  & 3.52  & 5.75  & 3.51  & 4.54  & 6.48  & 9.33  & 3.50 & 4.72  & 6.45  & 7.20  &  0.66 &  {\bf 2.06} & 4.02  & 7.05 \\ 
\multicolumn{1}{ l| }{Octave-band (wavelet) ideal filters} & 0.96 & 1.82  &  3.61 &  6.12 & 3.67 &  4.81& 6.97 & 10.16 &  4.37 & 5.81 & 8.23  & 11.61  &  0.44 & 1.49  & 3.62  & 6.55 \\
\multicolumn{1}{ l| }{DCT with frequency conversion \cite{sakiyama2016spectral}} & {\bf 1.77} & {\bf 2.30} & 3.94  & 6.26  & 4.87 & 5.88 & 7.61 & 10.32 &  4.86 & 5.83 & 7.51  & 8.73  & 0.63 & 1.89 & {\bf 4.14}  & {\bf 7.10} \\ 
\multicolumn{1}{ l| }{Log-warped DCT with frequency conversion \cite{sakiyama2016spectral}} &  1.39 & 2.07 &  3.85 & 6.26 & 4.51 & 5.76 & 7.60  & 10.37  &  {\bf 6.29} & {\bf 7.08} & {\bf 9.10} & {\bf 12.06}  & 0.41 & 1.39  & 3.59  & 6.31 \\ 
\multicolumn{1}{ l| }{Uniform translates: Itersine kernel \cite{shuman2013spectrum}} & 1.33  & 2.14  & 3.93  & 6.37  & 4.61  & 5.69  & 7.36  & 10.41  & 4.63 & 5.89  & 7.94  & 9.57 & {\bf 0.69}  & 2.03  & 4.10  & 7.09 \\
\multicolumn{1}{ l| }{Log-warped translates (wavelets): Itersine kernel \cite{shuman2013spectrum}} &  1.42 & 2.15 & 3.98 & 6.38 & 4.40 & 5.76 & 7.48 & 10.44  & 5.81 & 6.76 & 8.86  & 11.94  & 0.55 & 1.70  & 3.90  & 6.99 \\
\cline{1-17}
\multicolumn{1}{ l| }{Spectrum-adapted uniform translates: Itersine kernel \cite{shuman2013spectrum}} & 1.48 & 2.11  & 3.97  & 6.27  & 4.35 & 5.46 & 7.14  & 10.32  &  5.11 &  6.23 &  8.48 & 11.22  & 0.68 & 2.05  & 4.05  & 7.03 \\ 
\multicolumn{1}{ l| }{Spectrum-adapted wavelets: Itersine kernel \cite{shuman2013spectrum}} &  1.32 & 2.12 & 4.00 & {\bf 6.61}  & 4.46 & 5.60 & 7.59  & 10.45  &  5.69 & 6.65  & 8.94  & 12.02  & 0.63  & 1.95  & 4.05  & 7.10 \\
\cline{1-17}
\multicolumn{1}{ l| }{Signal-adapted Meyer-type \cite{behjat2016signal}} &  &  &  &  &  &  &  &  &  &  &  &  &  0.39 & 1.38  & 3.53 & 6.85 \\ 
\clineB{1-17}{2.5}
\end{tabular}
\vspace{.25cm}
}
\caption{\change{Denoising performance for different graph signals, noise levels, and filter designs, shown in SNR improvement: 
$\Delta_{\hbox{SNR}}=10\log_{10}\left(\frac{||{\bf f}||_2^2}{||{\bf f}_{\hbox{denoised}} - {\bf f}||_2^2}\right)-10\log_{10}\left(\frac{||{\bf f}||_2^2}{||{\boldsymbol \xi}||_2^2}\right)=10\log_{10}\left(\frac{||{\boldsymbol \xi}||_2^2}{||{\bf f}_{\hbox{denoised}} - {\bf f}||_2^2}\right)$
}} 
\label{Ta:denoising}
\vspace{-.5cm}
\end{table*}

We set the soft thresholds $\varUpsilon_{i,j}$ in \eqref{Eq:soft} 
to 0 for the scaling functions (atoms generated from filters satisfying $\hat{g}_j(0)>0$) since these coefficients are not expected to be sparse. For the other thresholds, as in \cite{gobel2018construction}, to account for the fact that the dictionary atoms have different norms, we use atom-adapted thresholds of the form $\varUpsilon_{i,j}=\varUpsilon_j\sigma ||{\boldsymbol \varphi}_{i,j}||_2$, where $\{\varUpsilon_j\}$  
are the $J-1$ scalar parameters. 
The optimal value of each $\varUpsilon_j$  
is estimated with Stein's Unbiased Risk Estimator (SURE). As detailed in \cite{de2019data}, for soft thresholding on a dictionary with a single lowpass filter ($j=1$), this amounts to solving the following for each $j=2,3,\ldots,J$:
\begin{align}\label{Eq:sure}
\argmin_{\varUpsilon_j}  \sum_{i=1}^N \left\{ 
\begin{array}{l}
\min\bigl\{|\langle {\bf y}, {\boldsymbol \varphi}_{i,j}\rangle|^2 ,\varUpsilon_j^2\sigma^2 ||{\boldsymbol \varphi}_{i,j}||_2^2\bigr\} \vspace{.03in} \\
+ 2 \sigma^2   ||{\boldsymbol \varphi}_{i,j}||_2^2 \Indicator_{\{|\langle {\bf y}, {\boldsymbol \varphi}_{i,j}\rangle| \geq \varUpsilon_j\sigma ||{\boldsymbol \varphi}_{i,j}||_2\}}
\end{array}
\right\}.
\end{align}
Importantly, the objective in \eqref{Eq:sure} does not depend on the unknown signal ${\bf f}$. 
The thresholds $\{\varUpsilon_{i,j}\}$ and the objective in  \eqref{Eq:sure} depend on the atom norms $\{||{\boldsymbol \varphi}_{i,j}||_2\}$. For small to medium graphs, these can be computed exactly; for polynomial filter designs on large, sparse graphs, they can be efficiently estimated as
$||{\boldsymbol \varphi}_{i,j}||_2 \approx \frac{1}{\sigma} \hbox{s.d.}\bigl(\{{\boldsymbol \delta}_i^*\hat{p}_{j,K}(\L){\boldsymbol \eta_l}\}_{l=1}^L\bigr),
$
where $\hat{p}_{j,K}$ is a polynomial approximation to $\hat{g}_j$ and $\{{\boldsymbol \eta_l}\}_{l=1}^L$ is a sequence of i.i.d. random vectors, each with i.i.d. entries normally distributed with mean 0 and variance $\sigma^2$.
}

\change{
\subsection{Non-linear approximation}
One approach to compression of smooth and piecewise-smooth graph signals is to represent them as sparse linear combinations of LSGFD atoms. To find such sparse representations for a graph signal  
${\bf f}$, 
the sparse coding optimization  
\begin{align*}  
\argmin_{\boldsymbol \alpha}   || {\bf f}-{\boldsymbol \Phi}{\boldsymbol \alpha} ||_2^2 \hbox{~~~subject to } ||{\boldsymbol \alpha}||_0 \leq T_0,
\end{align*}
where $T_0$ is a predefined sparsity level,
can be approximately solved, e.g., by normalizing the dictionary atoms and then applying the greedy orthogonal matching pursuit (OMP) algorithm \cite{tropp}. When the graph is very large and OMP becomes impractical computationally, a common approximation method is to hard threshold the analysis coefficients (normalized by an estimate of the corresponding atom norm), and then resynthesize the signal from the $T_0$ largest magnitude coefficients, via one of the inverse transform methods described in Example \ref{Ex:poly}.
}

\begin{figure*}[t]
\begin{minipage}{.2\linewidth}
\includegraphics[width=\linewidth]{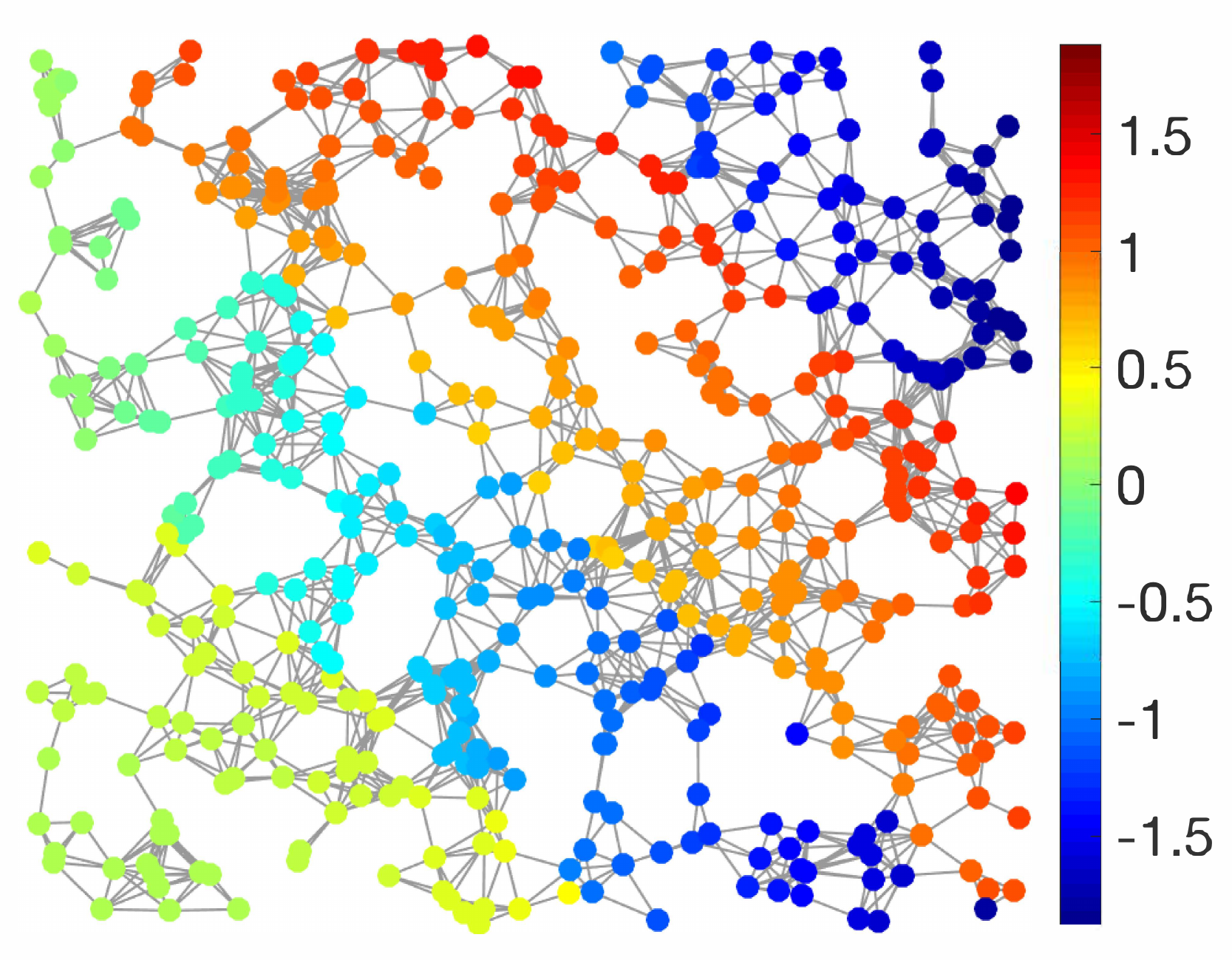}  \\
\vspace{-.08in}

\end{minipage}
\hfill
\begin{minipage}{.21\linewidth}
\includegraphics[width=\linewidth]{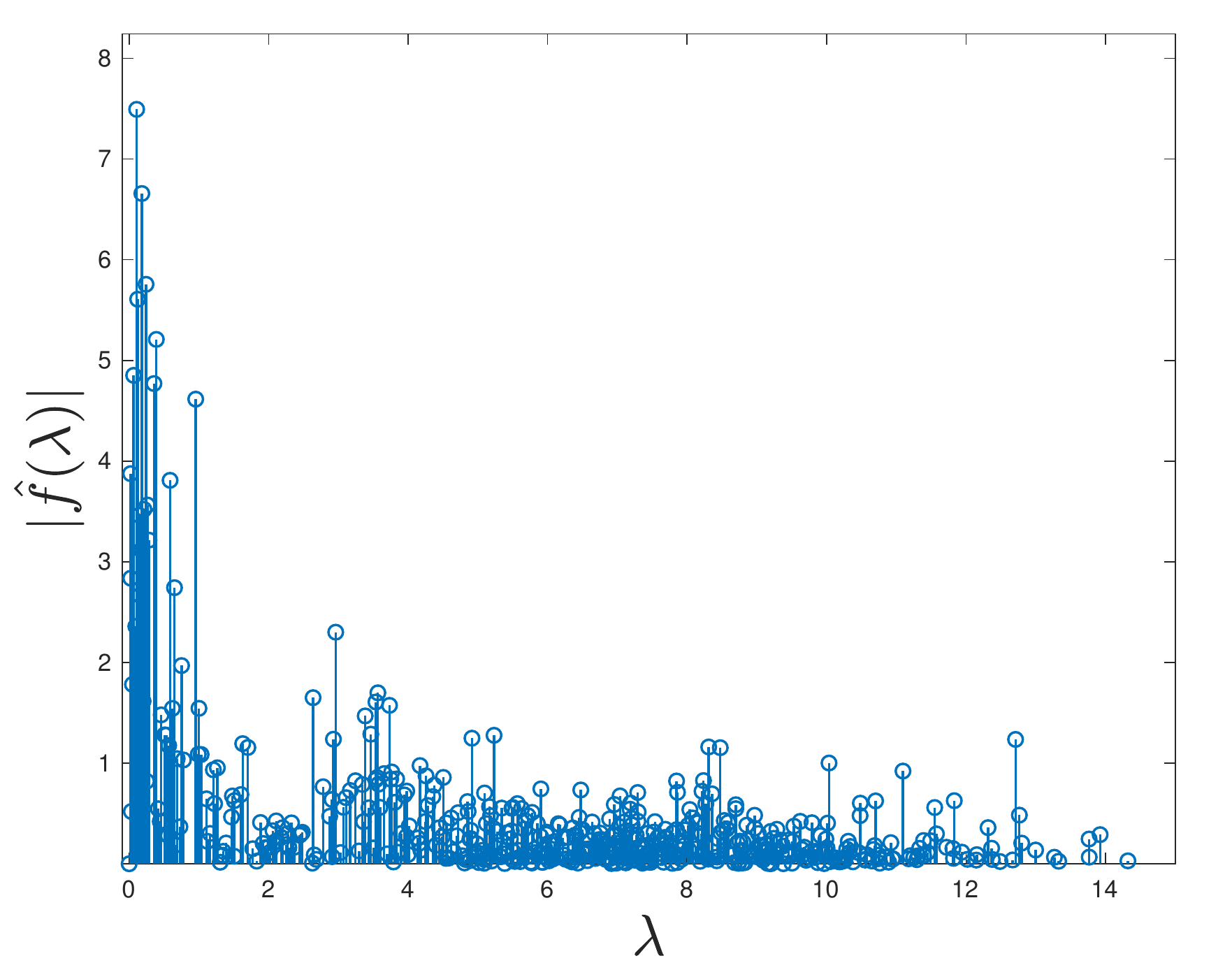} 
\end{minipage}
\hfill
\begin{minipage}{.31\linewidth}
\includegraphics[width=\linewidth]{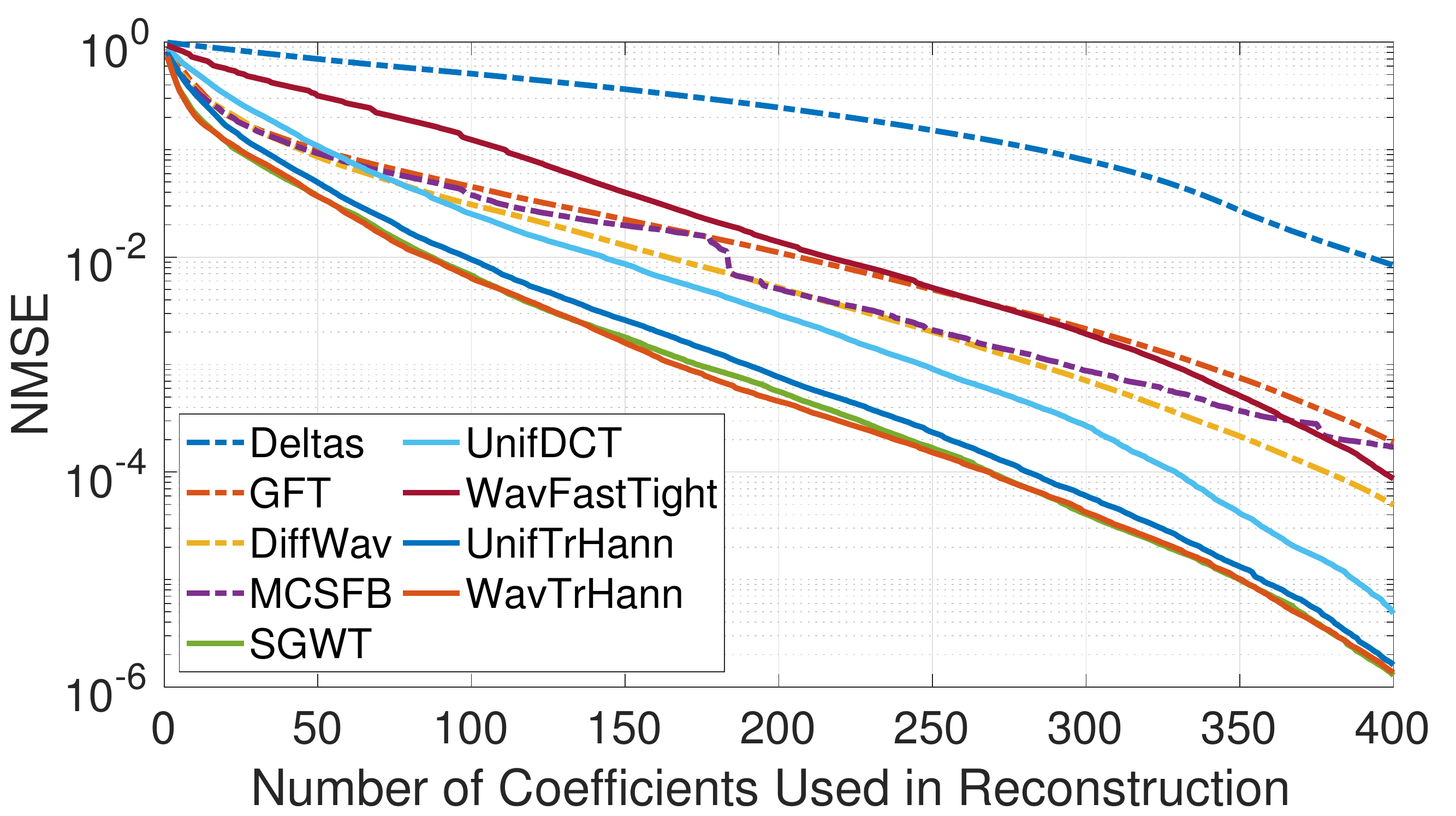}
\end{minipage}
\hfill
\begin{minipage}{.245\linewidth}
\includegraphics[width=\linewidth]{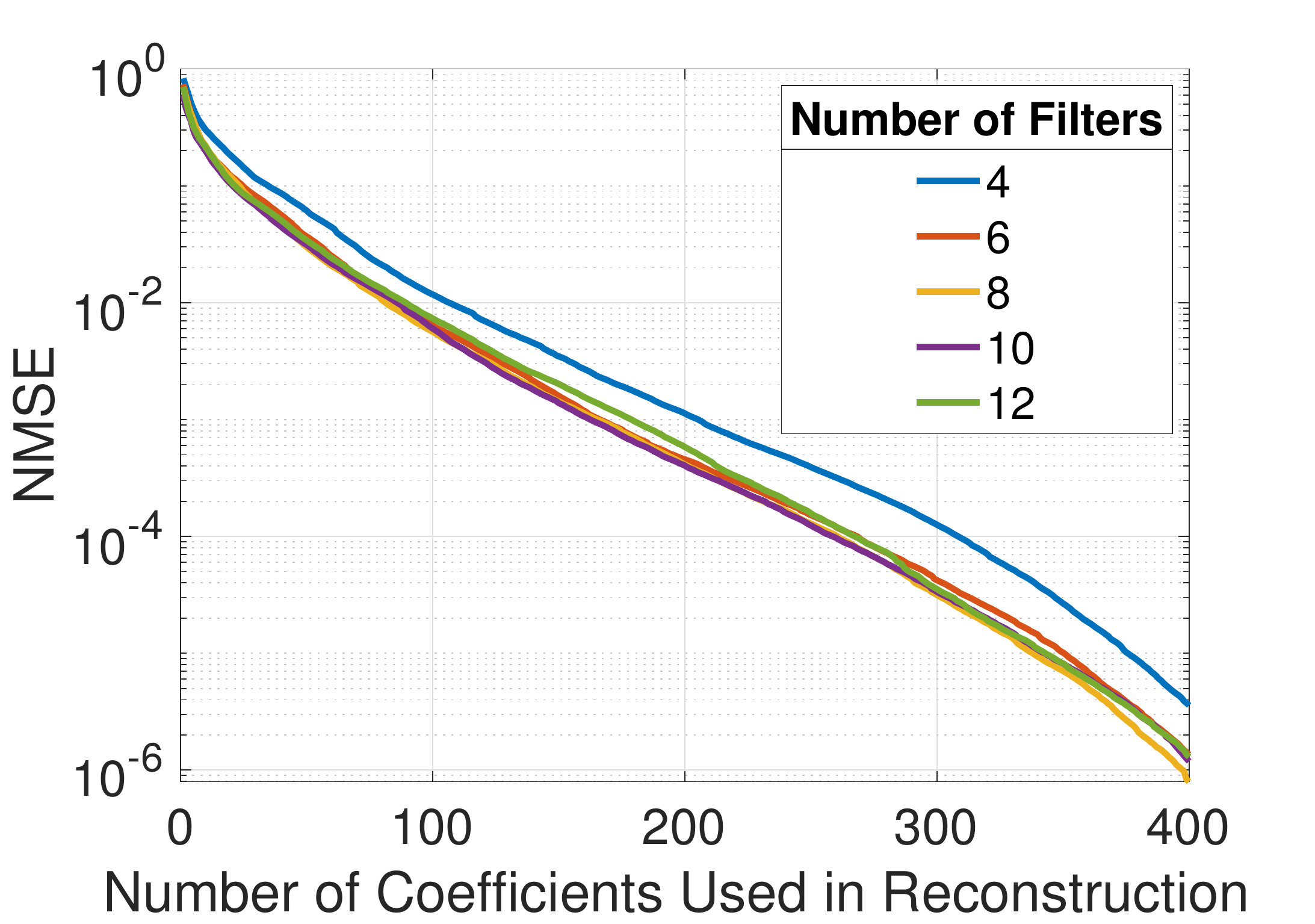}
\end{minipage}
\caption{\change{Non-linear approximation of the piecewise-smooth signal on a random sensor network with 500 vertices from \cite{shuman_TSP_multiscale}. The left two images show the signal in the vertex domain and spectral domain, respectively.
The third image shows the normalized mean square reconstruction error incurred by applying the OMP algorithm to the analysis coefficients from nine different dictionary transforms. The first four are bases, with errors shown in dashed lines, and the next five are frames generated from six filters, with all but the spectral graph wavelet frame being Parseval frames. The final image shows the same error for tight frames with different numbers of uniform translates of the Hann kernel as the filters. 
}
}\label{Fig:compression_sensor}
\end{figure*}

\change{
\subsection{Design considerations revisited}
We use five different test signals in this section: the piecewise-smooth signal on a sensor network shown in Fig. \ref{Fig:compression_sensor} \cite{shuman_TSP_multiscale}; the piecewise-smooth signal on the bunny graph shown in Fig. \ref{Fig:equivalence} \cite{shuman_TSP_multiscale}; the piecewise constant signal on the Minnesota road network from \cite{narang_bipartite_prod}; the average of 292 fMRI signals on the cerebellum graph \cite{behjat2015anatomically,behjat2016signal}; and the average temperatures for March 2018 at $N=469,404$ locations in the United States on an eight-neighbor local graph from \cite{li_mcsfb_2018}. We preprocess all signals by normalizing them to have mean zero.
}

\change{
\subsubsection{ 
How should we design the filters? 
Are there dictionary metrics that 
can inform this decision?} 
In Table \ref{Ta:denoising}, we examine the denoising performance for different graph signals, noise levels ($\sigma_{{\bf f}}$ is the standard deviation of the signal values and $\sigma$ is the standard deviation of the noise values), and filter designs. For each filter design type, we report the best SNR improvement 
over the range of 3-12 filters, each localized to every vertex in the graph to create the dictionary. In Fig. \ref{Fig:compression_sensor}, we use OMP to compress a piecewise-smooth signal on a sensor network with nine different dictionaries. For both of these application examples, we use exact computations throughout.

What are the design takeaways from these examples? First, there is no clear ``best'' filter design method across applications and setup parameters, which is not surprising but worth stating. Second, using redundant dictionaries generally enables sparser representations of the data and better compression performance than the bases considered in this example. Third, with exact computations, there is not a significant drawback from using a non-tight frame; e.g., the spectral graph wavelets have excellent performance in many of the denoising setups and the compression example. Fourth, in this setting, the cumulative coherence is not necessarily a good prediction of compression (sparse approximation) performance. For example, $\mu_1(25)$ is equal to 24.96,  24.77, 24.57, 15.71, 13.12, and 4.12 for the translated Hann wavelets, uniform Hann translates, spectral graph wavelets, uniform DCT filters, fast tight wavelet frame, and $M$-channel critically sampled filter bank, respectively; that is, higher, not lower, cumulative coherence is correlated with better performance. Indeed, the identification of dictionary metrics that correlate with application performance metrics is very much an open issue.
}

\begin{table}[b]
\centering
{\footnotesize
\tabcolsep=0.045cm
\begin{tabular}{l|cccc|cccc|cccc|cccc}
\cline{2-17}
 & \multicolumn{4}{ c| }{Sensor Network} & \multicolumn{4}{ c| }{Bunny} & \multicolumn{4}{ c| }{Minnesota} & \multicolumn{4}{ c }{Cerebellum}\\ 
\cline{2-17}
\multicolumn{1}{r|}{$\sigma$/$\sigma_{\bf f}$} & 1/8 & 1/4 & 1/2 & 1 & 1/8 & 1/4 & 1/2 & 1& 1/8 & 1/4 & 1/2 & 1 & 1/8 & 1/4 & 1/2 & 1 \\ 
\clineB{1-17 \vspace{.02in} }{2.5}
\multicolumn{1}{ l| }{SGWT} & 7 & 12 & 8 & 10 & 9 & 10  & 8  & 8 & 12 & 6 & 6 & 12  & 4 & 4  & 8   & 9 \\  
\multicolumn{1}{ l| }{CDF 9/7} & 12 & 8 & 10 & 6 & 7 & 12 & 9 & 6 & 9 & 8 & 12 & 9  & 12  & 9   & 12  & 5 \\
\cline{1-17}
\multicolumn{1}{ l| }{UnifIdeal}  & 6 &  10  & 11  & 12  & 6 & 7 & 11 & 12  & 9   & 12  &12  & 12  & 11  & 11  & 11  & 11 \\ 
\multicolumn{1}{ l| }{WavIdeal} & 4 & 8  & 4  & 6  & 6 & 11 & 9 & 6 & 9 & 8 & 12 & 9  & 12 & 9  & 6   & 5 \\
\multicolumn{1}{ l| }{UnifDCT} &  4 & 6 & 6  & 11  & 4 & 8 & 12 & 12 & 12 & 12 & 12 & 11  & 12 & 12  & 12 & 12 \\
\multicolumn{1}{ l| }{WavDCT} &  6 & 12 & 6  & 12  & 6 & 8 & 6 & 8 &  4 & 6 & 6 & 10  & 10 &  12 & 12  & 12 \\
\multicolumn{1}{ l| }{UnifTr}        &  4 & 6  & 6  & 10  & 4 & 5 & 8 & 12 & 10 & 12 & 12 & 12  & 11  & 11  & 12  & 9 \\
\multicolumn{1}{ l| }{WavTr}       &  3 & 4 & 12  & 4  & 4 & 4 & 5 & 6 &  3 & 4  & 5 & 9  & 7  & 11  & 12  & 12 \\
\cline{1-17}
\multicolumn{1}{ l| }{SpAUnifTr} & 3 & 4 & 6 & 7  & 5 & 5 & 9 & 11 & 6 & 11 & 12 & 12  & 10 & 12  & 10  & 11 \\ 
 \multicolumn{1}{ l| }{SpAWavTr} & 4 & 4 & 7  & 4 & 4 & 4 & 4 & 7 & 4 & 4 & 4 & 5  &  12 & 12  & 12  & 9 \\
 \cline{1-17}
\multicolumn{1}{ l| }{SigAUnif} &  &  &  &  &  &  &  &  &  &  &  &   &  12 &  12 &  12 & 12 \\
 \clineB{1-17}{2.5}
\end{tabular}
\vspace{.25cm}
}
\caption{\change{Number of filters that achieves the best denoising result shown in the corresponding entry in Table \ref{Ta:denoising}. The range considered for each design is 3 to 12 filters.}} 
\label{Ta:denoising_num_filters}
\vspace{-.5cm}
\end{table}

\begin{figure}[b]
\centering
\includegraphics[width=.8\linewidth]{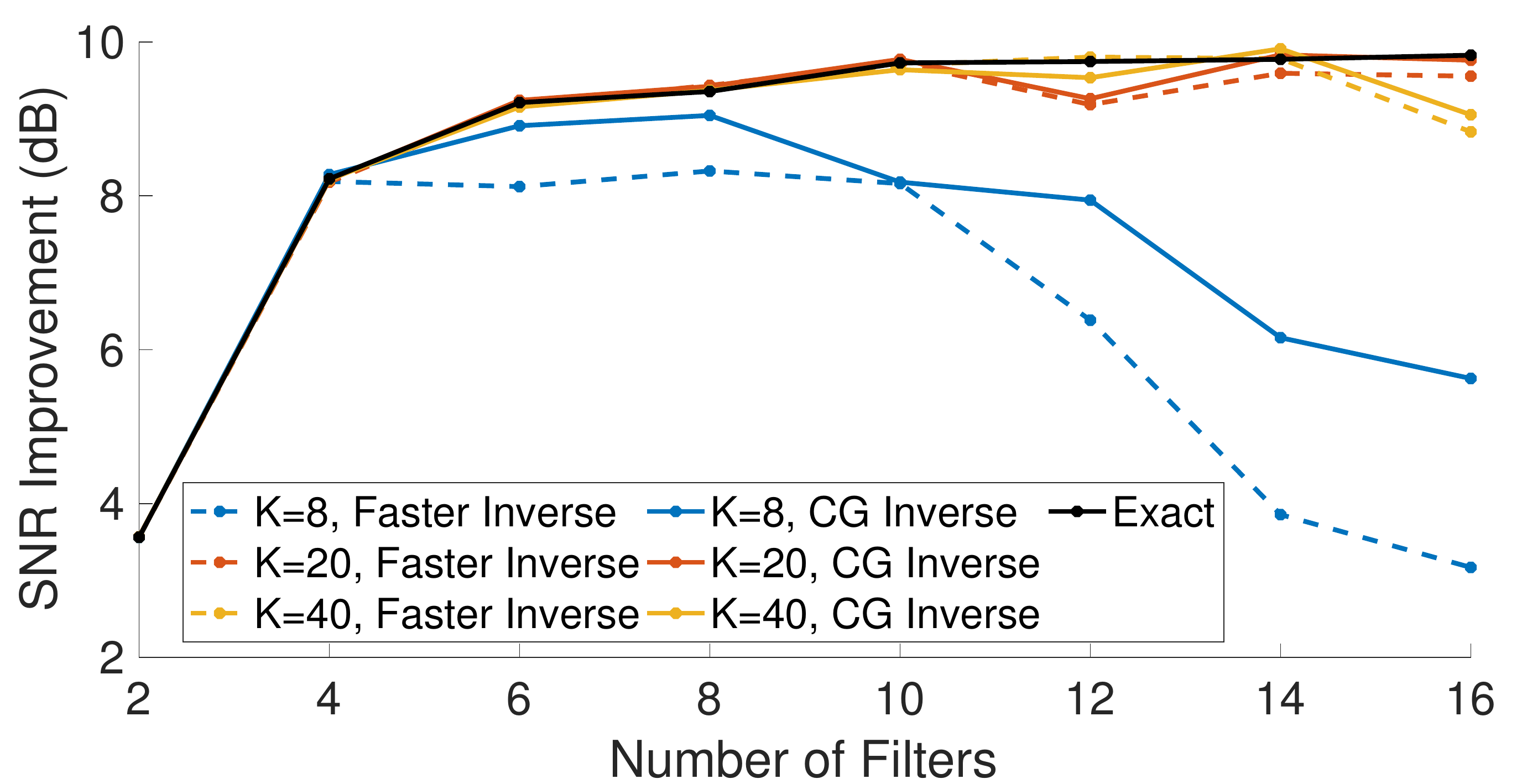} 
\vspace{-.1in}
\caption{\change{The wider filters arising in the filter designs with a lower number of filters are more amenable to approximation by low order polynomials, which in turn keeps the ratio of frame bounds closer to 1 and can lead to better reconstruction and denoising performance when fast numerical approximations are used.}}\label{Fig:denoising_num_filters}
\end{figure}

\change{
\subsubsection{How should we select the number of filters?}
There is not an easy answer for best practices in choosing the number of filters in an LSGFD. For small and medium graphs where exact filtering is tractable, the choice that yields the best performance usually depends on the specific application and signal model. For example, in Table \ref{Ta:denoising_num_filters}, for each type of filter design, we provide the number of filters that leads to the best performance in the corresponding entry of the denoising results of Table \ref{Ta:denoising}. The best number of filters to use varies widely depending on the type of filters, level of noise, and underlying graph signal, most likely according to how well some of the resulting atoms can capture the energy of the signal. Similarly, the last image in Fig. \ref{Fig:compression_sensor} shows that the choice of the number of filters in the compression of the sensor signal does not have an outsized impact when using an exactly computed dictionary generated (in this case, based on octave-band translates of a Hann kernel).

The drawbacks of using more filters are exacerbated when the data resides on a large graph, necessitating approximations for computational tractability. In Fig. \ref{Fig:denoising_num_filters}, we show the denoising performance on the bunny graph signal of tight wavelet frames generated from different numbers of uniform translates. With exact computations, the performance continues to slowly increase with additional filters (although it has nearly saturated at  $J=10$ filters); however, as the number of filters increases, they become narrower and more difficult to approximate by low degree polynomials, leading to worse performance with approximate computations. This is part of the reason a general rule of thumb in practice has been to use 4-8 filters.  
}
 
 \begin{figure*}[tb]
\begin{minipage}[m]{0.016\linewidth}
\centering
\vspace{.14in}
\includegraphics[width=\linewidth, page=3]{figures/compression_temp_bars} 
\end{minipage}
\begin{minipage}[m]{0.185\linewidth}
\centering
\centerline{\footnotesize{Original signal~~~}}\vspace{.12in}
\includegraphics[width=\linewidth, page=1]{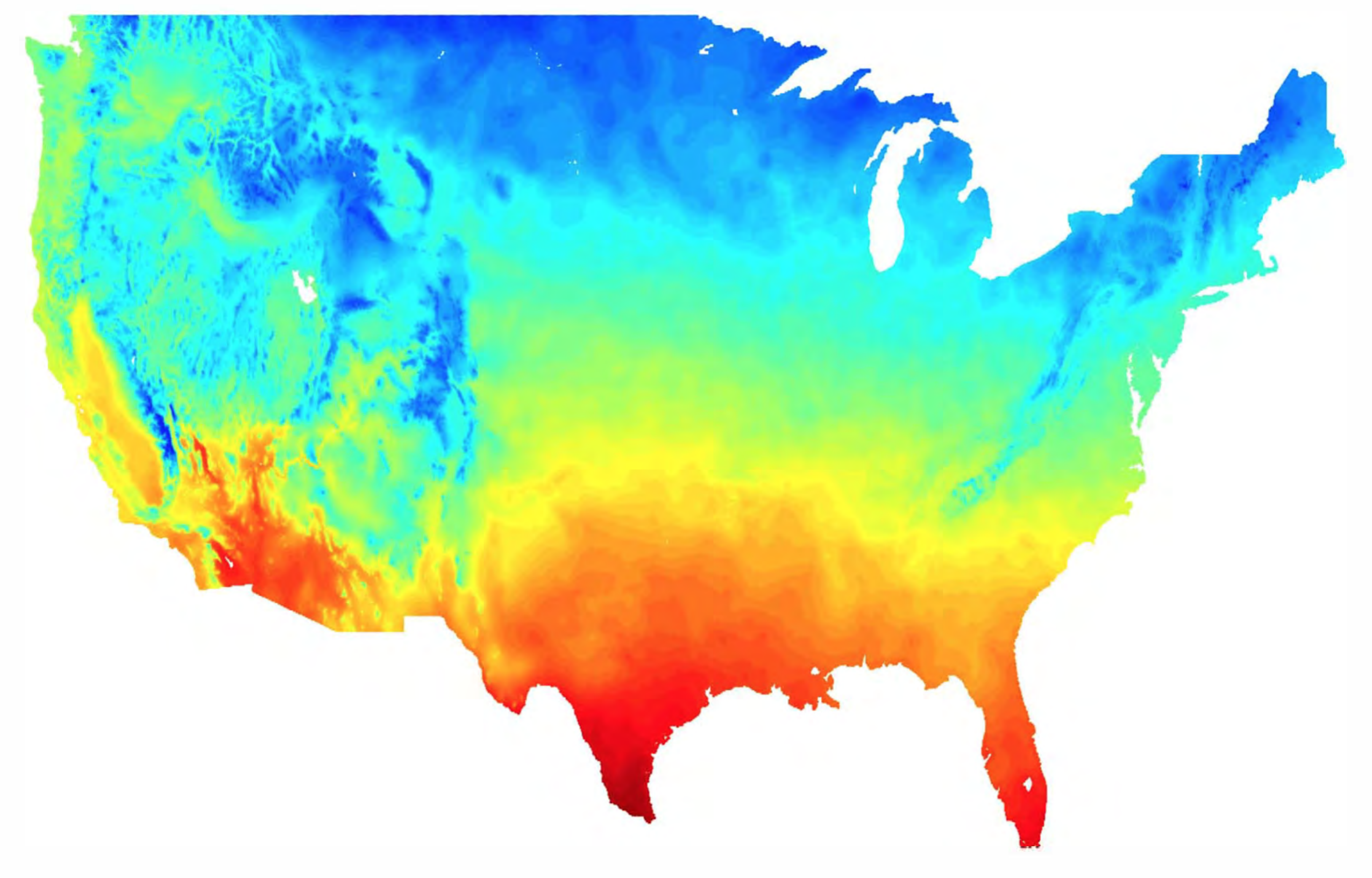} 
\end{minipage}
\begin{minipage}[m]{0.185\linewidth}
\centering
\centerline{\footnotesize{Reconstruction: Spectrum-~}}
\vspace{-.06in}
\centerline{\footnotesize{adapted tight wavelet frame~}} \vspace{.02in}
\includegraphics[width=\linewidth, page=5]{figures/compression_temp} 
\end{minipage}
\begin{minipage}[m]{0.185\linewidth}
\centering
\centerline{\footnotesize{Reconstruction: M-channel~}}
\vspace{-.06in}
\centerline{\footnotesize{critically sampled filter bank~}} \vspace{.02in}
\includegraphics[width=\linewidth, page=3]{figures/compression_temp} 
\end{minipage}
\begin{minipage}[m]{0.185\linewidth}
\centering
\centerline{\footnotesize{Error: Spectrum-}}
\vspace{-.06in}
\centerline{\footnotesize{adapted tight wavelet frame}} \vspace{.02in}
\includegraphics[width=\linewidth, page=4]{figures/compression_temp} 
\end{minipage}
\begin{minipage}[m]{0.185\linewidth}
\centering
\centerline{\footnotesize{Error: M-channel}}
\vspace{-.06in}
\centerline{\footnotesize{critically sampled filter bank}} \vspace{.02in}
\includegraphics[width=\linewidth, page=2]{figures/compression_temp} 
\end{minipage}
\begin{minipage}[m]{0.017\linewidth}
\centering
\vspace{.2in}
\includegraphics[width=\linewidth, page=2]{figures/compression_temp_bars} 
\end{minipage}
\caption{\change{Compression of the average temperature signal from \cite{li_mcsfb_2018}: the initial signal, the reconstructions from $\frac{N}{2}=234,702$ analysis coefficients of two different dictionaries, and the absolute values of the corresponding reconstruction errors. The NMSEs are 0.099 and 0.001 for the redundant tight wavelet frame and the critically sampled filter bank dictionary, respectively.}}\label{Fig:compression_temp}
\end{figure*}

 \change{
\subsubsection{When is it beneficial to use fewer center vertices?}
It appears to be most beneficial to include fewer center vertices in applications where either (i) memory is a critical issue, or (ii) the graph signals under consideration reside on very large, sparse graphs, necessitating approximations at all stages of the signal processing or machine learning pipelines. As an example, in Fig. \ref{Fig:compression_temp}, we compress the average temperature signal, reconstructing it from $\frac{N}{2}$ of the analysis coefficients. We compare the hard thresholding method using a five filter spectrum-adapted tight wavelet dictionary with complete sampling ($\V_j=\V$ for all $j$) to the band-by-band reconstruction method outlined in Section \ref{Se:band_by_band}, using the 5-channel critically sampled filter bank ($M$-CSFB) design detailed in \cite{li_mcsfb_2018} with the center vertices for each filter selected via signal-adapted nonuniform random sampling. The hard thresholding method tends to keep the coefficients associated with the scaling functions ($j=1$) and center vertices in the northern and southern parts of the country. Combined with the fact that the atoms are localized within $K=50$ hops of the center vertices due to the necessary polynomial approximation of the filters on a graph this size, it incurs more reconstruction error in areas of the graph where either the center vertices were not selected or there are sharper changes in the signal values, which would be captured by the discarded wavelet coefficients ($j>1$). Although it is possible to mitigate this issue to some extent via more clever reconstruction methods, the problem also becomes more pronounced as the compression ratio increases and fewer analysis coefficients are stored. The $M$-CSFB dictionary, on the other hand, only has 28,022 atoms generated from the scaling filter, leading to many of the wavelet coefficients being used in the reconstruction. Moreover, the random sampling selection of center vertices leads to preserved coefficients associated with atoms centered across the country. A  
high-level difference between these two approaches is that the $M$-CSFB already includes some of the compression in the process of choosing the strict subsets $\V_j \subset \V$ to be the center vertices.
}

\change{
\section{Summary and Future Directions}\label{Se:conclusion}
In summary, localized spectral graph filter frames feature structured atoms with analytically tractable properties such as localization around the center vertex and spectral patterns that carry a notion of smoothness with respect to the graph. At the same time, efficient numerical approximations exist to the forward and inverse transforms, rendering these dictionaries useful in myriad applications on large, sparse graphs. For small and medium graphs, it is typical to perform an exact eigendecomposition, localize each filter to every vertex, and choose the filters to satisfy the condition of Theorem \ref{Th:frame}, guaranteeing that the frame is tight. For large, sparse graphs, we reviewed fast techniques to approximate or bound the maximum graph Laplacian eigenvalue and the cumulative spectral density function, and subsequently, how to  leverage these approximations in the design and application of the spectral filters, the selection of center vertices via signal-adapted non-uniform random sampling, and fast reconstruction methods. 

Open issues and future directions in this line of research include:
\begin{enumerate}
\item Continued investigation of algorithms that use LSGFDs to efficiently extract information in the context of specific application domains and signal processing and machine learning tasks; as just one example, \cite{shuman_SSL_SAMPTA_2011} begins to investigate how to leverage the structured sparsity of LSGFD analysis coefficients to recover 
piecewise-smooth graph signals in the semi-supervised learning problem.
\item New connections between theoretical metrics, dictionary design, and applications. Many unanswered or partially answered questions remain on this front: What mathematical classes of graph signals are sparsely represented by LSGFDs with specific spectral patterns? Are there easily computable metrics on the dictionary that are good predictors of performance in application tasks, as demonstrated either empirically or via mathematical analysis? In what applications is it most beneficial to adapt the filters to the spectrum or the energy density of training signals?
\item The development of faster greedy or hybrid greedy/random graph sampling methods that are applicable to both smooth and non-smooth graph signals, as well as accompanying fast, scalable interpolation methods.
\item Extensions of the dictionary design principles reviewed here to the settings of data on directed graphs, time series data on graphs, and deep learning / convolutional neural networks on graphs.
\end{enumerate}
}

\section*{Acknowledgments}
\addcontentsline{toc}{section}{Acknowledgments}
The author would like to thank the anonymous reviewers and Hamid Behjat for constructive feedback on earlier versions of this article.

\bibliographystyle{IEEEtran}
\bibliography{lsgff_refs}

\begin{thebibliography}{10}
\providecommand{\url}[1]{#1}
\csname url@samestyle\endcsname
\providecommand{\newblock}{\relax}
\providecommand{\bibinfo}[2]{#2}
\providecommand{\BIBentrySTDinterwordspacing}{\spaceskip=0pt\relax}
\providecommand{\BIBentryALTinterwordstretchfactor}{4}
\providecommand{\BIBentryALTinterwordspacing}{\spaceskip=\fontdimen2\font plus
\BIBentryALTinterwordstretchfactor\fontdimen3\font minus
  \fontdimen4\font\relax}
\providecommand{\BIBforeignlanguage}[2]{{%
\expandafter\ifx\csname l@#1\endcsname\relax
\typeout{** WARNING: IEEEtran.bst: No hyphenation pattern has been}%
\typeout{** loaded for the language `#1'. Using the pattern for}%
\typeout{** the default language instead.}%
\else
\language=\csname l@#1\endcsname
\fi
#2}}
\providecommand{\BIBdecl}{\relax}
\BIBdecl

\bibitem{shuman2013emerging}
{D. I Shuman}, S.~K. Narang, P.~Frossard, A.~Ortega, and P.~Vandergheynst,
  ``The emerging field of signal processing on graphs: {E}xtending
  high-dimensional data analysis to networks and other irregular domains,''
  \emph{IEEE Signal Process. Mag.}, vol.~30, no.~3, pp. 83--98, May 2013.

\bibitem{ortega2018graph}
A.~Ortega, P.~Frossard, J.~Kova{\v{c}}evi{\'c}, J.~M. Moura, and
  P.~Vandergheynst, ``Graph signal processin{g: O}verview, challenges, and
  applications,'' \emph{Proc. IEEE}, vol. 106, no.~5, pp. 808--828, 2018.

\bibitem{shuman2015vertex}
{D. I Shuman}, B.~Ricaud, and P.~Vandergheynst, ``Vertex-frequency analysis on
  graphs,'' \emph{Appl. Comput. Harmon. Anal.}, vol.~40, no.~2, pp. 260--291,
  Mar. 2016.

\bibitem{Crovella2003}
M.~Crovella and E.~Kolaczyk, ``Graph wavelets for spatial traffic analysis,''
  in \emph{Proc. {IEEE INFOCOM}}, vol.~3, Mar. 2003, pp. 1848--1857.

\bibitem{wang}
W.~Wang and K.~Ramchandran, ``Random multiresolution representations for
  arbitrary sensor network graphs,'' in \emph{Proc. IEEE Int. Conf. Acc.,
  Speech, and Signal Process.}, vol.~4, May 2006, pp. 161--164.

\bibitem{gavish}
M.~Gavish, B.~Nadler, and R.~R. Coifman, ``Multiscale wavelets on trees, graphs
  and high dimensional data: {T}heory and applications to semi supervised
  learning,'' in \emph{Proc. Int. Conf. Mach. Learn.}, Jun. 2010, pp. 367--374.

\bibitem{narang_lifting_graphs}
S.~K. Narang and A.~Ortega, ``Lifting based wavelet transforms on graphs,'' in
  \emph{Proc. {APSIPA ASC}}, Oct. 2009, pp. 441--444.

\bibitem{szlam}
A.~D. {Szlam}, M.~{Maggioni}, R.~R. {Coifman}, and J.~C. {Bremer}, Jr.,
  ``{Diffusion-driven multiscale analysis on manifolds and graphs: top-down and
  bottom-up constructions},'' in \emph{Proc. {SPIE} Wavelets}, vol. 5914, Aug.
  2005, pp. 445--455.

\bibitem{irion}
J.~Irion and N.~Saito, ``Hierarchical graph {Laplacian} eigen transforms,''
  \emph{JSIAM Letters}, vol.~6, pp. 21--24, 2014.

\bibitem{diffusion_wavelets}
R.~R. Coifman and M.~Maggioni, ``Diffusion wavelets,'' \emph{Appl. Comput.
  Harmon. Anal.}, vol.~21, no.~1, pp. 53--94, 2006.

\bibitem{hammond2011wavelets}
D.~K. Hammond, P.~Vandergheynst, and R.~Gribonval, ``Wavelets on graphs via
  spectral graph theory,'' \emph{Appl. Comput. Harmon. Anal.}, vol.~30, no.~2,
  pp. 129--150, Mar. 2011.

\bibitem{shuman_TSP_multiscale}
{D. I Shuman}, M.~J. Faraji, and P.~Vandergheynst, ``A multiscale pyramid
  transform for graph signals,'' \emph{IEEE Trans. Signal Process.}, vol.~64,
  no.~8, pp. 2119--2134, Apr. 2016.

\bibitem{narang_icip}
S.~K. Narang and A.~Ortega, ``Local two-channel critically sampled filter-banks
  on graphs,'' in \emph{Proc. Int. Conf. Image Process.}, Sep. 2010, pp.
  333--336.

\bibitem{narang_bipartite_prod}
------, ``Perfect reconstruction two-channel wavelet filter banks for graph
  structured data,'' \emph{IEEE Trans. Signal Process.}, vol.~60, no.~6, pp.
  2786--2799, Jun. 2012.

\bibitem{ekambaram2013critically}
V.~N. Ekambaram, G.~Fanti, B.~Ayazifar, and K.~Ramchandran,
  ``Critically-sampled perfect-reconstruction spline-wavelet filterbanks for
  graph signals,'' in \emph{Proc. IEEE Glob. Conf. Signal and Inform.
  Process.}, 2013, pp. 475--478.

\bibitem{li_mcsfb_2018}
S.~Li, Y.~Jin, and {D. I Shuman}, ``Scalable {$M$}-channel critically sampled
  filter banks for graph signals,'' \emph{IEEE Trans. Signal Process.},
  vol.~67, no.~15, pp. 3954--3969, Aug. 2019.

\bibitem{narang_bior_filters}
S.~K. Narang and A.~Ortega, ``Compact support biorthogonal wavelet filterbanks
  for arbitrary undirected graphs,'' \emph{IEEE Trans. Signal Process.},
  vol.~61, no.~19, pp. 4673--4685, Oct. 2013.

\bibitem{tay2015design}
D.~B. Tay and Z.~Lin, ``Design of near orthogonal graph filter banks,''
  \emph{IEEE Signal Process. Lett.}, vol.~22, no.~6, pp. 701--704, Jun. 2015.

\bibitem{chen2015discrete}
S.~Chen, R.~Varma, A.~Sandryhaila, and J.~Kova\v{c}evi{\'{c}}, ``Discrete
  signal processing on graphs: {S}ampling theory,'' \emph{IEEE Trans. Signal
  Process.}, vol.~63, no.~24, pp. 6510--6523, Dec. 2015.

\bibitem{sakiyama2016spectral}
A.~Sakiyama, K.~Watanabe, and Y.~Tanaka, ``Spectral graph wavelets and filter
  banks with low approximation error,'' \emph{IEEE Trans. Signal Inf. Process.
  Netw.}, vol.~2, no.~3, pp. 230--245, 2016.

\bibitem{anis2017critical}
A.~Anis and A.~Ortega, ``Critical sampling for wavelet filterbanks on arbitrary
  graphs,'' in \emph{Proc. IEEE Int. Conf. Acoust., Speech and Signal
  Process.}, 2017, pp. 3889--3893.

\bibitem{tay2017critically}
D.~B. Tay, Y.~Tanaka, and A.~Sakiyama, ``Critically sampled graph filter banks
  with polynomial filters from regular domain filter banks,'' \emph{Signal
  Process.}, vol. 131, pp. 66--72, 2017.

\bibitem{tay2017almost}
------, ``Almost tight spectral graph wavelets with polynomial filters,''
  \emph{IEEE J. Sel. Topics Signal Process.}, vol.~11, no.~6, pp. 812--824,
  2017.

\bibitem{teke2016extending}
O.~Teke and P.~P. Vaidyanathan, ``Extending classical multirate signal
  processing theory to graphs -- {Part II: M-c}hannel filter banks,''
  \emph{IEEE Trans. Signal Process.}, vol.~65, no.~2, pp. 423--437, 2016.

\bibitem{kotzagiannidis2017splines}
M.~S. Kotzagiannidis and P.~L. Dragotti, ``Splines and wavelets on circulant
  graphs,'' \emph{Appl. Comput. Harmon. Anal.}, vol.~47, no.~2, pp. 481--515,
  Sep. 2019.

\bibitem{pesenson_splines}
I.~Pesenson, ``Variational splines and {Paley-Wiener} spaces on combinatorial
  graphs,'' \emph{Constr. Approx.}, vol.~29, no.~1, pp. 1--21, Feb. 2009.

\bibitem{erb2019graph}
W.~Erb, ``Graph signal interpolation with positive definite graph basis
  functions,'' \emph{arXiv preprint arXiv:1912.02069}, 2019.

\bibitem{thanou_learning_TSP_2014}
D.~Thanou, {D. I Shuman}, and P.~Frossard, ``Learning parametric dictionaries
  for signals on graphs,'' \emph{IEEE Trans. Signal Process.}, vol.~62, no.~15,
  pp. 3849--3862, Aug. 2014.

\bibitem{behjat2016signal}
H.~Behjat, U.~Richter, D.~Van De~Ville, and L.~S{\"o}rnmo, ``Signal-adapted
  tight frames on graphs,'' \emph{IEEE Trans. Signal Process.}, vol.~64,
  no.~22, pp. 6017--6029, Nov. 2016.

\bibitem{leonardi_multislice}
N.~Leonardi and D.~{Van De Ville}, ``Tight wavelet frames on multislice
  graphs,'' \emph{IEEE Trans. Signal Process.}, vol.~61, no.~13, pp.
  3357--3367, Jul. 2013.

\bibitem{shuman2013spectrum}
{D. I Shuman}, C.~Wiesmeyr, N.~Holighaus, and P.~Vandergheynst,
  ``Spectrum-adapted tight graph wavelet and vertex-frequency frames,''
  \emph{IEEE Trans. Signal Process.}, vol.~63, no.~16, pp. 4223--4235, Aug.
  2015.

\bibitem{gobel2018construction}
F.~G{\"o}bel, G.~Blanchard, and U.~von Luxburg, ``Construction of tight frames
  on graphs and application to denoising,'' in \emph{Handbook of Big Data
  Analytics}.\hskip 1em plus 0.5em minus 0.4em\relax Springer, 2018, pp.
  503--522.

\bibitem{dong2017sparse}
B.~Dong, ``Sparse representation on graphs by tight wavelet frames and
  applications,'' \emph{Appl. Comput. Harmon. Anal.}, vol.~42, no.~3, pp.
  452--479, 2017.

\bibitem{behjat2013statistical}
H.~Behjat, N.~Leonardi, and D.~Van De~Ville, ``Statistical parametric mapping
  of functional{ MRI }data using wavelets adapted to the cerebral cortex,'' in
  \emph{Proc. IEEE Int. Symp. Biomed. Imag.}, 2013, pp. 1070--1073.

\bibitem{behjat2014canonical}
H.~Behjat, N.~Leonardi, L.~S{\"o}rnmo, and D.~Van De~Ville, ``Canonical
  cerebellar graph wavelets and their application to {fMRI} activation
  mapping,'' in \emph{Proc. IEEE Int. Conf. Eng. Med. Biol. Soc.}, 2014, pp.
  1039--1042.

\bibitem{behjat2015anatomically}
------, ``Anatomically-adapted graph wavelets for improved group-level {fMRI}
  activation mapping,'' \emph{NeuroImage}, vol. 123, pp. 185--199, 2015.

\bibitem{tremblay2014graph}
N.~Tremblay and P.~Borgnat, ``Graph wavelets for multiscale community mining,''
  \emph{IEEE Trans. Signal Process.}, vol.~62, pp. 5227--5239, 2014.

\bibitem{saitoNatural}
N.~Saito, ``The first steps toward building natural graph wavelets.''\hskip 1em
  plus 0.5em minus 0.4em\relax Presented at the Graph Signal Processing
  Workshop, Jun. 2019, \url{http://math.ucdavis.edu/\textasciitilde
  saito/talks/gsp19.pdf}.

\bibitem{bunny}
{Stanford University Computer Graphics Laboratory}, ``{The Stanford 3D Scanning
  Repository},'' http://graphics.stanford.edu/data/3Dscanrep/.

\bibitem{dong2019learning}
X.~Dong, D.~Thanou, M.~Rabbat, and P.~Frossard, ``Learning graphs from data:
  {A} signal representation perspective,'' \emph{IEEE Signal Process. Mag.},
  vol.~36, no.~3, pp. 44--63, May 2019.

\bibitem{mateos2019connecting}
G.~Mateos, S.~Segarra, A.~G. Marques, and A.~Ribeiro, ``Connecting the dots:
  {I}dentifying network structure via graph signal processing,'' \emph{IEEE
  Signal Process. Mag.}, vol.~36, no.~3, pp. 16--43, May 2019.

\bibitem{lanczos_bound}
Y.~Zhou and R.~C. Li, ``Bounding the spectrum of large {H}ermitian matrices.''
  \emph{Linear Algebra Appl.}, vol. 435, no.~3, pp. 480--493, 2011.

\bibitem{anderson_morley}
W.~N. Anderson and T.~D. Morley, ``Eigenvalues of the {L}aplacian of a graph,''
  \emph{Linear Multilinear Algebra}, vol.~18, no.~2, pp. 141--145, 1985.

\bibitem{das}
K.~C. Das and R.~P. Bapat, ``A sharp upper bound on the largest {L}aplacian
  eigenvalue of weighted graphs,'' \emph{Lin. Alg. Appl.}, vol. 409, pp.
  153--165, Nov. 2005.

\bibitem{lin_spectral_density}
L.~Lin, Y.~Saad, and C.~Yang, ``Approximating spectral densities of large
  matrices,'' \emph{SIAM Review}, vol.~58, no.~1, pp. 34--65, 2016.

\bibitem{silver1996kernel}
R.~N. Silver, H.~R{\"o}der, A.~F. Voter, and J.~D. Kress, ``Kernel polynomial
  approximations for densities of states and spectral functions,'' \emph{J.
  Comput. Phys.}, vol. 124, no.~1, pp. 115--130, 1996.

\bibitem{gleich}
D.~Gleich, ``The {MatlabBGL} {Matlab} library,''
  http://www.cs.purdue.edu/homes/dgleich/packages/matlab{\_}bgl/index.html.

\bibitem{suitesparse}
T.~A. Davis and Y.~Hu, ``The {University of Florida} sparse matrix
  collection,'' \emph{ACM Trans. Math. Softw.}, vol.~38, no.~1, pp. 1:1--1:25,
  2011.

\bibitem{kovacevic_frames1}
J.~Kova\v{c}evi{\'{c}} and A.~Chebira, ``Life beyond bases: {T}he advent of
  frames (part {I}),'' \emph{IEEE Signal Process. Mag.}, vol.~24, no.~4, pp.
  86--104, Jul. 2007.

\bibitem{frames}
O.~Christensen, \emph{Frames and Bases}.\hskip 1em plus 0.5em minus 0.4em\relax
  Birkhauser, 2008.

\bibitem{handscomb}
J.~C. Mason and D.~C. Handscomb, \emph{Chebyshev Polynomials}.\hskip 1em plus
  0.5em minus 0.4em\relax Chapman and Hall, 2003.

\bibitem{atap}
L.~N. Trefethen, \emph{Approximation Theory and Approximation Practice}.\hskip
  1em plus 0.5em minus 0.4em\relax {SIAM}, 2013.

\bibitem{ward2018interpolating}
J.~P. Ward, F.~J. Narcowich, and J.~D. Ward, ``Interpolating splines on graphs
  for data science applications,'' \emph{arXiv preprint arXiv:1806.10695}, Oct.
  2018.

\bibitem{kondor2002diffusion}
R.~I. Kondor and J.~Lafferty, ``Diffusion kernels on graphs and other discrete
  structures,'' in \emph{Proc. Int. Conf. Mach. Learn.}, Jul. 2002, pp.
  315--22.

\bibitem{saito2018can}
N.~Saito, ``How can we naturally order and organize graph {Laplacian}
  eigenvectors?'' in \emph{Proc. IEEE Stat. Signal Process. Wkshp.}, 2018, pp.
  483--487.

\bibitem{cloninger2018dual}
A.~Cloninger and S.~Steinerberger, ``On the dual geometry of {Laplacian}
  eigenfunctions,'' \emph{Exp. Math.}, pp. 1--11, 2018.

\bibitem{li2019metrics}
H.~Li and N.~Saito, ``Metrics of graph {Laplacian} eigenvectors,'' in
  \emph{{SPIE} Wavelets and Sparsity}, Aug. 2019.

\bibitem{gspbox}
N.~Perraudin, J.~Paratte, {D. I Shuman}, V.~Kalofolias, P.~Vandergheynst, and
  D.~K. Hammond, ``{GSPBOX:} {A} toolbox for signal processing on graphs,''
  \url{https://lts2.epfl.ch/gsp/}.

\bibitem{shi2015infinite}
X.~Shi, H.~Feng, M.~Zhai, T.~Yang, and B.~Hu, ``Infinite impulse response graph
  filters in wireless sensor networks,'' \emph{IEEE Signal Process. Lett.},
  vol.~22, no.~8, pp. 1113--1117, Aug. 2015.

\bibitem{liu2018filter}
J.~Liu, E.~Isufi, and G.~Leus, ``Filter design for autoregressive moving
  average graph filters,'' \emph{IEEE Trans. Signal Inf. Process. Netw.},
  vol.~5, no.~1, pp. 47--60, 2019.

\bibitem{higham}
N.~J. Higham, \emph{Functions of Matrices}.\hskip 1em plus 0.5em minus
  0.4em\relax Society for Industrial and Applied Mathematics, 2008.

\bibitem{frommer}
A.~Frommer and V.~Simoncini, ``Matrix functions,'' in \emph{Model Order
  Reduction: Theory, Research Aspects and Applications}.\hskip 1em plus 0.5em
  minus 0.4em\relax Springer, 2008, pp. 275--303.

\bibitem{shuman_distributed_sipn}
{D. I Shuman}, P.~Vandergheynst, D.~Kressner, and P.~Frossard, ``Distributed
  signal processing via {Chebyshev} polynomial approximation,'' \emph{IEEE
  Trans. Signal Inf. Process. Netw.}, vol.~4, no.~4, pp. 736--751, 2018.

\bibitem{daubechies1992ten}
I.~Daubechies, \emph{Ten Lectures on Wavelets}.\hskip 1em plus 0.5em minus
  0.4em\relax Society for Industrial and Applied Mathematics, 1992, vol.~61.

\bibitem{druskin}
V.~L. Druskin and L.~A. Knizhnerman, ``Two polynomial methods of calculating
  functions of symmetric matrices,'' \emph{{U.S.S.R.} Comput. Maths. Math.
  Phys.}, vol.~29, no.~6, pp. 112--121, 1989.

\bibitem{di2016efficient}
E.~Di~Napoli, E.~Polizzi, and Y.~Saad, ``Efficient estimation of eigenvalue
  counts in an interval,'' \emph{Numer. Linear Algebra Appl.}, vol.~23, no.~4,
  pp. 674--692, Aug. 2016.

\bibitem{golub}
G.~H. Golub and C.~F. {Van Loan}, \emph{Matrix Computations}.\hskip 1em plus
  0.5em minus 0.4em\relax Johns Hopkins University Press, 2013.

\bibitem{loukas2015distributed}
A.~Loukas, A.~Simonetto, and G.~Leus, ``Distributed autoregressive moving
  average graph filters,'' \emph{IEEE Signal Process. Lett.}, vol.~22, no.~11,
  pp. 1931--1935, Nov. 2015.

\bibitem{teke2018energy}
O.~Teke and P.~Vaidyanathan, ``Energy compaction filters on graphs,'' in
  \emph{Proc. IEEE Glob. Conf. Signal and Inform. Process.}, 2018, pp.
  783--787.

\bibitem{fan_saop_ICASSP_2019}
L.~Fan, {D. I Shuman}, S.~Ubaru, and Y.~Saad, ``Spectrum-adapted polynomial
  approximation for matrix functions,'' in \emph{Proc. IEEE Int. Conf. Acc.,
  Speech, and Signal Process.}, May 2019.

\bibitem{behjat2019spectral}
H.~Behjat and D.~Van De~Ville, ``Spectral design of signal-adapted tight frames
  on graphs,'' in \emph{Vertex-Frequency Analysis of Graph Signals}.\hskip 1em
  plus 0.5em minus 0.4em\relax Springer, 2019, pp. 177--206.

\bibitem{anis2016efficient}
A.~Anis, A.~Gadde, and A.~Ortega, ``Efficient sampling set selection for
  bandlimited graph signals using graph spectral proxies,'' \emph{IEEE Trans.
  Signal Process.}, vol.~64, no.~14, pp. 3775--3789, Jul. 2016.

\bibitem{sakiyama2019eigendecomposition}
A.~Sakiyama, Y.~Tanaka, T.~Tanaka, and A.~Ortega, ``Eigendecomposition-free
  sampling set selection for graph signals,'' \emph{IEEE Trans. Signal
  Process.}, vol.~67, no.~10, pp. 2679--2692, Mar. 2019.

\bibitem{lorenzo2018sampling}
P.~{Di Lorenzo}, S.~Barbarossa, and P.~Banelli, ``Sampling and recovery of
  graph signals,'' in \emph{Cooperative and Graph Signal Processing}, 2018, pp.
  261--282.

\bibitem{tanaka2020sampling}
Y.~Tanaka, Y.~C. Eldar, A.~Ortega, and G.~Cheung, ``Sampling on graphs: {F}rom
  theory to applications,'' \emph{arXiv preprint arXiv:2003.03957}, 2020.

\bibitem{pesenson_paley}
I.~Pesenson, ``Sampling in {Paley-Wiener} spaces on combinatorial graphs,''
  \emph{Trans. Amer. Math. Soc}, vol. 360, no.~10, pp. 5603--5627, 2008.

\bibitem{PuyTGV15}
G.~Puy, N.~Tremblay, R.~Gribonval, and P.~Vandergheynst, ``Random sampling of
  bandlimited signals on graphs,'' \emph{Appl. Comput. Harmon. Anal.}, vol.~44,
  no.~2, pp. 446--475, Mar. 2018.

\bibitem{bai2019fast}
Y.~Bai, F.~Wang, G.~Cheung, Y.~Nakatsukasa, and W.~Gao, ``Fast graph sampling
  set selection using {G}ershgorin disc alignment,'' \emph{IEEE Trans. Signal
  Process.}, vol.~68, pp. 2419--2434, Mar. 2020.

\bibitem{drineas2012fast}
P.~Drineas, M.~Magdon-Ismail, M.~W. Mahoney, and D.~P. Woodruff, ``Fast
  approximation of matrix coherence and statistical leverage,'' \emph{J. Mach.
  Learn. Res.}, vol.~13, no. Dec., pp. 3475--3506, 2012.

\bibitem{mahoney2009cur}
M.~W. Mahoney and P.~Drineas, ``{CUR} matrix decompositions for improved data
  analysis,'' \emph{Proc. Natl. Acad. Sci.}, vol. 106, no.~3, pp. 697--702,
  2009.

\bibitem{tropp}
J.~A. Tropp, ``Greed is good: {A}lgorithmic results for sparse approximation,''
  \emph{IEEE. Trans. Inform. Theory}, vol.~50, no.~10, pp. 2231--2242, Oct.
  2004.

\bibitem{tsitsvero2016signals}
M.~Tsitsvero, S.~Barbarossa, and P.~Di~Lorenzo, ``Signals on graph{s:
  Un}certainty principle and sampling,'' \emph{IEEE Trans. Signal Process.},
  vol.~64, no.~18, pp. 4845--4860, 2016.

\bibitem{perraudin_uncertainty_APSIPA_2018}
N.~Perraudin, B.~Ricaud, {D. I Shuman}, and P.~Vandergheynst, ``Global and
  local uncertainty principles for signals on graphs,'' \emph{APSIPA Trans.
  Signal Inf. Process.}, Apr. 2018.

\bibitem{van2017slepian}
D.~Van De~Ville, R.~Demesmaeker, and M.~G. Preti, ``When {Slepian meets
  Fiedler: Putting} a focus on the graph spectrum,'' \emph{IEEE Signal Process.
  Lett.}, vol.~24, no.~7, pp. 1001--1004, 2017.

\bibitem{teke2017uncertainty}
O.~Teke and P.~P. Vaidyanathan, ``Uncertainty principles and sparse
  eigenvectors of graphs,'' \emph{IEEE Trans. Signal Process.}, vol.~65,
  no.~20, pp. 5406--5420, 2017.

\bibitem{erb2019shapes}
W.~Erb, ``Shapes of uncertainty in spectral graph theory,'' \emph{arXiv
  preprint arXiv:1909.10865}, Sep. 2019.

\bibitem{donoho_theory}
D.~L. Donoho, ``Unconditional bases are optimal bases for data compression and
  for statistical estimation,'' \emph{Appl. Comput. Harmon. Anal.}, vol.~1,
  no.~1, pp. 100--115, Dec. 1993.

\bibitem{chen2018multiresolution}
S.~Chen, A.~Singh, and J.~Kova{\v{c}}evi{\'c}, ``Multiresolution
  representations for piecewise-smooth signals on graphs,'' \emph{arXiv
  preprint arXiv:1803.02944}, Mar. 2018.

\bibitem{ricaud_sparsity_SPIE_2013}
B.~Ricaud, {D. I Shuman}, and P.~Vandergheynst, ``On the sparsity of wavelet
  coefficients for signals on graphs,'' in \emph{{SPIE} Wavelets and Sparsity},
  Aug. 2013.

\bibitem{de2019data}
B.~de~Loynes, F.~Navarro, and B.~Olivier, ``Data-driven thresholding in
  denoising with spectral graph wavelet transform,'' \emph{arXiv preprint
  arXiv:1906.01882}, 2019.

\bibitem{shuman_SSL_SAMPTA_2011}
{D. I Shuman}, M.~J. Faraji, and P.~Vandergheynst, ``Semi-supervised learning
  with spectral graph wavelets,'' in \emph{Proc. Int. Conf. Samp. Theory and
  Appl.}, May 2011.

\end{thebibliography}

\newpage

\begin{IEEEbiography}[{\includegraphics[width=1in,height=1.25in,clip,keepaspectratio]{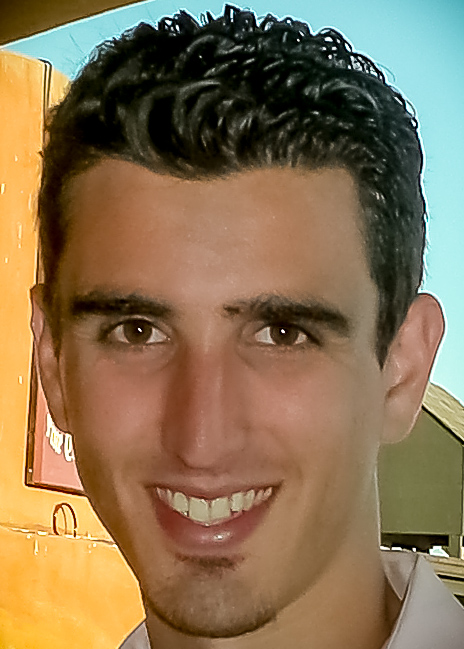}}]%
{David I Shuman}
received the B.A. degree in economics and the M.S. degree in engineering-economic systems and operations research from Stanford University, Stanford, CA, in 2001 and the M.S. degree in electrical engineering: systems, the M.S. degree in applied mathematics, and the Ph.D. degree in electrical engineering: systems from the University of Michigan, Ann Arbor, in 2006, 2009, and 2010, respectively.

He is currently an Associate Professor in the Department of Mathematics, Statistics, and Computer Science, Macalester College, St. Paul, Minnesota, which he joined in January 2014. From 2010 to 2013, he was a Postdoctoral Researcher at the Institute of Electrical Engineering, Ecole Polytechnique  F{\'e}d{\'e}rale de Lausanne (EPFL), Lausanne, Switzerland. His research interests include signal processing on graphs, computational harmonic analysis, and stochastic scheduling and resource allocation problems.

Dr. Shuman has served as an Associate Editor for the IEEE Transactions on Signal and Information Processing Over Networks (2019-) and the IEEE Signal Processing Letters (2017-2019), and on the Technical Program Committee for the IEEE Global Conference on Signal and Information Processing (2015-2018). He received the 2016 IEEE Signal Processing Magazine Best Paper Award, and was a 2014-2015 Project {NExT} Fellow of the Mathematical Association of America.
\end{IEEEbiography}


\end{document}